\documentclass[fleqn,usenatbib]{mnras}

\usepackage{newtxtext,newtxmath}

\usepackage[T1]{fontenc}

\usepackage[utf8]{inputenc}

\DeclareRobustCommand{\VAN}[3]{#2}
\let\VANthebibliography\thebibliography
\def\thebibliography{\DeclareRobustCommand{\VAN}[3]{##3}\VANthebibliography}

\usepackage{graphicx}	
\usepackage{amsmath}	
\usepackage{threeparttable}
\usepackage{outlines}
\usepackage{float}
\usepackage{orcidlink}

\title[Testing TDE models with rpTDEs]{Can tidal disruption event models reliably measure black hole masses?}

\author[C.~R.~Angus et al.]{C.~R.~Angus$^{1}$\orcidlink{0000-0002-4269-7999}\thanks{E-mail: c.angus@qub.ac.uk (CRA)},
A.~J.~Smith$^{1}$\orcidlink{0009-0007-9351-675X},
D.~Magill$^{1}$\orcidlink{0009-0000-6521-8842},
P.~Ramsden$^{2,1}$\orcidlink{0009-0009-2627-2884},
N.~Sarin$^{3,4}$\orcidlink{0000-0003-2700-1030},
M.~Nicholl$^{1}$\orcidlink{0000-0002-2555-3192},
B.~Mockler$^{5,6}$\orcidlink{0000-0001-6350-8168},
\newauthor
E.~Hammerstein$^{7}$\orcidlink{0000-0002-5698-8703},
R.~Stein$^{8,9,10}$\orcidlink{0000-0003-2434-0387},
Y.~Yao$^{11,7}$\orcidlink{0000-0001-6747-8509},
T.~de Boer$^{12}$\orcidlink{0000-0001-5486-2747},
K.~C.~Chambers$^{12}$\orcidlink{0000-0001-6965-7789},
M.~E.~Huber$^{12}$\orcidlink{0000-0003-1059-9603},
\newauthor
C.~-C.~Lin$^{12}$\orcidlink{0000-0002-7272-5129},
T.~B.~Lowe$^{12}$\orcidlink{0000-0002-9438-3617},
E.~A.~Magnier$^{12}$\orcidlink{0000-0002-7965-2815},
S.~J.~Smartt$^{13,1}$\orcidlink{0000-0002-8229-1731} and 
R.~J.~Wainscoat$^{12}$\orcidlink{0000-0002-1341-0952}
\\
$^{1}$ Astrophysics Research Centre, School of Mathematics and Physics, Queen’s University Belfast, Belfast BT7 1NN, UK\\
$^{2}$ School of Physics and Astronomy, University of Birmingham, Birmingham B15 2TT, UK\\
$^{3}$ Kavli Institute for Cosmology, University of Cambridge, Madingley Road, CB3 0HA, UK\\
$^{4}$ Institute of Astronomy, University of Cambridge, Madingley Road, CB3 0HA, UK\\
$^{5}$ The Observatories of the Carnegie Institute for Science, 813 Santa Barbara St., Pasadena, CA 91101, USA\\
$^{6}$ Department of Physics \& Astronomy, University of California, Davis, CA 95616, USA\\
$^{7}$ Department of Astronomy, University of California, Berkeley, CA 94720-3411, USA\\
$^{8}$ Department of Astronomy, University of Maryland, College Park, MD 20742, USA\\
$^{9}$ Joint Space-Science Institute, University of Maryland, College Park, MD 20742, USA\\
$^{10}$ Astrophysics Science Division, NASA Goddard Space Flight Center, Mail Code 661, Greenbelt, MD 20771, USA\\
$^{11}$ Miller Institute for Basic Research in Science, 206B Stanley Hall, Berkeley, CA 94720, USA\\
$^{12}$ Institute for Astronomy, University of Hawaii, 2680 Woodlawn Drive, Honolulu HI 96822\\
$^{13}$ Department of Physics, University of Oxford, Keble Road, Oxford, OX1 3RH, UK\\
}

\date{Accepted XXX. Received YYY; in original form ZZZ}

\pubyear{\the\year{}}

\begin{document}
\label{firstpage}
\pagerange{\pageref{firstpage}--\pageref{lastpage}}
\maketitle

\begin{abstract}
Tidal disruption event (TDE) light curves are increasingly used to infer the masses of quiescent supermassive black holes ($M_{\rm BH}$), offering a powerful probe of low-mass black hole demographics independent of host-galaxy scaling relations. However, the reliability of most semi-analytic TDE models assume full stellar disruption, despite theoretical expectations that partial disruptions dominate the TDE population. In this work we test the robustness of current TDE models using three repeating partial TDEs (rpTDEs), in which the multiple flares produced by the same surviving stellar core must yield consistent black hole masses. We present spectroscopic observations establishing AT\,2023adr as a rpTDE, making it the third such spectroscopically confirmed event. We independently model the flares of the three rpTDEs; 2020vdq, 2022dbl, and 2023adr, applying fallback-accretion fits, stream–stream collision scaling relations, luminosity-based empirical relations, and cooling-envelope fits. After accounting for statistical and model-specific systematics, we find that all TDE models generally return self-consistent $M_{\rm BH}$ values between flares, and are broadly consistent with host-galaxy $M_{\rm BH}$ proxies, recovering $M_{\rm BH}$ to within 0.3–0.5\,dex. However, the convergence of fallback models towards unphysical stellar masses and impact parameters reveals limitations in the existing fallback model grids. We also show that light curve coverage, particularly in the near-UV, is critical for constraining model parameters. This has direct implications for interpreting the thousands of TDE light curves expected from upcoming surveys such as the Rubin Observatory's Legacy Survey of Space and Time, where from simulations, we find that $M_{\rm BH}$ may be underestimated on average by 0.5\,dex without additional follow-up. 
\end{abstract}




\section{Introduction}


The growth and evolution of galaxies appears to be inextricably linked to the black holes that they host \citep{magorrian_demography_1998,ferrarese_fundamental_2000}, such that every massive galaxy is believed to contain a black hole at its centre. Despite this, our knowledge of the black hole population is severely impeded by our ability to detect them. With only a small fraction \citep[$\sim$10\%;][]{kewley_host_2006,aird_primus_2012} of galaxies in the local Universe hosting an Active Galactic Nucleus (AGN), the rest remain dormant and unseen. Tidal Disruption Events (TDEs), transients produced when an unfortunate star scattered within the tidal radius of a black hole \citep{hills_possible_1975}, can fleetingly illuminate hidden black holes. When this occurs, the black hole's gravitational force produces a strong gradient across the star, overwhelming its self gravity and producing streams of bound and unbound debris, resulting in a luminous flare. TDEs are thus encoded with information about the disrupting black hole, which if deciphered correctly, presents an opportunity to probe the black hole population in an unbiased, galaxy-independent way. 

TDEs emit across the electromagnetic spectrum, with detections spanning from the X-ray \citep[][]{komossa_x-ray_2001,auchettl_new_2017} to the radio at late times \citep[][]{cendes_ubiquitous_2024,dykaar_untargeted_2024}. However, the bulk of TDE identification takes place at optical wavelengths, with wide-field time-domain surveys such as the Zwicky Transient Facility \citep[ZTF; ][]{graham_zwicky_2019,bellm_zwicky_2019} enabling the systematic detection of new events, with over 100 TDEs discovered to date \citep[][]{van_velzen_seventeen_2021,gezari_tidal_2021,hammerstein_final_2023,yao_tidal_2023}, and multiple spectroscopic subgroups \citep{charalampopoulos_detailed_2022,hammerstein_final_2023}. 

From the optical light curve, we can potentially measure the mass, $M_{\rm BH}$ \citep[e.g. ][]{mockler_weighing_2019,ryu_measuring_2020} and spin of the disrupting black hole \citep[e.g. ][]{leloudas_superluminous_2016,mummery_maximum_2024}. 
However, the powering mechanisms behind the observed optical light curve are unclear. In canonical TDE models, energy is liberated from the stellar debris as it forms a compact accretion disk around the black hole, whose luminosity follows the fallback rate of material ($\propto\,t^{-5/3}$; \citealt{rees_tidal_1988,phinney_manifestations_1989}). This model requires efficient circularization and kinetic energy loss from the debris for the disk to form promptly \citep{rees_tidal_1988,dai_soft_2015,bonnerot_long-term_2017}. Alternatively, shocks between collisions of self-intersecting streams of debris may release energy and generate the observed TDE luminosities \citep{piran_disk_2015,jiang_prompt_2016}, possibly in combination with fallback accretion at later times \citep{steinberg_stream-disk_2024}. This picture is complicated further by the apparent presence of an optically thick reprocessing atmosphere \citep{loeb_optical_1997,guillochon_hydrodynamical_2013,roth_x-ray_2016,dai_unified_2018}, required to explain the large photospheric radii inferred from TDE spectral energy distributions (SEDs) and the weaker than expected soft X-rays from optically selected TDEs, both of which are inconsistent with hot accretion discs \citep[e.g.][]{arcavi_continuum_2014,hung_revisiting_2017,charalampopoulos_detailed_2022}. This reprocessing layer may be the result of outflows -- launched either promptly following accretion \citep{alexander_discovery_2016,van_velzen_radio_2016} or from material ejected by stream-stream collisions \citep{lu_self-intersection_2020,goodwin_radio-emitting_2023} -- or take the form of a pressure-supported envelope formed due to weakly bound stellar debris \citep{metzger_cooling_2022}. 

Currently these different theories remain in an observational deadlock. Though some TDEs have the $L\propto\,t^{-5/3}$ power-law light curve declines predicted by fallback models \citep{rees_tidal_1988,gezari_ps1-10jh_2015,brown_long_2017}, there is significant variation in the post-peak evolution of the population, with many showing shallower, or even exponential declines \citep{van_velzen_seventeen_2021,gezari_tidal_2021,hammerstein_final_2023,charalampopoulos_at_2023}, and some exhibiting rebrightenings at late times from the X-ray \citep[e.g.][]{gezari_x-ray_2017,wevers_live_2023} to the radio \citep[e.g.][]{malyali_rebrightening_2023,cendes_ubiquitous_2024}.
The absorption and re-emission of high-energy radiation produced by fallback accretion by an envelope at $\sim10^{14}$\,cm \citep{guillochon_hydrodynamical_2013} produces similar blackbody emission to colliding debris at the same radius \citep{piran_disk_2015}. And whilst compelling evidence for outflows has been observed in some TDEs \citep[e.g.][]{alexander_discovery_2016,nicholl_outflow_2020,goodwin_radio-emitting_2023}, in other events the implied outflow mass would exceed that of a typical star \citep{uno_application_2020}. 
Thus whilst there are now several publicly available semi-analytical models to constrain the physical parameters of TDEs from their optical light curves \citep{mockler_weighing_2019,kovacs-stermeczky_fitting_2023,sarin_tidal_2024} or from the scaling of these key parameters with the luminosity \citep{ryu_measuring_2020,mummery_maximum_2024}, it is unclear which of these models offers the most reliable way of probing the black hole properties. Recent work from \cite{mummery_tidal_2025} has challenged the key underlying assumptions of many TDE parameter inference frameworks. Comparing the observed scaling of radiated energy and peak luminosity with $M_{\rm BH}$ against predictions from emission models, they find that models in which the optical/UV light curve closely follows the fallback rate are strongly disfavoured, while reprocessing-based or disk-formation models provide a better, though still incomplete, description of the observed emission. Further to this, \cite{guolo_compact_2025} have shown that simultaneously fitting the late-time exposed accretion-disk SEDs with multi-band light curves yields black hole mass estimates more closely aligned with established host-galaxy scaling relations. 

Many models assume that a \lq\,full\rq\, disruption takes place, whereby the star is on a plunging orbit, whose pericentre lies within the tidal radius of the black hole, leaving around half the mass of the original star within the bound debris \citep[][]{hills_possible_1975,rees_tidal_1988,kochanek_aftermath_1994,guillochon_hydrodynamical_2013}. However, the majority of black hole encounters are predicted to be grazing, where the star wanders within the vicinity of the tidal radius, but does not cross it. These are partial TDEs (pTDEs), as only a fraction of star is disrupted, leaving a surviving core \citep{guillochon_hydrodynamical_2013}. Due to loss cone dynamics (the mechanisms through which stars are scattered within the region of parameter space where they can be disrupted), pTDEs should be at least as common as full ones \citep{stone_rates_2016,krolik_tidal_2020,chen_light_2021,zhong_revisit_2022}, potentially even exceeding the rate of full disruptions by a factor of ten \citep{bortolas_partial_2023}. 

Crucially, with only part of the stellar material stripped during a pTDE encounter, increased degeneracies are introduced between the key model parameters  \citep[$M_{\rm BH}$, stellar mass and impact parameter; ][]{guillochon_hydrodynamical_2013,mainetti_fine_2017}, making their accurate recovery with analytical models complex. Though the properties of pTDEs are predicted to deviate from full disruptions, potentially exhibiting multiple peaks \citep{chen_light_2021}, and faster decay rates in their post-peak evolution $\propto\,t^{-9/4}$ \citep{guillochon_hydrodynamical_2013,coughlin_partial_2019}, the dependency of the fallback rate on more realistic stellar structure \citep{lodato_stellar_2009,law-smith_stellar_2020}, combined with the effects of reprocessing \citep{roth_x-ray_2016,metzger_bright_2016} makes pTDEs difficult to observationally distinguish from full disruptions. Combined with the higher pTDE rate, this calls into question whether we are actually able to use TDEs as reliable probes of black hole properties. 

Only as the temporal baselines of transient surveys have extended to several years have they become long enough to begin identifying potential {\textit{repeating}} partial disruptions (rpTDEs), whereby the partially disrupted star's elliptical orbit brings the surviving stellar core back within the vicinity of the black hole on successive orbits \citep{kiroglu_partial_2023,liu_tidal_2023,liu_repeating_2025}, producing multiple flares from the same black hole. Candidate events have now been identified in both the X-ray \citep[e.g.][]{wevers_live_2023,liu_deciphering_2023,hampel_new_2022,malyali_rebrightening_2023,evans_monthly_2023,guolo_x-ray_2024}, and optical \citep[e.g.][]{payne_asassn-14ko_2021,somalwar_first_2025,lin_unluckiest_2024,hinkle_double_2024,sun_at2021aeuk_2025,bao_gleeoks_2024,makrygianni_double_2025,sun_at2021aeuk_2025,langis_repeating_2025}. Because the central black hole mass cannot change on such short (year–decade) timescales, each flare must encode the same black hole mass, regardless of differences in luminosity, temperature, or geometry between passages. This makes rpTDEs uniquely powerful for testing TDE models, as any robust modelling framework must recover a consistent $M_{\rm BH}$ when fitting each flare independently. Thus the ability of a TDE model to replicate $M_{\rm BH}$ when fitting the flares independently provides a strong indicator of model performance. Given their known partial nature, rpTDEs also allow us to test the reliability of parameters derived from TDE models -- usually designed with full disruptions in mind -- when describing partial disruptions. This is particularly important for upcoming time domain surveys, such as the Vera C. Rubin Observatory's Legacy Survey of Space and Time (LSST; \citealt{ivezic_lsst_2019}), which is set to discover tens of thousands of TDEs \citep{bricman_prospects_2020} during its lifetime. With less than 0.1\% of events expected to be spectroscopically classified \citep{villar_superraenn_2020}, the overwhelming majority of TDEs will be photometrically selected. Though a significant fraction of these will be targeted for host galaxy spectra \citep{soumagnac_most_2024,frohmaier_tides_2025} from which a virial measurement of $M_{\rm BH}$ may be obtained, there will still be a heavy reliance upon optical photometric modelling to derive this property. Further, black hole mass estimates derived directly from TDE light curves are particularly valuable for calibrating observed scaling relations between the central black hole and the properties of its host galaxy \citep[e.g][]{kormendy_coevolution_2013,reines_relations_2015,ramsden_bulge_2022}, particularly in the low-black hole mass regime where constraints on the slopes of these relations are less well constrained \citep{angus_fast-rising_2022}. Ensuring that light curve models can dependably reproduce TDE parameters will thus be essential for maximise the scientific output of these events.  

In this paper we test current TDE light curve models using the a sample of confirmed rpTDEs from the literature. Throughout this work we assume a standard $\Lambda$CDM cosmology of H$_{0}$=70\,km\,s$^{-1}$\,Mpc$^{-1}$, $\Omega_{M}$=0.3 and $\Omega_{\Lambda}$=0.7. 


\section{Repeating Partial TDE Sample}

Whilst there are a number of candidate rpTDEs that have been identified within the literature \citep[e.g.][]{wevers_live_2023,somalwar_first_2025,lin_unluckiest_2024,hinkle_double_2024,makrygianni_double_2025,langis_repeating_2025,quintin_lost_2025}, to ensure fair testing of the TDE light curve models, we select events which possess:

\begin{itemize}
    \item Spectroscopic confirmation of one, or both, peaks as TDEs.
    \item Temporally distinct peaks.
    \item Multi-band coverage of both peaks.
    \item A spectroscopic redshift. 
\end{itemize}

These criteria allow a fair and controlled test of TDE models, as they ensure well-constrained spectral energy distributions and enable clean, independent fits to each flare. However, they necessarily exclude many candidate rpTDEs that either lack multiband coverage for both peaks or do not exhibit temporally distinct flares - cases where the second brightening could instead be a re-brightening of the first \citep[e.g.][]{yao_tidal_2023,guo_reverberation_2025,zhong_modeling_2025}. Although such systems are not definitively ruled out as rpTDEs, their light-curve morphology prevents robust single-flare modelling \cite[although see ][for simultaneous modelling of re-brightening TDEs]{zhong_modeling_2025}. 

Thus in this work, we use two previously confirmed optical rpTDEs: TDE\,2020vdq \citep[previously AT\,2020vdq;][]{somalwar_first_2025} and TDE\,2022dbl \citep[previously AT\,2022dbl;][]{lin_unluckiest_2024,hinkle_double_2024,makrygianni_double_2025}. To these we add the repeating nuclear transient AT\,2023adr \cite[][hereafter TDE\,2023adr]{llamas_lanza_identification_2024, quintin_lost_2025}, whose TDE nature we spectroscopically confirm in the next section.

We note that we do not include the multiply flaring nuclear transient ASASSN-14ko in ESO 253-G003 \citep{payne_asassn-14ko_2021} within our analysis. Whilst ASASSN-14ko exhibits multiple peaks whose inferred temperatures and radii are consistent with the broad TDE population \citep{somalwar_first_2025}, and a decaying orbital period consistent with orbital energy losses following each disruption \citep{payne_asassn-14ko_2021}, the inferred mass of ESO 253-G003's black hole ($\log{M_{\mathrm{BH}}}$ = 7.8) and multiple flares requires the disruption of a higher mass, evolved star \citep{payne_asassn-14ko_2021,bandopadhyay_repeating_2024}. As the majority of TDE models assume the disruption of a low mass main sequence star (see Section \ref{sect:models}), we exclude ASASSN-14ko from our analysis. 

\subsection{TDE\,2023adr Classification}

TDE\,2023adr was first detected as an optical transient at RA=14:36:19.830, Dec=+32:23:16.48 by the Asteroid Terrestrial Impact Last Alert System \citep[ATLAS; ][]{tonry_atlas_2018} on 2023 January 22 (MJD 59967). The transient is located with 0.1'' of the nucleus of SDSS J143619.83+322316.5, a $m_{r}=18.6$ galaxy from the Sloan Digital Sky Survey \citep[SDSS; ][]{aihara_eighth_2011}. From an ESO Faint Object Spectrograph and Camera (EFOSC) spectrum obtained on 2024 April 10 (MJD 60410) by the extended Public European Southern Observatory Spectroscopic Survey of Transient Objects \citep[ePESSTO+; ][]{smartt_pessto_2015}, the redshift of the host galaxy of TDE\,2023adr was measured to be $z=0.131$ \citep{dalen_epessto_2024}. Initially identified as a candidate superluminous supernova \citep{perley_ztf_2023}, 2023adr was spectroscopically monitored by the Zwicky Transient Facility \citep[ZTF;][]{bellm_zwicky_2019} using the Spectral Energy Distribution Machine \citep[SEDM;][]{blagorodnova_sed_2018}, on the Palomar 60-inch telescope (P60) on 2023 January 25 and February 09 (MJDs 59969 and 59984). These initial low-resolution spectra exhibit blue continua, and evidence of some broad hydrogen emission. Combined with its sustained blue color \citep[as noted by ][]{aleo_anomaly_2024}, this is consistent with a TDE origin. 

Further motivated by photometric classification of the first peak as a TDE candidate by {\tt tdescore} \citep[][]{stein_tdescore_2024}, we obtained a spectrum with the Keck Low-Resolution Imaging Spectrometer \citep[LRIS;][]{oke_keck_1995} on 2024 February 02 (MJD 60351), originally intended to probe the late-time evolution of the first flare. However, this observation was serendipitously obtained during the very early rise of the second flare, preceding the first ZTF detection by several days and corresponding to a phase of $\approx20$\,days before peak brightness. To our knowledge, this represents the earliest spectroscopic observation of a repeating TDE to date, providing a rare benchmark for assessing whether distinctive spectroscopic signatures are present at early times in repeating partial disruptions. Further follow-up spectra were obtained with the DeVeny Spectrograph mounted on the Lowell Discovery Telescope on 2024 March 09 (MJD 60376), and with Keck/LRIS on 2024 July 07 (MJD 60498). We present the spectroscopic evolution, spanning both flares of TDE\,2023adr in Figure \ref{fig:23ar_class}, and a log of all spectroscopic observations and observing setups are presented in the Appendix. From the higher resolution spectra obtained of the second peak, we see a much stronger similarity to classical TDEs from the literature, in particular ASASSN-15oi \citep{holoien_asassn-15oi_2016}, with clear broad \citep[widths $\sim15,000$ km\,s$^{-1}$;][]{charalampopoulos_detailed_2022} hydrogen, helium and nitrogen emission features, placing TDE\,2023adr in the \lq H+He\rq\,spectral class \citep{hammerstein_final_2023}. Given the persisting broad hydrogen emission features and blue colour evolution, we therefore confirm that 2023adr is indeed a rpTDE, and only the second rpTDE to have spectroscopic confirmation of both peaks. 


\begin{figure}
	\includegraphics[width=\columnwidth]{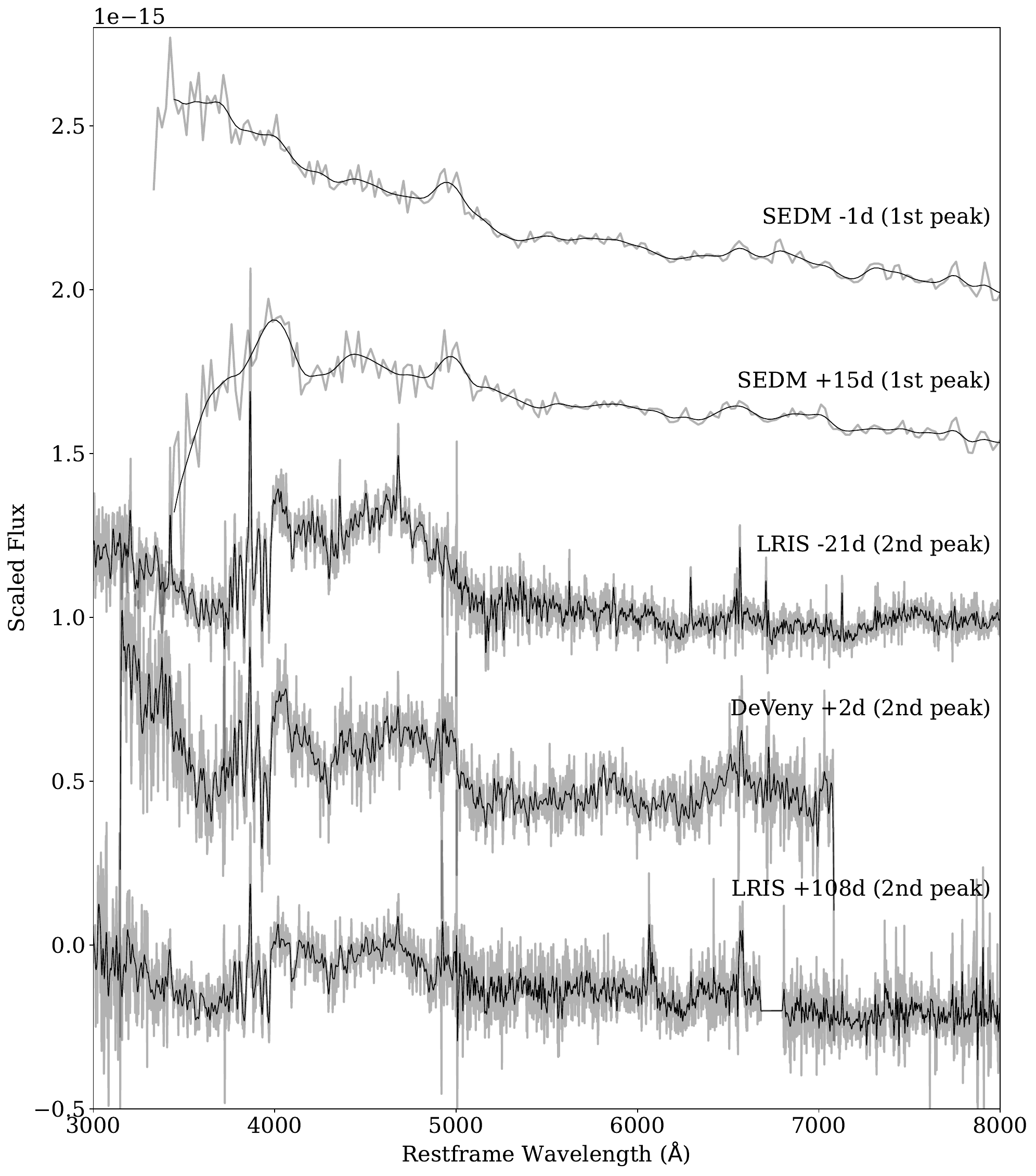}
         \includegraphics[width=\columnwidth]{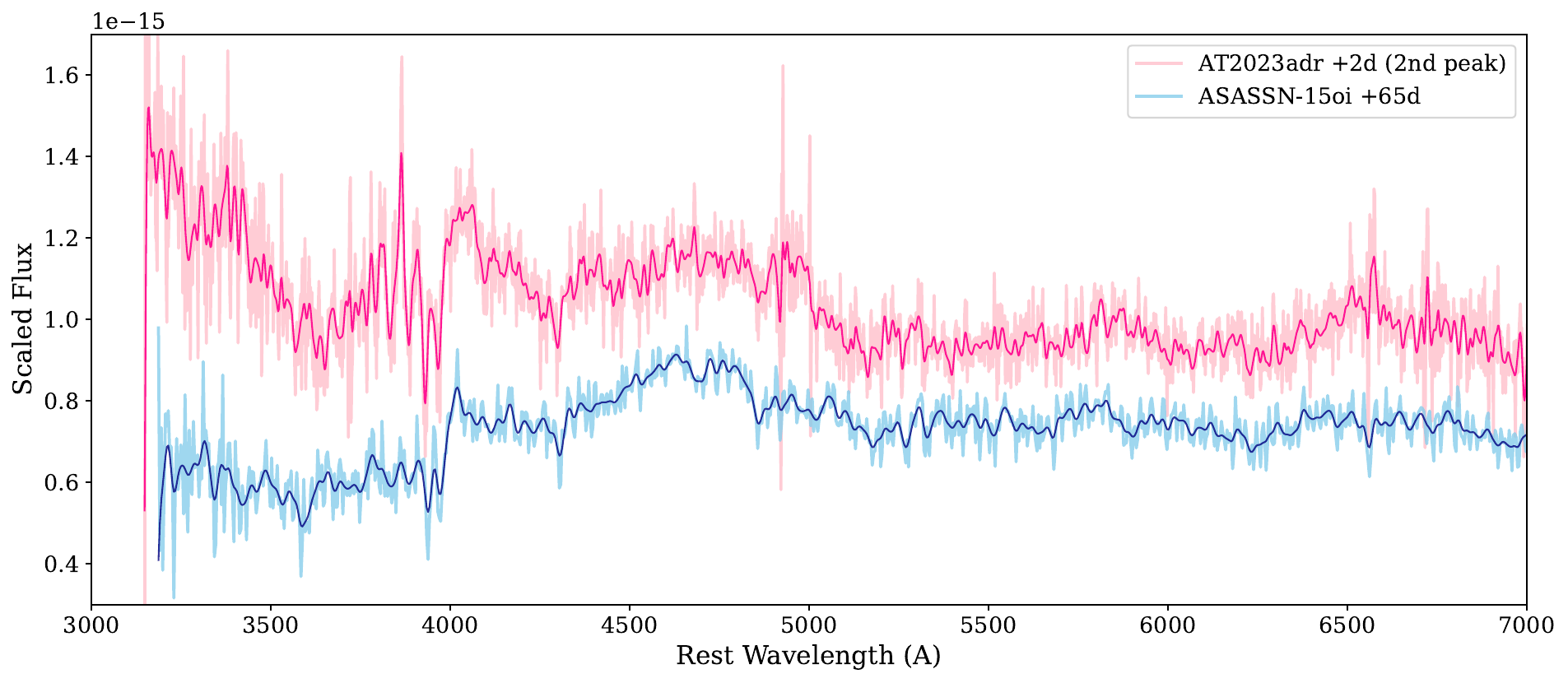}
    \caption{{\it{Top:}}Spectroscopic evolution of TDE\,2023adr. Low resolution spectra during the first peak show a strong blue contiuum with some broad balmer emission, whilst the second peak exhibits broaded, blended helium and nitrogen emission. {\it{Bottom:}} Comparison of spectrum during the second peak with the well observed TDE, ASASSN-15oi \citep{holoien_asassn-15oi_2016}.}
    \label{fig:23ar_class}
\end{figure}

\subsection{Photometric Data}

We collect photometry for TDEs 2020vdq, 2022dbl, and 2023adr from ZTF, the Asteroid Terrestrial Impact Last Alert System  \citep[ATLAS;][]{tonry_atlas_2018}, and the Panoramic Survey Telescope and Rapid Response System (Pan-STARRS) Survey for Transients \citep[PSST;][]{huber_pan-starrs_2015}, alongside targeted follow-up observations with the ultraviolet (UV) and optical photometry from the UV-Optical Telescope \citep[UVOT;][]{roming_swift_2005} on board the Neil Gehrels Swift Observatory \citep[{\it{Swift}};][]{gehrels_swift_2004}. Additionally, follow-up observations were obtained of TDE\,2022dbl with Liverpool Telescope (LT). 

We use the ZTF forced photometry server \citep{masci_new_2023} to recover public $g$- and $r$-band photometry of all three rpTDEs, alongside the ATLAS forced photometry server \citep{tonry_atlas_2018,smith_design_2020,shingles_release_2021} for difference-image photometry in the $o$ and $c$ bands. We also retrieve Pan-STARRS $i$-, $z$-, $y$- and $w$-band data from the PSST survey. PSST images were processed in real time as described in \cite{magnier_pan-starrs_2020}, with forced photometry performed at the transient location using reference 3$\pi$ sky survey images \citep{huber_pan-starrs_2015}. 

We use the {\it{Swift}} $uvw2$, $uvm2$, and $uvw1$ host-subtracted UVOT photometry from \cite{somalwar_first_2025} and \cite{hinkle_double_2024} for TDEs 2020vdq and 2022dbl respectively. We do not include the UVOT $U,B,V$ data in our analysis of these sources, as the late time color evolution in these filters deviates from other bands (particularly for TDE\,2022dbl) suggesting some level of host emission remains within these bands. For TDE\,2023adr we retrieve UVOT $uvm2$, and $uvw1$ imaging, measuring the source flux within a 5'' aperture and background within a 10'' aperture using the \texttt{uvotsource} tool. To estimate the host galaxy contamination, we compute synthetic photometry from the best-fit host galaxy spectral energy distribution (SED) to archival photometry (see Section \ref{sec:hostproxies}), which we subtract from the measured flux before converting to AB magnitudes using the photometric zero-points \citep{breeveld_updated_2011}. For our LT photometry, we perform aperture photometry upon the $u, g, r, i$ data using the {\texttt{PSF}} routine \citep{nicholl_at_2023}, extracting using an optimized aperture and archival Sloan Digital Sky Survey images \citep{albareti_13th_2017} for template subtraction and calibration.

\begin{figure}
	\includegraphics[width=\columnwidth]{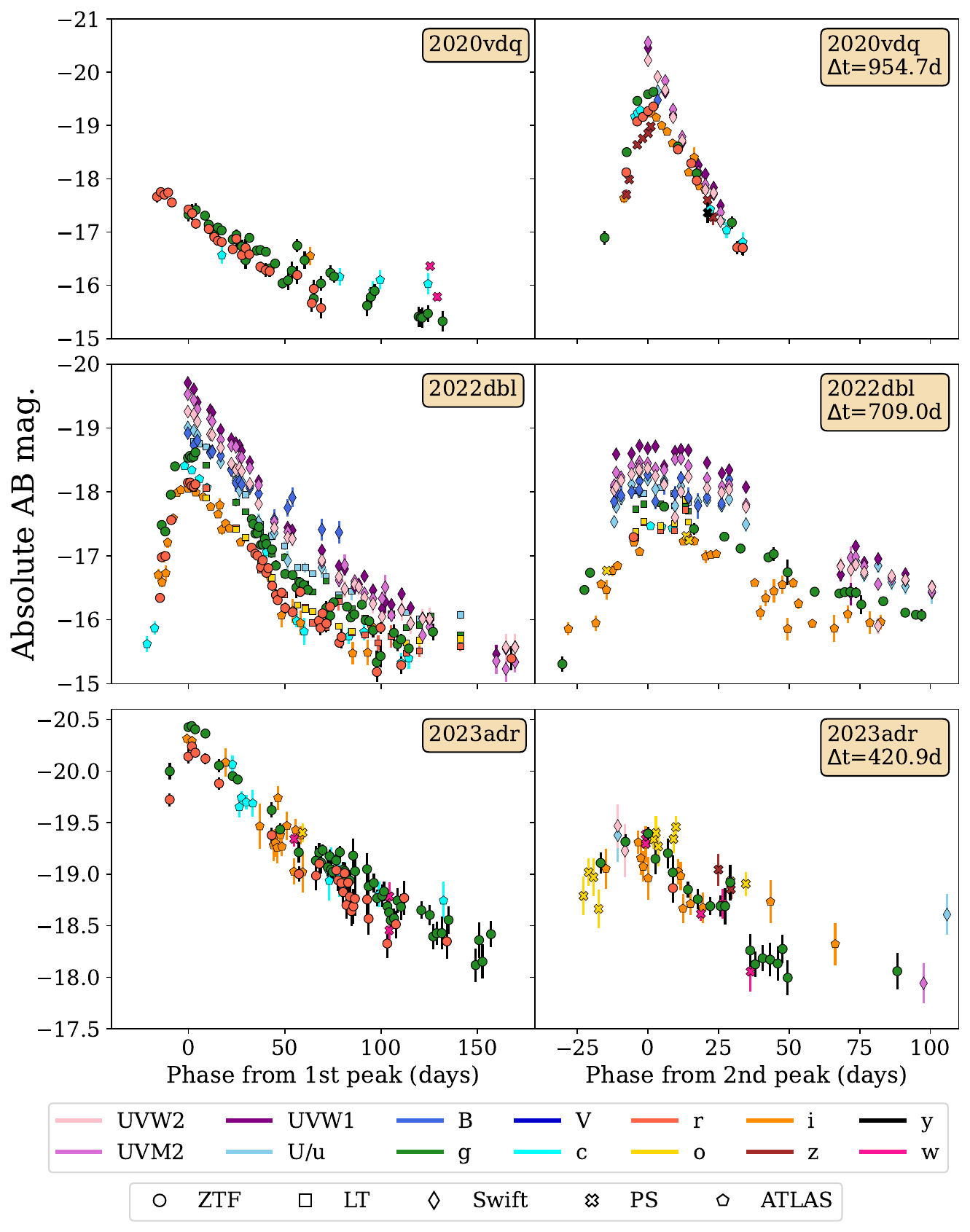}

    \caption{UV-Optical light curves of the first ({\textit{left}}) and second ({\textit{right}}) flares of the confirmed rpTDEs, with survey data from ZTF, ATLAS, PanSTARRS, alongside targeted follow-up observations from {\it{Swift}} and Liverpool Telescope. All magnitudes are host subtracted, corrected for galactic reddening and presented in the AB system. Phases are given in the rest frame of the TDE with respect to each flare.}
    \label{fig:rpTDE_LCs}
\end{figure}

All photometry is corrected for the Milky Way foreground extinction using the dust maps of \cite{schlafly_measuring_2011}. We do not correct for internal host galaxy extinction. We present the optical light curves of the flares of the three rpTDE in Figure~\ref{fig:rpTDE_LCs}. We take the peak epochs of the first and second flares respectively to be MJD = 59126.5, 60081.2 for TDE\,2020vdq, MJD = 59637.6, 60346.6 for TDE\,2022dbl, and MJD = 59967.5, 60388.4 for TDE\,2023adr. 

\section{Models and TDE Scaling Relations}\label{sect:models}

Here we briefly summarise key details of the different semi-analytical TDE models and scaling relations that we will consider in this work, to infer $M_{\rm BH}$ including any assumptions either inherent to the model or implemented within the fitting process, and report our final fitted parameters for each peak of the three rpTDEs in our sample. 

\subsection{Fallback Model - \texttt{MOSFiT}}\label{sec:mosfit}

In the fallback model, the luminosity of a TDE is directly linked to the rate of return of the stellar debris to the black hole, such that the fallback rate and the evolutionary timescales of TDE light curve are dependent upon $M_{\rm BH}$ \citep{guillochon_hydrodynamical_2013,mockler_weighing_2019}. Any delays between the fallback timescale and the emission timescale are accounted for through a viscous timescale parameter, which can slow the rise of the observable emission in time while largely preserving the functional form of the fallback-powered light curve. To account for the discrepancy between the predicted temperatures from a hot source (such as a compact accretion disk) and the cooler inferred temperatures from typical optical TDE colours ($10^{5}-10^{6}$K vs a few $\times10^{4}$K), a `reprocessing' layer is invoked, composed of an optically thick medium with a characteristic size $\gtrsim100$ times the disk scale, thus moving the peak of the emission to the UV/optical regime. The radius of this layer is assumed to evolve as a power-law function of the luminosity, enabling the model to reproduce the observed evolution in temperature and radius. The fallback TDE model implemented in the Modular Open Source Fitter for Transients \citep[{\texttt{MOSFiT}};][]{guillochon_mosfit_2018} is based on hydrodynamically simulated fallback rates of stellar debris, computed across a range of black hole and stellar parameters. Importantly, {\tt{MOSFiT}} assumes the disrupted stars lie on the main-sequence, with stellar structure taken from polytropic models \citep{guillochon_hydrodynamical_2013,mockler_weighing_2019}. This allows the model to account for how variations in stellar mass and polytropic index influence the fallback rate, which in turn determines the disruption outcome - whether it is partial or full.

In general, the shape of the fallback light curves are primarily controlled by three parameters: the black hole mass ($M_{\rm BH}$), the mass of the disrupted star ($M_{\star}$), and the penetration depth of the encounter relative to the pericentre distance of the orbit (the `impact parameter', $\beta=R_{t}/R_{p}$). Increasing $M_{\rm BH}$ lowers the peak fallback rate, thus broadening the light curve ($t_{fb} \propto M_{\rm BH}\,^{0.5}$). In contrast, increasing $M_{\star}$ raises the overall normalization of the fallback rate and thus the light curve peak luminosity, whilst also steepening the early decline because more mass is returned over a shorter dynamical timescale. Finally, increasing $\beta$, corresponding to deeper encounters, increases the energy spread of the debris, producing an earlier peak and faster evolving light curve.

The {\texttt{MOSFiT}} fallback model treats several quantities as free parameters: $M_{\rm BH}$, $M_{\star}$, the normalized photospheric radius and its luminosity scaling index ($R_{ph,0}$ and $\ell_{ph}$), the radiative efficiency ($\epsilon$), the emission onset time relative to first detection ($t_{0}$), and host galaxy extinction, which is scaled with hydrogen column density ($n_{H,host}$). Because the degree of stellar disruption depends on the star’s internal density profile, the model defines a ``scaled'' impact parameter, $b$, rather than $\beta$ to represent penetration relative to the tidal radius. Here $b \ge 1$ corresponds to full disruptions and $b = 0$ to no disruption.  The scaled impact parameter encodes both stellar mass and internal compactness of the disrupted star. Because lower-mass stars are less centrally concentrated, they are more easily fully disrupted in shallower encounters, whereas more massive, radiation-pressure–dominated stars require significantly deeper encounters to achieve the same outcome. The {\tt{MOSFiT}} fallback models assume a polytropic stellar structure, where the internal density profile varies with the polytropic index $\gamma$. Low-mass main-sequence stars with $M < 0.3,M_\odot$ and very massive stars with $M > 22,M_\odot$ are treated as $\gamma = 5/3$ polytropes, whilst stars with masses $1 \lesssim M/M_\odot \lesssim 15$ are modelled using $\gamma = 4/3$. Stars falling within the transition regimes ($0.3$–$1\,M_\odot$ and $15$–$22\,M_\odot$) are treated with hybrid fallback functions, ensuring a smooth interpolation between $4/3$ and $5/3$ polytropic behavior. 
The viscous timescale parameter ($T_{\nu}$) is also included, to capture the delay between debris fallback and emission onset. Additionally, the model fits a white noise term ($\sigma$) to account for observational scatter. For full model details, we refer the reader to \cite{guillochon_hydrodynamical_2013} and \cite{mockler_weighing_2019}.

\begin{table}
\centering
\caption{Priors assumed for all rpTDE peaks in the {\tt{MOSFiT}} TDE model. Prior distributions are flat unless stated.}
\renewcommand{\arraystretch}{1.15}
   \begin{threeparttable}
\begin{tabular}{llc}
\hline
Parameter & Prior / Range & Units \\
\hline
$\log_{10}(M_{\rm BH}/M_\odot)$ & $[4.7,\,8.7]$                   & - \\
$M_\star$                      & Kroupa IMF,\ $\,[0.01,\,30]$ & $M_\odot$ \\
$b$                            & $[0,\,2]$                   & - \\
$\log_{10}\epsilon$            & $[-4,\,-0.4]$               & - \\
$\log_{10} R_{ph,0}$       & $[-4,\,4]$                  & (model units)~\tnote{*}  \\
$\ell_{\rm ph}$                & $[0,\,2]$                   & - \\
$\log_{10} T_\nu$              & $[-3,\,3]$                  & d \\
$t_0$                          & $[-500,\,0]$                & d \\
$\log_{10} N_{H,host}$     & $[14,\,25]$                 & cm$^{-2}$ \\
$\log_{10}\sigma$              & $[-3,\,2]$                  & - \\
\hline
\end{tabular}
\begin{tablenotes}
      \footnotesize
      \item[*]{$R_{\rm ph,0}$ is parameterized as a luminosity-dependent photosphere with normalization $R_{\rm ph,0}$ and index $\ell_{\rm ph}$.}
\end{tablenotes}
\end{threeparttable}
\label{tab:mosfit_tde_priors}
\end{table}

We evaluate the posterior distributions of the model parameters using dynamic nested sampling with {\tt{DYNESTY}} \citep{speagle_dynesty_2020}. When fitting each peak, we assume the same broad priors for each event, which we present within Table \ref{tab:mosfit_tde_priors}. Because our goal was to test how well existing TDE models recover parameters in partial encounters, we imposed deliberately broad priors on $b$, enabling the sampler to explore solutions corresponding to both partial and full disruptions. 
We present the resulting distributions for the key parameters ($M_{\rm BH}$,$M_{\star}$ and $b$) for the peaks of TDE\,2022dbl, TDE\,2020vdq and TDE\,2023adr from {\tt{MOSFiT}} modelling within Figure \ref{fig:fits_mosfit} and in Table \ref{tab:fits_mosfit}. Our fits to the light curves are presented in the Appendix.

\begin{figure}
    \includegraphics[width=0.88\columnwidth]{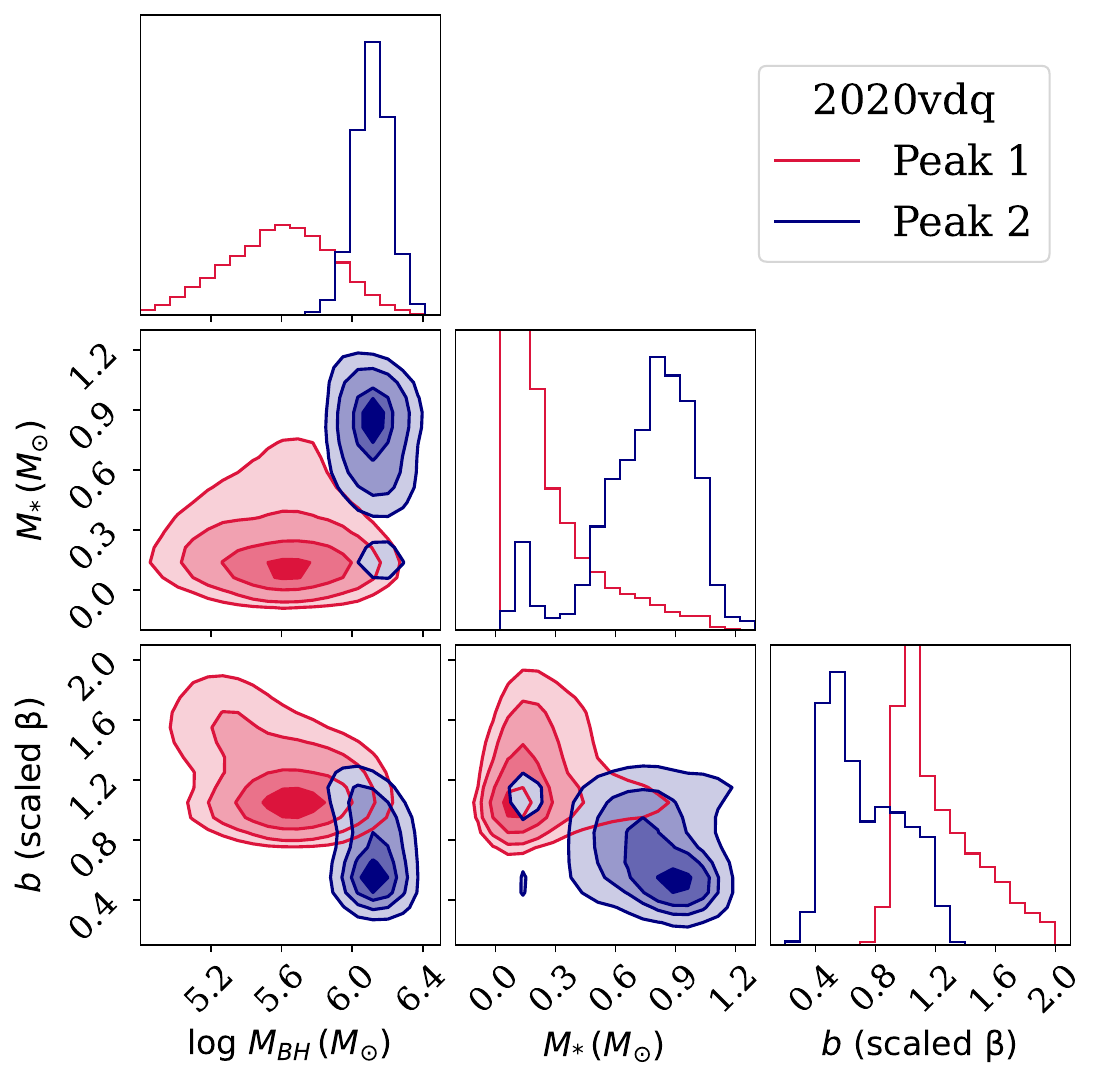}
    \includegraphics[width=0.88\columnwidth]{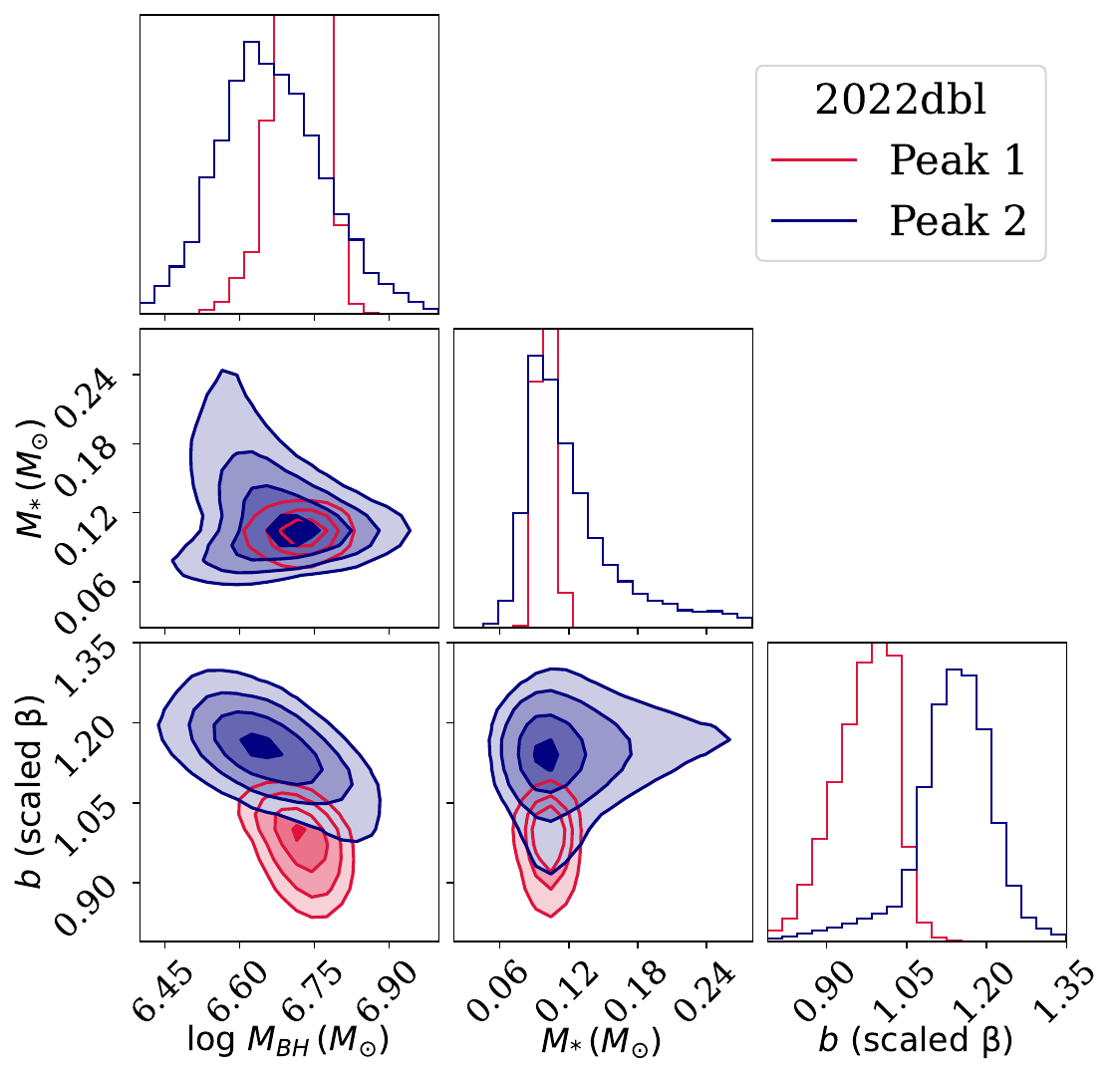}
    \includegraphics[width=0.88\columnwidth]{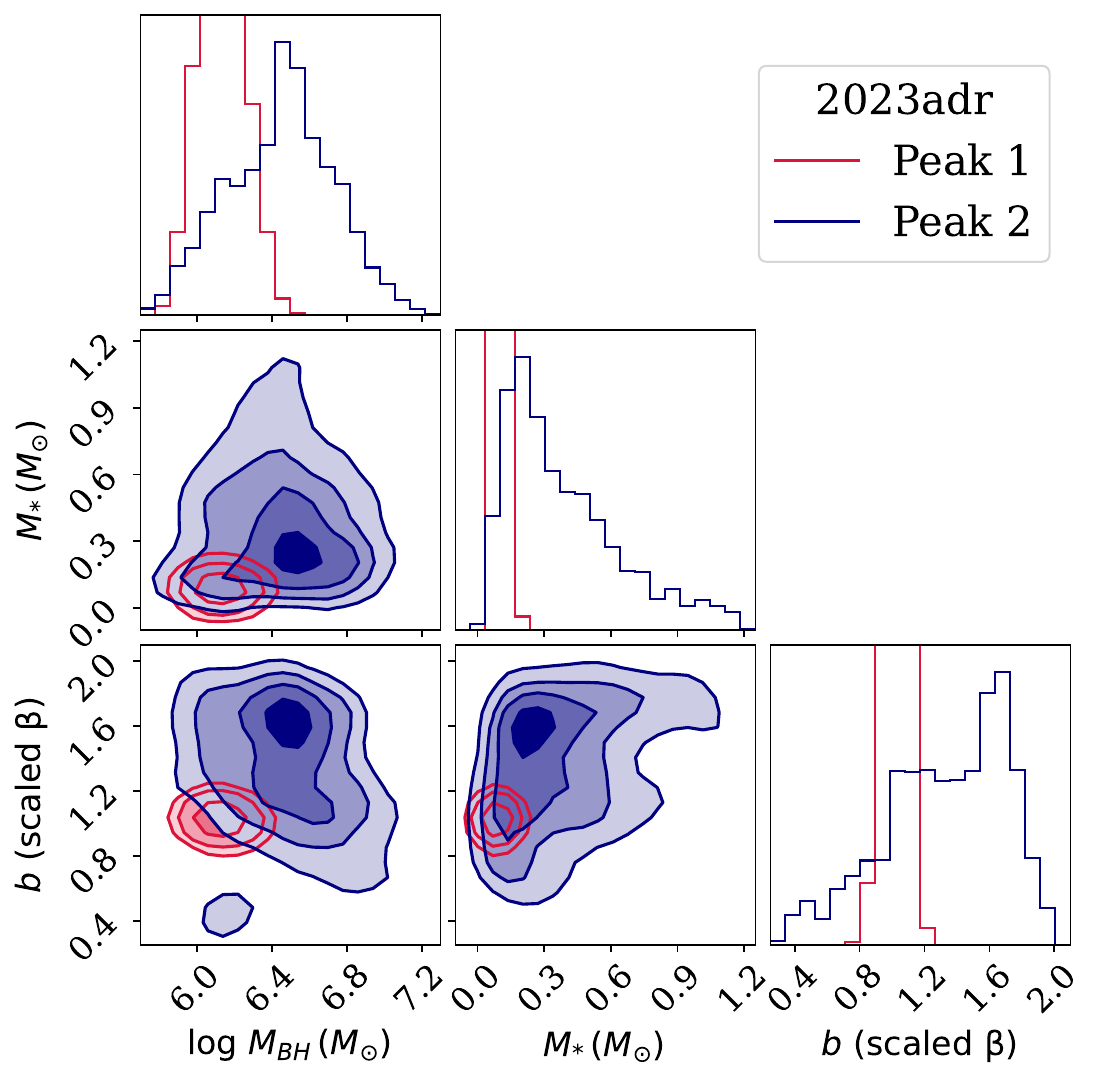}
    \caption{Resulting distributions of key parameters ($M_{\rm BH}$, $M_{\star}$ and scaled impact parameter, $b$) from {\texttt{MOSFiT}} fallback modelling of the rpTDEs 2020vdq ({\it{top}}), 2022dbl ({\it{middle}}), and 2023adr ({\it{bottom}}). Results from independently fitting the first peak are shown in red, and in blue for the second peak. }
    \label{fig:fits_mosfit}
\end{figure}

\begin{table}
    \centering
    \begin{tabular}{l|l|l|l}
    \hline
        TDE & $\log_{10}M_{\rm BH}$ & $M_{\star}$ & $b$ \\
            \hline
         2020vdq (Peak 1) &  $5.71^{+0.19}_{-0.44}$  & $0.21^{+0.21}_{-0.14}$ & $1.25^{+0.27}_{-0.26}$ \\[0.3em]
         2020vdq (Peak 2) &  $6.23^{+0.05}_{-0.12}$  & $1.02^{+0.08}_{-0.30}$ & $0.46^{+0.14}_{-0.37}$ \\[0.3em]
         \hline
         2022dbl (Peak 1) &  $6.74^{+0.03}_{-0.06}$  & $0.101^{+0.002}_{-0.004}$ & $1.00^{+0.03}_{-0.09}$ \\[0.3em]
         2022dbl (Peak 2) &  $6.69^{+0.07}_{-0.14}$  & $0.12^{+0.04}_{-0.04}$ & $1.16^{+0.03}_{-0.09}$ \\[0.3em]
         \hline
         2023adr (Peak 1) &  $6.17^{+0.08}_{-0.15}$  & $0.09^{+0.02}_{-0.02}$ & $1.04^{+0.04}_{-0.08}$ \\[0.3em]
         2023adr (Peak 2) &  $6.55^{+0.19}_{-0.41}$  & $0.43^{+0.23}_{-0.29}$ & $1.54^{+0.16}_{-0.60}$ \\[0.3em]
       \hline
    \end{tabular}
    \caption{Resulting key parameters from fallback modelling in {\tt{MOSFiT}}. Uncertainties correspond to statistical errors, and do not include the model systematic errors determined within \citet{mockler_weighing_2019}.}
    \label{tab:fits_mosfit}
\end{table}

From fallback modelling, we find that for events with data spanning the entirety of each flare (TDEs 2022dbl and 2023adr), where peaks have a higher data coverage (in both cases, the first peak), this yields significantly tighter constraints on $M_{\rm BH}$, $M_{\star}$, and $b$ than the second, as can be seen in Figure \ref{fig:fits_mosfit}. For both sources, the inferred black hole masses from independent fits to each peak agree within $1\sigma$. In contrast, the absence of observations covering the rise of the first peak of TDE\,2020vdq leads to weaker constraints across all parameters, making it difficult to assess the relative performance of the model across both flares.

The posteriors on $M_{\rm BH}$, $M_{\star}$, and $b$ inferred with \texttt{MOSFiT} often underestimate the true uncertainty, since these parameters are strongly degenerate and depend on assumptions about radiative efficiency and emission geometry. Thus all key parameters are subject to large systematic uncertainties. \citet{mockler_weighing_2019} infer systematic uncertainties of $\pm0.2$ for $M_{\rm BH}$, $\pm0.66$ for $M_{\star}$ and $\pm0.66$ for the impact parameter $\beta$. 

A notable outcome of the {\tt{MOSFiT}} fitting is that for both 2022dbl and 2023adr, without constricting priors, the model prefers a nearly full or full disruption for the first peaks ($b\approx1$). The way in which $b$ is defined implies that for $b>1$, there should be no surviving remnant bound to the system (regardless of the polytropic index adopted for the disrupted star), but in a rpTDE, there is by definition a remaining core. Given the encoding of $M_{\star}$ in $b$, the fact that the first flares of 2022dbl and 2023adr don't have a strong preference for partial disruptions may arise due {\tt{MOSFiT}} attempting to accommodate for the typically lower luminosities of rpTDEs \citep{makrygianni_double_2025,hinkle_double_2024,somalwar_first_2025}, by lowering $M_{\star}$ and thus increasing $b$. It may also be a consequence of the stellar models the {\tt{MOSFiT}} TDE model adopts from the hydrodynamical simulations of \cite{guillochon_hydrodynamical_2013}. The assumption of a polytropic structure may be invalid for repeating partial disruptions, as on the relatively short ($\sim$ few year) timescales of their orbits, the core may still not have relaxed to hydrostatic or thermal equilibrium by the second passage. Further, several studies have shown that following initial tidal stripping, the surviving remnant will be more compact with a shallower outer envelope \citep{guillochon_hydrodynamical_2013,nixon_partial_2021,sharma_partial_2024}, making it more resistant to further stripping during repeat encounters. Thus a wider suite of stellar models may enable more realistic recovery of $M_{\star}$ and $b$ parameters within the {\tt{MOSFiT}} fallback model.  

\begin{figure}
    \includegraphics[width=0.88\columnwidth]{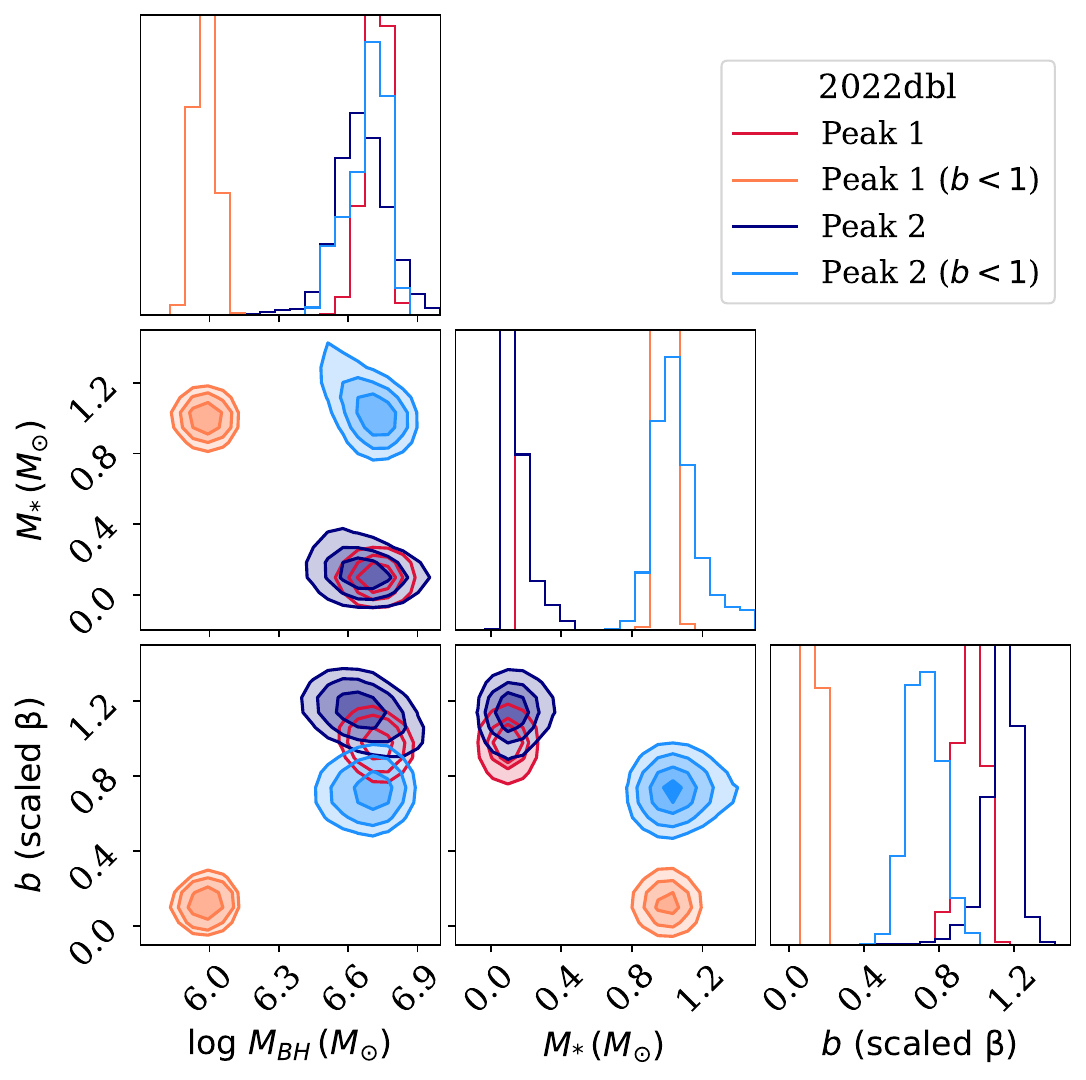}
    \caption{Comparison of {\tt{MOSFiT}} parameter distributions when modelling with constrained priors on scaled impact parameter, $b$ for TDE\,2022dbl.}
    \label{fig:22dbl_b_comp}
\end{figure}

To test this, we re-run {\tt{MOSFiT}} for 2022dbl and 2023adr, restricting the $b$ parameter between [0,1]. We compare the parameter distributions with unconstrained and constrained $b$ for TDE\,2022dbl in Figure \ref{fig:22dbl_b_comp}. With tighter priors on the encounter geometry, {\tt{MOSFiT}} finds solutions at lower $b$ for both peaks (0.13 and 0.75 respectively), and stellar masses of $\sim\,1\,M_{\odot}$. However this leads to a tension in the black hole masses, producing $\log{M_{\rm BH}}=6.0^{+0.02}_{-0.05}$ and $6.77^{+0.03}_{-0.14}$ (statistical uncertainties) for the first and second peaks respectively. This likely reflects the degeneracies between $M_{\star}$, $M_{\rm BH}$, and $b$ discussed in \cite{mockler_weighing_2019}, which become apparent when fitting the first peak of TDE\,2022dbl with a restricted impact parameter. Smaller values of $b$ reduce the debris energy spread, creating lower, more delayed peak fallback rates \cite{guillochon_hydrodynamical_2013}. Given the dependency of the fallback rate upon both black hole mass and stellar mass, ${\dot{M}}\propto\,M_{\rm BH}^{-1/2}M_{\star}^{2}R_{\star}^{-3/2}$, {\tt{MOSFiT}} has to compensate by increasing $M_{\star}$ (which boosts the fallback rate but slows the rise to peak) and by decreasing $M_{\rm BH}$ (which also increases the fallback but shortens the time to peak), enabling it to reproduce the observed light curve luminosity and evolutionary timescale. For 2023adr, we find less constrained fits, with the first and second peak pushing towards the higher values of $b$ within the allowed priors (0.95 and 0.83, corresponding to $\beta=0.89$ and $0.96$ respectively), and low, poorly constrained stellar masses. that a substantial fraction ($\sim$40–60\%) of the stellar core can survive encounters involving $0.1$--$0.5\,M_{\odot}$ stars, indicating that partial disruptions remain possible even at small pericentre distances.

In summary, fallback-based modelling with {\tt MOSFiT} yields broadly consistent black hole mass estimates between the two flares of each rpTDE, although in several cases this agreement is driven in part by the limited data quality of one peak, which results in broad and weakly informative posteriors. By contrast, the stellar mass and encounter geometry are far less robustly constrained: the inferred $M_{\star}$ values are highly degenerate and should not be over-interpreted, consistent with recent findings that stellar masses derived from TDE light curves are particularly unreliable \citep[e.g.][]{mummery_tidal_2025}. Finally, the preference of the model for solutions with near-full disruptions is in tension with the known physical nature of rpTDEs, and as such should be viewed as indicative rather than definitive for these events. Together, these considerations imply that while fallback-based fits can provide useful, order-of-magnitude constraints on $M_{\rm BH}$, inferences regarding stellar properties and disruption depth in rpTDEs should be treated with appropriate caution.


\subsection{Photometric Scaling Relations}\label{sec:TDEscaling}

Scaling relations that link observable light-curve properties (e.g. peak or late-time luminosities, characteristic timescales) to the underlying physical parameters of the system provide a fast and efficient alternative for estimating black hole masses. Their reduced reliance on high-quality or densely sampled data makes them particularly appealing for application to large TDE samples, such as those to be identified within LSST.

Here we examine two commonly used scaling relations to measure the black hole masses of rpTDEs: the relationship between peak luminosity and colour implemented within {\tt{TDEmass}} \citep{ryu_measuring_2020} and the correlation between black hole mass and the late-time TDE luminosity, and the peak $g-$band luminosty, determined by \cite{mummery_maximum_2024}. 

\subsubsection{Luminosity-color relationships: {\tt{TDEmass}}}

Under the assumption that circularization takes place slowly, {\tt{TDEmass}} \citep{ryu_measuring_2020} builds upon the theory proposed by \cite{piran_disk_2015} in which the UV/optical emission in TDEs is the result of shocks between intersecting debris streams near their orbital apocenters. Using the peak luminosity and color-temperature of the flare, {\tt{TDEmass}} determines likely ranges of $M_{\rm BH}$ and $M_{\star}$ \citep{ryu_measuring_2020}. 

To calculate the peak luminosities and temperatures of each flare, we first construct the bolometric light curves of the rpTDEs, grouping their multi-band photometry at 2-day intervals before fitting a simple black body function to each epoch, from which we determine the photospheric temperature and radii before integrating across the full blackbody curve to find the bolometric luminosity. We present our inferred {\tt{TDEmass}} parameters for each rpTDE peak in Table \ref{tab:tdemass}. For the first peak of TDE\,2020vdq, we treat the first detection following its emergence from solar conjunction as an upper limit on the light-curve peak.

\begin{table}
    \centering
    \begin{threeparttable}
    \begin{tabular}{l|c|l}
    \hline
        TDE & $\log_{10}(M_{\rm BH}/M_{\odot})$ & $M_{\star}$ \\
            &  & ($M_{\odot}$) \\
            \hline
         2020vdq (Peak 1)~\tnote{*} & $6.78^{+0.75}_{-0.29}$  & $0.093^{+0.63}_{-0.092}$ \\[0.3em]
         2020vdq (Peak 2) & $6.79^{+0.23}_{-0.67}$  & $1.1^{+2.4}_{-0.58}$ \\[0.3em]
         \hline
         2022dbl (Peak 1) & $6.26^{+0.24}_{-0.28}$  & $0.69^{+0.14}_{-0.17}$ \\[0.3em]
         2022dbl (Peak 2) & $5.97^{+0.27}_{-0.24}$  & $0.41^{+0.14}_{-0.17}$ \\[0.3em]
         \hline
         2023adr (Peak 1) & $7.30^{+0.14}_{-0.82}$  & $1.5^{+7.6}_{-1.2}$ \\[0.3em]
         2023adr (Peak 2) & $6.62^{+0.29}_{-0.51}$  & $0.78^{+0.31}_{-0.46}$ \\[0.3em]
       \hline
    \end{tabular}
    \begin{tablenotes}
      \footnotesize
      \item[*]{``Peak'', of the flare taken as first epoch following emergence from solar conjunction}
\end{tablenotes}
\end{threeparttable}
    \caption{Input parameters and inferred black hole and stellar masses for each TDE peak using the {\tt TDEmass} model. Uncertainties correspond to 68\% confidence intervals.}
    \label{tab:tdemass}
\end{table}

For the rpTDEs within our sample, {\tt TDEmass} generally returns self-consistent (within $\sim2\sigma$) estimates of $M_{\rm BH}$ and $M_{\star}$. In the case of TDE\,2022dbl, both parameters are relatively well constrained, with uncertainties comparable to those obtained from {\tt MOSFiT} once systematic errors are taken into account. By contrast, larger discrepancies and weaker constraints are evident for TDE\,2020vdq and TDE\,2023adr. In particular, for TDE\,2020vdq the stellar mass inferred from the first and second peaks differs significantly, reflecting the strong sensitivity of the method to the poorly constrained peak luminosity of the first flare. For TDE\,2023adr, the black hole mass posterior for the second peak is especially broad, likely driven by uncertainties in the inferred temperature and the lack of near-UV coverage. These examples highlight the dependence of the {\tt TDEmass} constraints on both peak epoch data coverage and robust temperature estimates.

\subsubsection{Luminosity scaling relationships}\label{sec:plateaus}

It has demonstrated that the luminosity reached during the late–time ``plateau'' phase of TDE light curves, $L_{\rm plat}$, shows a tight correlation with host galaxy stellar mass, used as a proxy for the central black hole mass. This plateau emission arises from the combined cooling and radial expansion of the newly formed accretion disc, and the observed luminosity–mass relation is a natural consequence of time–dependent disc accretion theory \citep{mummery_spectral_2020, mummery_fundamental_2024,mummery_optical_2025}. UV plateau luminosities in particular may provide a more robust scaling with $M_{\rm BH}$, as emission in this regime is disc-dominated and comparatively insensitive to reprocessing, winds, or geometry of the system. It has been shown that by fitting the full SED from the X-ray through to UV-optical wavelengths, plateau luminosities can measure $M_{\rm BH}$ for TDEs with uncertainties of order $\lesssim0.3$dex \citep{guolo_compact_2025}. However, where the full SED or UV data is unavailable, \cite{mummery_fundamental_2024} have shown that correlations with the rest frame late-time luminosity in the $g$-band may also be used to estimate $M_{\rm BH}$, albeit with larger scatter. 

We take the published UV plateau luminosities (measured at $\nu=10^{15}$\,Hz) and corresponding black hole masses from \cite{mummery_optical_2025} for TDEs 2020vdq and 2022dbl. The exact epoch of these observations used to derive the masses in this work is unknown, but is assumed to be $\gtrsim$1 year post maximum light from the first flares of these rpTDEs (i.e. prior to the second peak). For TDE\,2023adr, we use the late time $g$-band photometry from Pan-STARRS to compute the plateau luminosity, obtained following the second peak on MJD=60885 and 61033 (corresponding to rest frame phases +811d/+439d and +942d/+570d with respect to the first and second flares), from which we measure an average apparent brightness of $m_{g}=21.36\pm0.18$. We show this late-time plateau measurement with respect to the main light curve in the Appendix. The plateau luminosity black hole masses are reported in Table \ref{tab:scaling}.

Given the lack of corresponding data following each flare from which a plateau could be measured, we do not directly compare the performance of plateau-luminosity scaling $M_{\rm BH}$ between the flares of each rpTDE. Additionally, as plateau-luminosity scaling implicitly assumes a single, self-contained accretion episode in which the disc evolves from peak to a steady late-time cooling phase, it may not be expected that reproducible black hole masses can be obtained from the light curves of rpTDEs. With multiple passes of the star, the fallback rate, disc mass, and radiative output will be replenished rather than declining monotonically. Further, with shorter orbital periods, the flares of successive passes may overlap within the light curve, such that a late-time plateau regime is never truly reached. 

\begin{table}
    \centering
    \begin{threeparttable}
    \begin{tabular}{l|l|l|l}
    \hline
         & Plateau-scaling&    & Peak-scaling \\
        TDE & $\log_{10}M_{\rm BH}$ & $L_{\mathrm{peak,g}}$ & $\log_{10}M_{\rm BH}$  \\
         &  & $\times10^{43}$erg\,s$^{-1}$ &  \\
            \hline
         2020vdq (Peak 1) & $5.52^{+0.59}_{-0.46}$~\tnote{\emph{*}} & $>0.08$ & $>5.5$    \\[0.3em]
         2020vdq (Peak 2) & - & $1.76\pm0.02$ & $6.76 \pm 0.40$   \\[0.3em]
         \hline
         2022dbl (Peak 1) & $5.68^{+0.60}_{-0.44}$~\tnote{{*}} & $0.69\pm0.01$ &$6.36\pm0.40$   \\[0.3em]
         2022dbl (Peak 2) & - & $0.33\pm0.12$ &$6.04\pm0.41$   \\[0.3em]
         \hline
         2023adr (Peak 1) & - & $3.80\pm0.14$ & $7.08\pm0.41$   \\[0.3em]
         2023adr (Peak 2) & $8.2\pm0.32$~\tnote{{\textdagger}}& $1.43\pm0.08$ & $6.67\pm0.42$    \\[0.3em]
      \hline
    \end{tabular}
    \begin{tablenotes}
      \footnotesize
      \item[*]{Black hole masses from UV plateau luminosities presented within \citealt{mummery_optical_2025}. The precise epoch of these observations is unknown, but assumed to occur before the second peak of the light curve.}
      \item[\textdagger]{Plateau measured in the rest-frame $g$-band.}
\end{tablenotes}
\end{threeparttable}
    \caption{Black hole masses inferred from plateau luminosity scaling, alongside peak $g$-band luminosities and corresponding black hole masses determined using the \citealt{mummery_fundamental_2024} scaling relationship. }
    \label{tab:scaling}
\end{table}

Fortunately there is another, less well understood, empirical correlation between the observed peak $g$-band luminosities of TDEs and $M_{\rm BH}$ \citep{mummery_fundamental_2024}, which can be used as a proxy for black hole mass in cases where a plateau luminosity cannot be secured (or in this case, may not be physically applicable). Though this correlation provides typically less constraining measurements of $M_{\rm BH}$ \citep[scatter of 0.4 dex; ][]{mummery_fundamental_2024}, it is observationally inexpensive and therefore well suited for application to large samples of TDEs. 

For all of the rpTDEs, we determine the peak $g$-band luminosities from available photometry, and report these luminosities alongside their inferred black hole masses using the scaling relationship of \cite{mummery_fundamental_2024} in Table \ref{tab:scaling}. With peak luminosity scaling we find self-consistent black hole masses for TDEs 2022dbl and 2023adr. Once again the lack of data covering the peak of TDE\,2020vdq makes it difficult to check model consistency between peaks.

\subsection{Cooling Envelope Model}\label{sec:redback}

While classical treatments emphasise emission tracking the fallback rate of stellar debris onto the SMBH (often approximated by a $t^{-5/3}$ decay), many well-observed events exhibit shallower declines, large photospheric radii, and nearly constant effective temperatures that are difficult to reconcile with direct disc accretion alone. To address this, the cooling-envelope model invokes a pressure-supported, quasi-spherical envelope formed from the returning stellar debris following super-Eddington fallback onto the black hole \citep{metzger_cooling_2022}. The optical/UV emission is then powered by the cooling and Kelvin–Helmholtz contraction of this envelope, with a compact accretion disc forming only at later stages, which may explain the delayed X-ray or radio signatures observed in some TDEs \citep[e.g.][]{yao_tidal_2022,cendes_ubiquitous_2024}. 

We model the peaks of the rpTDEs with a modified implementation of the cooling envelope model of \cite{sarin_tidal_2024} in the {\tt{Redback}} software package \citep{sarin_redback_2024}. In this implementation, the treatment of the early-time, pre-envelope emission is adjusted to be less tightly coupled to the instantaneous fallback rate, instead adopting a more phenomenological description of the initial energy injection that sets the envelope’s mass and thermal state. This approach preserves the physical interpretation of the subsequent envelope cooling and disc formation phases, while allowing greater flexibility in how the early optical/UV luminosity is generated. Specifically, we adopt a power-law rise and exponential decay which smoothly transitions to the emission from the cooling envelope. In doing so, it is expected to yield black hole mass estimates that are more closely aligned with those inferred from late-time plateau luminosity scaling relations \citep[][]{mummery_fundamental_2024,mummery_optical_2025}, without imposing those relations directly within the model. We further modify the model presented and applied to a sample of TDEs in~\citet{sarin_tidal_2024} to include partial disruptions, by assuming that the disrupted material is some fraction of $M_{\star}$ for penetration factors $\beta \leq 1$ following numerical simulations~\citep{ryu_tidal_2020-1}. 

Following \cite{sarin_tidal_2024, wise_at2019cmw_2025}, we fit the multiband light curves of each peak with the adapted cooling envelope model, using the Dynesty sampler \citep{speagle_dynesty_2020} wrapped with Bilby \citep{ashton_bilby_2019} to sample the parameter space. We present the posterior distributions to $M_{\rm BH}$ and $M_{\star}$ in Table \ref{tab:fits_redback}. 

\begin{table}
    \centering
    \begin{tabular}{l|l|l}
     \hline
        TDE & $\log_{10}M_{\rm BH}$ & $M_{\star}$  \\
            \hline
         2020vdq (Peak 1) &  $5.41_{-0.47}^{+0.68}$  & $1.17_{-0.75}^{+1.11}$ \\[0.3em]
         2020vdq (Peak 2) &  $5.25_{-0.39}^{+0.58}$  & $1.10_{-0.75}^{+1.13}$ \\[0.3em]
         \hline
         2022dbl (Peak 1) &  $5.41_{-0.48}^{+0.70}$  & $1.17_{-0.79}^{+1.13}$ \\[0.3em]
         2022dbl (Peak 2) &  $5.08_{-0.26}^{+0.38}$  & $1.02_{-0.76}^{+1.17}$ \\[0.3em]
         \hline
         2023adr (Peak 1) &  $5.16_{-0.34}^{+0.74}$  & $0.76_{-0.50}^{+1.19}$ \\[0.3em]
         2023adr (Peak 2) &  $6.38_{-0.07}^{+0.09}$  & $0.19_{-0.03}^{+0.04}$ \\[0.3em]
       \hline
    \end{tabular}
    \caption{Fitted parameters from the cooling envelope modelling implemented in {\tt{Redback}}.} 
    \label{tab:fits_redback}
\end{table}

We find that the adapted cooling envelope model implemented in {\tt Redback} does produce self-consistent values of $M_{\rm BH}$ and $M_{\star}$ for all three rpTDEs. These masses are also all systematically lower than those obtained from fallback-based light-curve fits and peak-luminosity scaling relations. This behaviour is qualitatively consistent with recent studies showing that UV plateau–based scaling relations tend to yield lower, and often more host-consistent, $M_{\rm BH}$ estimates than methods that assume the optical/UV emission directly traces the fallback rate \citep[e.g.][]{mummery_fundamental_2024,guolo_compact_2025}. Given that the modified cooling envelope model relaxes the tight coupling between early-time luminosity and fallback rate, it is perhaps not surprising that it  also produces lower $M_{\rm BH}$ values. However, given the potentially complicated context of black hole mass inference within repeating partial TDE systems (see earlier discussion in Section \ref{sec:plateaus}, it is unclear whether for these systems this reflects an improved fidelity of the model, or a limitation of applying plateau-calibrated frameworks to systems that may never reach a true plateau phase.


\section{Host Proxy Measurements}\label{sec:hostproxies}

We compare the TDE lightcurve $M_{\rm BH}$ estimates to those derived from the properties of the host galaxy. Scaling relations between SMBHs and their hosts are thought to arise from co-evolutionary processes such as AGN feedback \citep[e.g.][]{silk_quasars_1998,fabian_observational_2012}, merger-driven bulge growth \citep[e.g.][]{kauffmann_unified_2000,di_matteo_energy_2005}, and the central limit imposed by the gravitational binding of the stellar potential \citep[e.g.][]{ferrarese_fundamental_2000,gebhardt_relationship_2000}. Agreement between light curve and host derived masses would strengthen confidence in both the modelling and the assumed emission mechanism, while discrepancies may point to systematic uncertainties or reveal unusual properties of the host–black hole system \citep[e.g.][]{wevers_black_2017,mockler_weighing_2019}. However, as TDE hosts often occupy the low-mass, post-starburst, and/or bulge-poor regime where these relations are least well constrained \citep[e.g.]{arcavi_continuum_2014,french_tidal_2016,graur_dependence_2018,wevers_black_2019}, a careful comparison is warranted. In this regime, intrinsic scatter, selection effects, and potential deviations from canonical scaling relations may bias host-based $M_{\rm BH}$ estimates, complicating direct one-to-one comparisons. Moreover, if TDEs preferentially occur in galaxies with atypical nuclear structures or recent dynamical disturbances, the underlying assumptions of virial equilibrium and bulge–SMBH co-evolution may not hold. Consequently, systematic offsets between light-curve-inferred and host-derived masses could reflect either limitations of the scaling relations themselves or genuine differences in the growth histories of TDE-hosting black holes, rather than failures of the TDE modelling alone. 

Here we use three commonly used host galaxy properties known to correlate with black hole mass; the stellar velocity dispersion, total stellar mass and galaxy bulge mass. 

\subsection{Stellar and Bulge Masses}

For the rpTDE host galaxies, we use the {\sc{GALFETCH}} pipeline \citep{ramsden_evidence_2025} to collect broad-band photometry from the UV \citep[GALEX][]{martin_galaxy_2005}, optical \citep[SDSS, PanSTARRS, DECAM;][]{honscheid_dark_2008}, NIR \citep[2MASS;][]{skrutskie_two_2006}, mid-IR \citep[unWISE][]{schlafly_unwise_2019} to construct the host SED. We then model host galaxy photometry with the stellar population synthesis code {\sc{Prospector}} \citep{leja_deriving_2017,johnson_stellar_2021}. As per \cite{ramsden_bulge_2022,ramsden_evidence_2025}, we invoke a nine-free parameter model, including a six-component non-parametric star formation history, stellar mass, metallicity and a dust parameter controlling interstellar extinction within the host, sampling the posterior probabilities with {\tt{dynesty}}. From the resulting stellar masses, $M_{*}$, we determine black hole masses using the scaling relationship of \cite{reines_relations_2015}. Uncertainties on these inferred $M_{\rm BH}$ values incorporate typical uncertainties introduced to stellar masses via assumptions made within the SED fitting process regarding star formation and dust attenuation \citep[0.11\,dex for non-parametric star formation histories;][]{lower_how_2020}, alongside 0.24\,dex intrinsic scatter from the $M_{*}-M_{\rm BH}$ relationship \citep{reines_relations_2015}. 

From the stellar masses, we also determine the galaxy bulge mass, $M_{\rm bulge}$, using the bulge-to-total mass ratio, $(B/\,T)_{g}$, determined using the ratio of PanSTARRS PSF and Kron fluxes in the $g$-band, as per \cite{wevers_evidence_2019}. We use the TDE-specific black hole-bulge mass scaling relationship determined by \cite{ramsden_evidence_2025} to infer black hole masses. This relationship introduces an intrinsic uncertainty of 0.24\,dex on $M_{\rm BH}$. We present the best fitting stellar masses for each galaxy, their inferred bulge masses, alongside corresponding black hole mass inferences for each host galaxy proxy in Table \ref{tab:hosts}. 

\begin{table*}
\begin{threeparttable}
\centering
\begin{tabular}{lcccccc}
\hline
TDE Host &
$\log(M_*/M_\odot)$ &
$\log\,M_{\rm BH}(M_*)$ &
$(B/T)_{g}$ &
$\log\,M_{\rm BH}(M_{\rm bulge})$ &
$\sigma_{*}$ (km s$^{-1}$) &
$\log\,M_{\rm BH}(\sigma_{*})$ \\
\hline
2020vdq & $9.22^{0.16}_{0.17}$ & $5.58^{0.37}_{0.38}$ & $0.784\pm0.007$ & $5.96^{0.25}_{0.25}$ & $44\pm±3$~\tnote{\emph{a}}  & $5.59\pm0.29$ \\[0.3em]
2022dbl & $10.29^{0.06}_{0.14}$ & $6.70^{0.29}_{0.32}$ & $0.256\pm0.001$ & $6.65^{0.24}_{0.24}$ & $66.92\pm2.71$~\tnote{\emph{b}}   & $6.40\pm0.33$  \\[0.3em]
2023adr & $10.15^{0.14}_{0.15}$ & $6.56^{0.32}_{0.33}$ & $0.744\pm0.008$ & $7.03^{0.24}_{0.24}$ & $113\pm10.0$ & $6.95\pm0.23$ \\[0.3em]
\hline
    \end{tabular}
    \begin{tablenotes}
      \footnotesize
      \item[\emph{a}]{As reported in \citealt{somalwar_first_2023}.}
      \item[\emph{b}]{As reported in \citealt{lin_unluckiest_2024}.}
\end{tablenotes}
\end{threeparttable}
\caption{Galaxy properties and inferred black hole masses from different host-proxies.}
\label{tab:hosts}
\end{table*}

\subsection{Stellar Velocity Dispersion}

We also compare to black hole masses inferred from the stellar velocity dispersion, $\sigma_{*}$ using the relationship of \cite{kormendy_coevolution_2013}. We take the reported $\sigma_{*}$ and $M_{\rm BH}$ values for TDEs 2020vdq and 2022dbl from \citet{somalwar_first_2023} and \citet{lin_unluckiest_2024} respectively. 

For TDE\,2023adr we measure $\sigma_{*}$ from host galaxy features in the +108d LRIS spectrum.  We use the Penalized Pixel Fitting software \citep[pPXF; ][]{cappellari_full_2023} to determine the stellar kinematics of the galaxy, convolving high-resolution (R$\sim$10,000) spectral templates from the X-shooter Spectral Library \citep{gonneau_x-shooter_2020} to the resolution of the LRIS spectrum (R$\sim5000$) to determine the first two moments of the stellar line-of-sight velocity distribution. We fit to wavelength regions covering the Mg Ib triplet at $5160 - 5190$\AA\, and the Na-D doublet at $5890, 5895$\AA, masking the TDE spectrum, and use a Monte Carlo bootstrap method \citep{geha_least-luminous_2009} to resample the 1D spectrum over 1000 noise realizations to determine the median and uncertainties. We measure a velocity dispersion of 113$\pm$10 km\,s$^{-1}$, and use the scaling relationship of \cite{kormendy_coevolution_2013} to determine the black hole mass. The host galaxy velocity dispersions and black hole masses from the $M_{\rm BH}-\sigma_{*}$ relation are also presented in Table \ref{tab:hosts}. 


\section{Discussion}
\label{sec:discuss}

\begin{figure}
        \includegraphics[width=0.99\columnwidth]{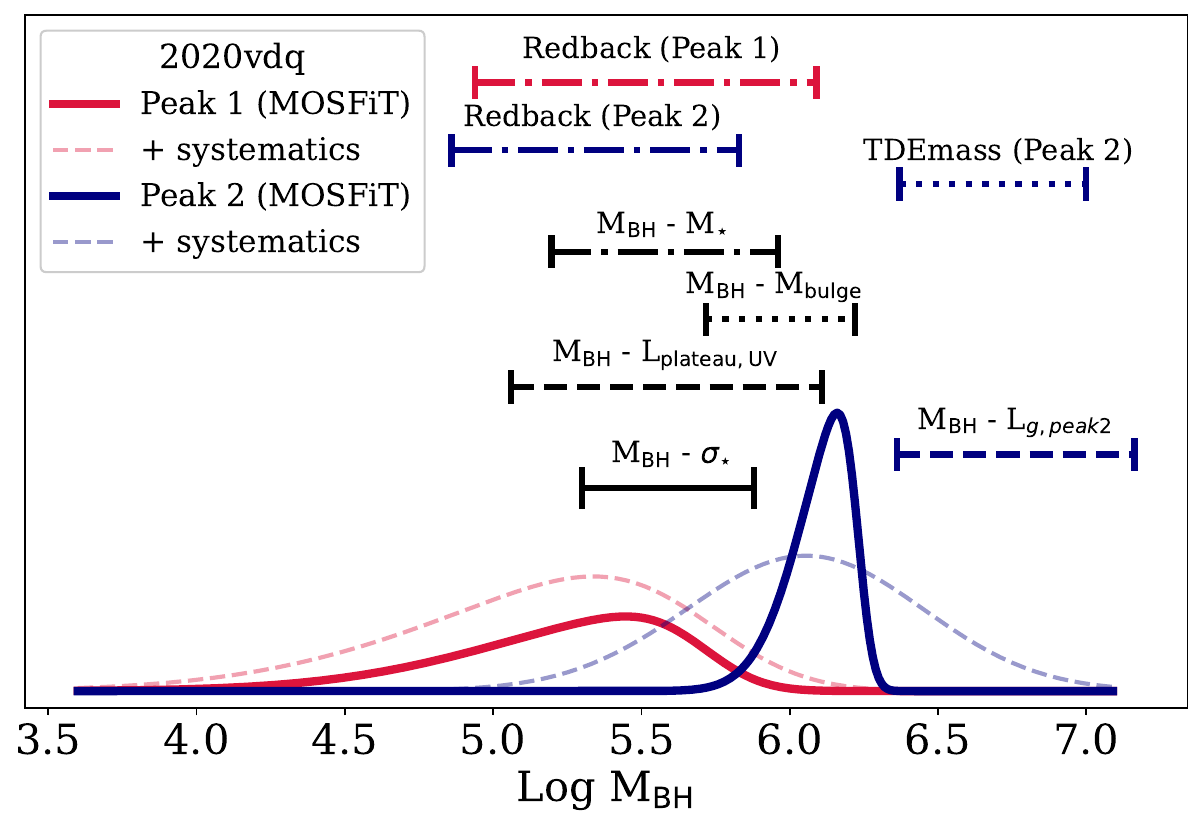}
        \includegraphics[width=0.99\columnwidth]{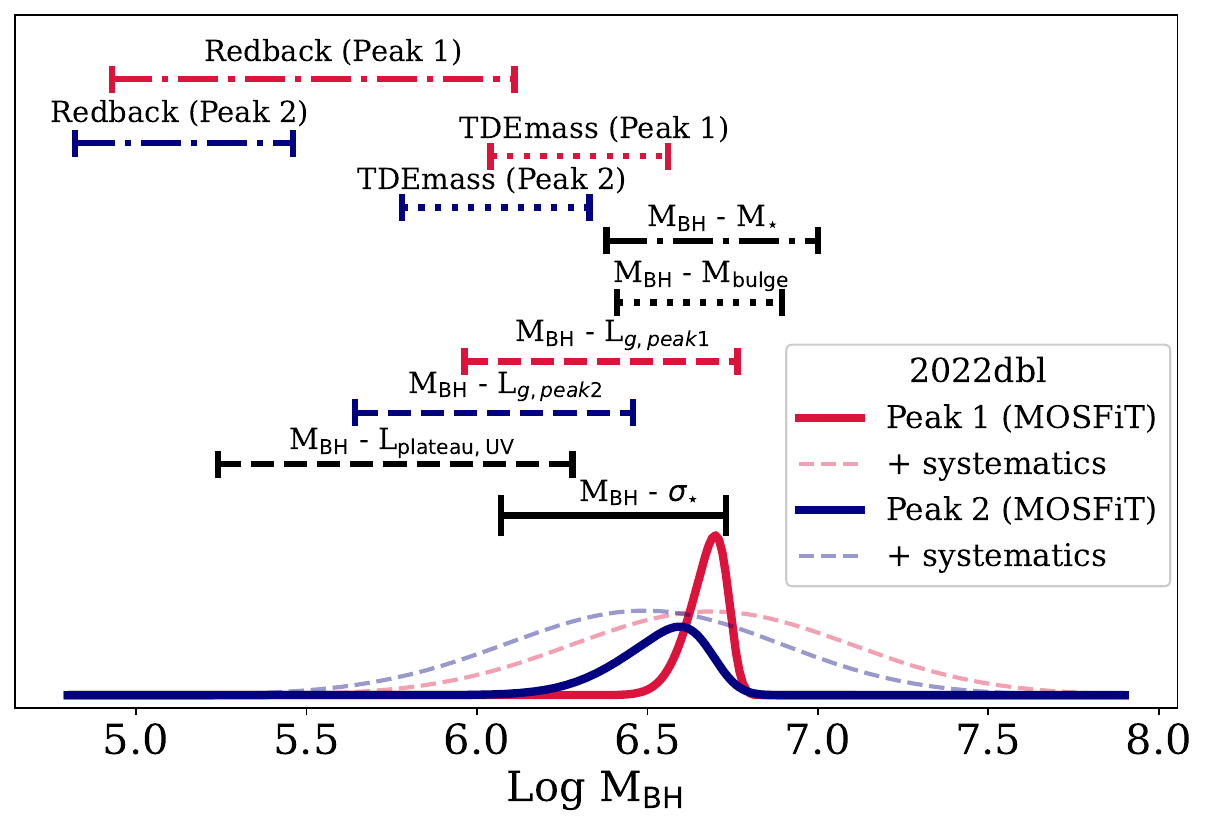}
        \includegraphics[width=0.99\columnwidth]{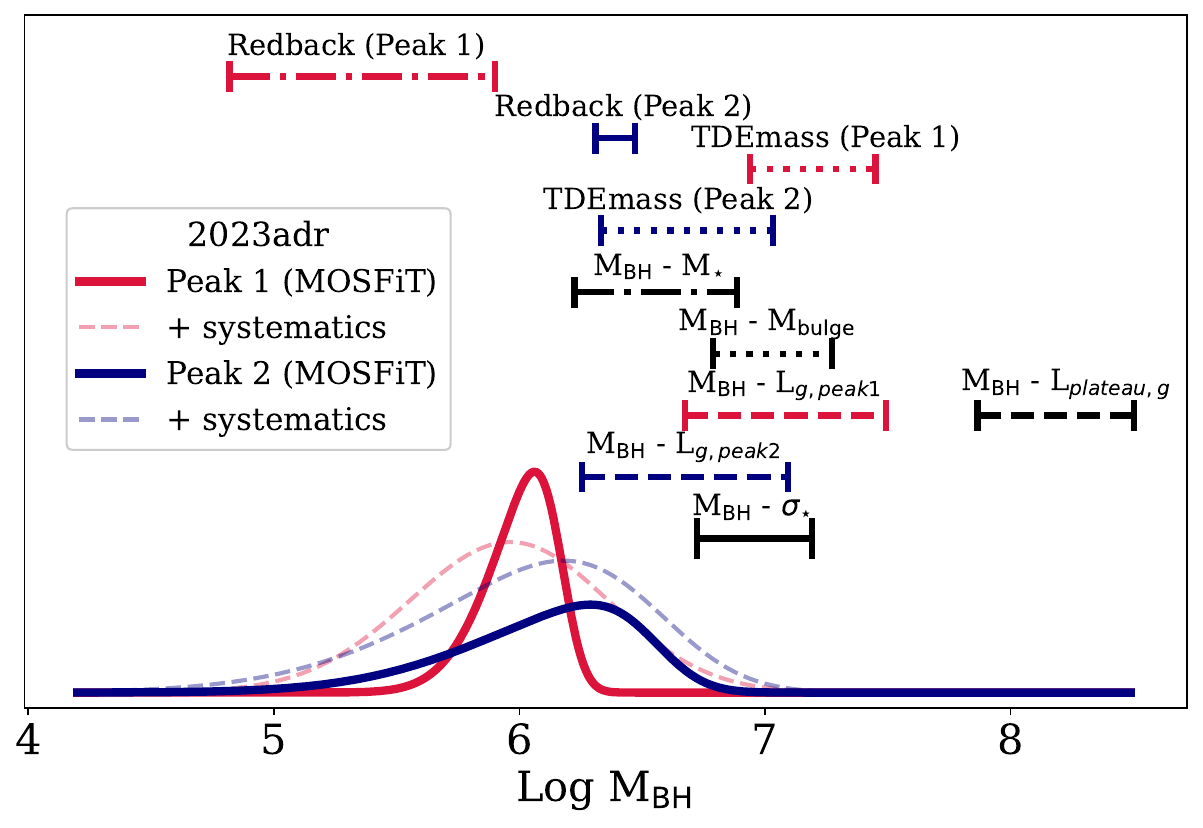}
    \caption{Comparison of $M_{\rm BH}$ 1$\sigma$ posterior distributions from fitting the individual peaks of the rpTDEs (top: TDE\,2020vdq, middle: TDE\,2022dbl, bottom TDE\,2023adr) with the {\tt{MOSFiT}} fallback model, {\tt{TDEmass}} stream collision model, the cooling envelope model implimented in {\tt{Redback}}, plateau- and peak-luminosity scaling relations, and from host galaxy $M_{\rm BH}$ approximations (stellar mass, velocity dispersion, bulge mass fitting). For {\tt{MOSFiT}}, we show fallback posterior distributions with (dashed lines) and without (solid lines) systematics uncertainties  \citep{mockler_weighing_2019}.}
    \label{fig:BH_mass_comp}
\end{figure}

\subsection{Model comparison}
\label{sec:model_comparison}

The three rpTDEs analysed here provide an opportunity to evaluate the reliability of black hole mass measurements inferred from TDE light-curve modelling, by comparing masses inferred from each peak, and also comparing to those derived from host galaxy properties. To illustrate this, in Figure \ref{fig:BH_mass_comp} for each rpTDE we display the posterior $M_{\rm BH}$ distributions from light curve modelling ({\tt{MOSFiT}}), colour-luminosity scaling ({\tt{TDEmass}}), cooling envelope modelling ({\tt{Redback}}), light-curve scaling from disk models, and different host galaxy proxies.

Across the three rpTDEs analysed here, we find that the different methods of measuring $M_{\rm BH}$ using the TDE properties generally produce self-consistent (within $\sim1\sigma$ agreement) black hole mass estimates between rpTDE flares, once both statistical and model-specific systematic uncertainties are accounted for. Further, the different light-curve–based methods yield broadly similar $M_{\rm BH}$ estimates for individual events, and these values are typically consistent within uncertainties with those derived from host-galaxy scaling relations, as has been shown within previous studies \citep[e.g.][]{mockler_weighing_2019,ryu_measuring_2020}. We note, however, that a more stringent test of any TDE-based $M_{\rm BH}$ estimator is whether a population of black hole masses inferred using a single method can reproduce observed host-galaxy scaling relations \citep[e.g.][]{ramsden_bulge_2022, mummery_fundamental_2024,ramsden_evidence_2025}. While such population-level validation is beyond the scope of this work given our small sample size, this remains an important consideration when interpreting light-curve–derived black hole masses.

\begin{figure}
	\includegraphics[width=\columnwidth]{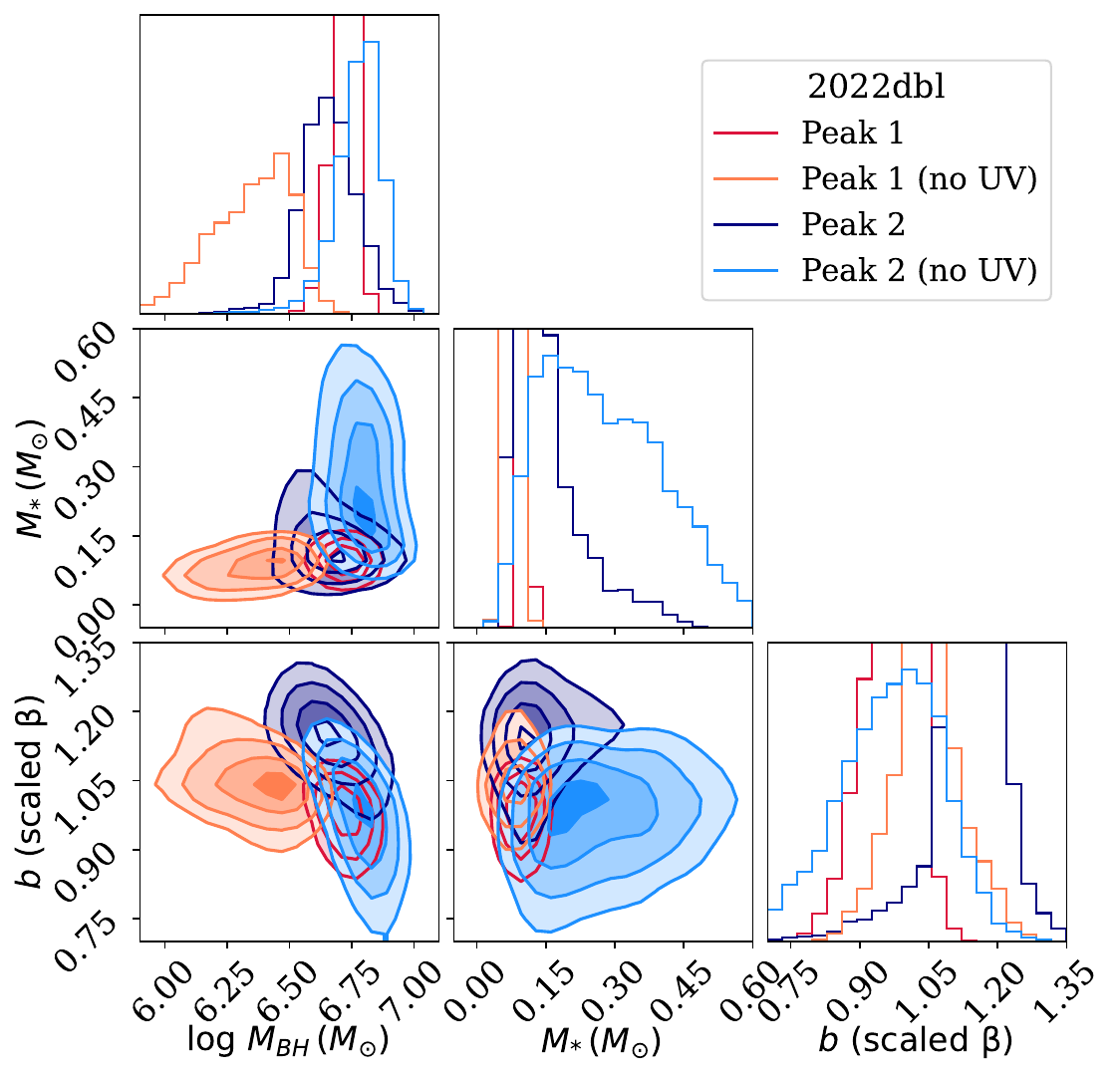}
    \caption{Comparison of {\tt{MOSFiT}} posteriors derived with and without the inclusion of UVOT photometry for TDE\,2022dbl. Fitting with near-UV produces more stringent estimates of key TDE parameters.}
    \label{fig:uv_comp}
\end{figure}

Of the three rpTDEs, TDE\,2022dbl yields by far the most consistent and tightly constrained black hole masses across all methods considered in this work. This is primarily a consequence of its excellent light-curve coverage in both the near-UV and optical during each flare. The quality, cadence, and wavelength range of the photometry used to infer TDE black hole masses exert a strong influence on the resulting constraints. For TDE\,2020vdq, the absence of data around the first peak, arising from the temporal edge of survey coverage, eliminates several estimators ({\tt TDEmass}, peak-luminosity scaling) and leads to weaker {\tt{MOSFiT}} constraints due to the poorly determined disruption epoch. Similarly, for events with limited near-UV coverage (TDE\,2023adr and the first peak of TDE\,2020vdq), uncertainties in the photospheric temperature remain large because the data no longer probe the peak of the spectral energy distribution. This directly affects methods such as {\tt MOSFiT} and {\tt TDEmass}, which rely on accurate temperature and bolometric corrections to recover the fallback rate and therefore $M_{\rm BH}$. Near-UV coverage is particularly important because, despite significant reprocessing, TDEs still radiate most strongly in the ultraviolet. Sampling the SED without the near-UV can lead to underestimated temperatures and peak luminosities, introducing larger uncertainties and potentially biases in the inferred $M_{\rm BH}$. To test the impact of no near-UV data upon light curve modelling, we re-run {\tt{MOSFiT}} for TDE\,2022dbl excluding the UVOT $UVW2$, $UVM2$ and $UVW1$ photometry. We compare the resulting posteriors to the original posteriors in Figure \ref{fig:uv_comp}. Though the non-UV inferred $M_{\star}$, $b$ and $M_{\rm BH}$ values are consistent with our initial results, these posteriors are notably less well-constrained, highlighting the importance of UV photometry in TDE $M_{\rm BH}$ inference methods. 

Given that a significant fraction (or even the majority) of TDEs are expected to be partial disruptions \citep[][]{stone_rates_2016,krolik_tidal_2020,chen_light_2021,zhong_revisit_2022,bortolas_partial_2023},  a potential source of systematic uncertainty in TDE-based black hole mass estimates where the assumption of a full disruption is encoded within the underlying light-curve models (e.g. the fallback model in {\tt{MOSFiT}}, the standard cooling envelope model in {\tt{Redback}}, {\tt{TDEmass}}). Hydrodynamic simulations show that partial encounters reduce the bound mass fraction and the debris energy spread relative to full disruptions \citep[e.g.][]{guillochon_hydrodynamical_2013,ryu_shocks_2023}, which in principle could bias the inferred fallback rate, or  alter the strength and timing of stream–stream collision-powered emission, and therefore affect the recovered $M_{\rm BH}$. In this work we do not find large discrepancies between the black hole masses inferred from the TDE models and those derived from independent galaxy scaling relations. The broad consistency across methods suggests that, for the rpTDEs in our sample, uncertainties related to the full-versus-partial assumption have only a modest impact on the final $M_{\rm BH}$ estimates. This may reflect the fact that the overall timescale of the light curve, which is primarily what constrains $M_{\rm BH}$, is relatively insensitive to the precise fraction of mass stripped, especially when other parameters (e.g.\ $M_{\star}$ and $\beta$) are allowed to vary. Thus, within the precision currently achievable for optical TDE data, the choice between full and partial disruption prescriptions do not appear to introduce the large systematic offsets, though we caution the small number statistics of the sample presented here.

On the other hand, empirical scaling methods, such as the plateau luminosity--black hole mass scaling relations \citep[e.g.][]{mummery_fundamental_2024,mummery_optical_2025}, do not implicitly include assumptions regarding the level of the disruption, and should be free of any systematic offset. Plateau luminosities in particular have been suggested as capable of providing some of the tightest empirical constraints on $M_{\rm BH}$, with recent work by \citet{guolo_compact_2025} showing that plateau-phase modelling can yield particularly precise black hole masses of order $\lesssim 0.3$\,dex when multi-wavelength data are available. However, it may not always be possible to use this method to measure $M_{\rm BH}$ in  rpTDE systems. These relations are calibrated on systems that exhibit a late-time, near-constant optical/UV plateau, interpreted as emission from a quasi-steady accretion flow that has had time to viscously relax after a single major disruption episode. In rpTDEs, however, the bound stellar core can return on an orbital timescale comparable to, or shorter than, the disc viscous time \citep[][]{wevers_live_2023,pasham_potential_2024,liu_rapid_2024,somalwar_first_2025}. In this regime, the accretion flow may never settle into a true plateau between pericentre passages, or the plateau may be truncated or strongly modulated by successive stripping events. Indeed, for TDEs\,2022dbl and 2023adr, $L_{\rm plat}$--$M_{\rm BH}$ scalings produce the most inconsistent black hole masses with respect to host galaxy estimates, which may be indicative of the system not having reached a steady plateau phase. 

Although these approaches consistently recover black hole masses of the same order of magnitude, for all TDE $M_{\rm BH}$ inferences considered here, the associated uncertainties are significant ($\sim$0.3-0.5 dex), even where data quality and coverage is good across both peaks in the case of TDE\,2022dbl. For most, the dominant contribution to the black hole mass error budget arises from systematic uncertainties inherent to the models themselves, which are often not accounted for when estimating black hole masses via light curve modelling in some TDE studies. Reducing these model-driven uncertainties will require improved constraints on the physical state of the disrupted star, better characterisation of the circularisation and emission processes, and more comprehensive multi-wavelength coverage, particularly in the UV, to break degeneracies in temperature, radius evolution, and fallback timescale. Until such systematics are narrowed, TDE-derived black hole masses should be viewed as approximate and model-dependent, even when the statistical posteriors appear precise.

\subsection{Implications for the TDE population}
\label{sec:TDEimplications}

A notable outcome of our modelling is that the fallback model implemented in {\tt{MOSFiT}} tends to favour disruption parameters consistent with full or near-full disruptions when allowed to explore the full parameter space. This behaviour is broadly consistent with the findings of \cite{makrygianni_double_2025}, who argue that a large fraction of currently observed TDEs may in fact originate from partial disruptions, with light curves similar to those of full disruptions. Given this, it is perhaps unsurprising that methods such as {\tt TDEmass} and empirical peak-luminosity scalings which implicitly assume a full disruption do not ``break'' when applied to events that may be partial in nature. This would support theoretical predictions that the optical TDE population may be dominated by partial disruptions \citep[e.g.][]{bortolas_partial_2023}, even if light-curve–based models cannot currently explicitly distinguish between the two scenarios.

While the inferred black hole masses appear reasonably robust across different light-curve models and host-galaxy scaling relations, parameters such as the penetration factor $\beta$ and stellar mass $M_{\star}$ remain highly degenerate. As shown in Section~\ref{sec:mosfit}, small changes in $\beta$ can be compensated by large shifts in $M_{\star}$ with only small changes in $M_{\rm BH}$, making the detailed physical interpretation of these parameter posteriors problematic. We note that similar behaviour has also been reported for the cooling-envelope framework: \citet{wise_at2019cmw_2025} find that fits to the TDE candidate AT\,2019cmw using the cooling-envelope model favour high stellar masses and impact parameters. This does suggests that degeneracies in stellar mass and impact parameter may be a more general challenge across current semi-analytic TDE models. 

This degeneracy does lend support to the recent results of \citet{pursiainen_muse_2025}, who used IFU spectroscopy of TDE host nuclei to show that the stellar population demographics inferred directly from the galaxy are often in tension with the masses of the disrupted stars implied by light-curve fits. Our results are consistent with this picture: while $M_{\rm BH}$ is primarily constrained by the global timescale of the flare, the stellar properties remain largely unconstrained within the current modelling framework, especially once uncertainties in the disruption depth are taken into account. Indeed, if modelling degeneracies allow partial disruptions to masquerade as full disruptions by favouring higher $\beta$ and lower $M_{\star}$, then, should a significant fraction of TDEs fall into this regime, the discrepancy highlighted by \citet{pursiainen_muse_2025} may be further exacerbated.

These limitations highlight the need for more physically realistic stellar models within TDE light-curve fitting frameworks. Current implementations, including {\tt{MOSFiT}}, rely on simplified mappings from stellar mass and radius to fallback rates based on a grid of hydrodynamic simulations. Improved stellar evolution models, a more complete treatment of partial disruptions, and better interpolation across stellar structure parameter space would allow for more reliable inference of the disrupted star’s properties and reduce the strong degeneracies currently present. Incorporating these advances into next-generation TDE modelling tools will be essential for linking observed light curves to the underlying physics of tidal stripping and for robustly interpreting the demographics of TDE progenitor stars. In this context, the ability of a given model to recover consistent physical parameters across multiple flares from the same rpTDEs provides an additional and valuable test of its underlying physics, complementary to -- but independent of -- recent population-level assessments based on reproducing host-galaxy scaling relations \citep{ramsden_bulge_2022,mummery_fundamental_2024,ramsden_evidence_2025}.

\subsection{Prospects for TDE $M_{\rm BH}$ Measurements in LSST}
\label{sec:LSST_TDEs}

Given the strong dependence of TDE-based black hole mass estimates on both cadence and wavelength coverage, our results highlight the challenges that will arise when interpreting events discovered by future wide-field time-domain surveys. Although the Rubin Observatory LSST will be a powerhouse for transient discovery, the vast majority of the TDEs it detects are unlikely to receive extensive dedicated follow-up due to limited observational resources, making it essential to assess what physical information can be reliably recovered from the LSST light curves alone. Recent population-level simulations by \citet{french_prospects_2025} demonstrate that LSST-quality light curves can recover black hole masses with typical uncertainties of a few tenths of a dex when well sampled, but also show that parameter recovery is highly sensitive to early-time coverage and survey cadence. Assessing the reliability of $M_{\rm BH}$, $M_{\star}$, and $\beta$ recovery from LSST data therefore requires controlled tests using realistic survey conditions that isolate the impact of observing strategy from intrinsic TDE diversity.

To quantify the performance of recovering parameters from LSST-quality light curves, we use {\tt{MOSFiT}} to simulate the two flares of TDE\,2022dbl, using the parameters and uncertainties from the posterior distributions reported in Table~\ref{tab:fits_mosfit}. Synthetic light curves are generated in the LSST $u$, $g$, $r$, $i$, $z$, and $y$ filters. To investigate the role of rest-frame near-UV coverage, which strongly influences temperature and bolometric corrections, we simulate each flare at two redshifts: the native redshift of TDE\,2022dbl ($z=0.0284$), and a higher redshift ($z=0.6$) chosen so that rest-frame emission corresponding to the UVOT $UVW2$, $UVM2$, and $UVW1$ bands is shifted into the LSST $u$ and $g$ filters while still yielding detectable flux in individual LSST visits (depths of $m \approx 24.5$).

To simulate LSST candence, we use the pipeline of \cite{magill_mallorn_2025}. For each simulation, we resample the model light curves using the LSST cadence Baseline v5.0.0 from the Rubin Survey Simulator \citep{peter_yoachim_lsstrubin_sim_2025,peter_yoachim_lsstrubin_scheduler_2025}. Each peak is assigned a random sky position within the LSST footprint and a random occurrence time within the 10-year survey. The corresponding visit times and filters are then drawn directly from the cadence simulation. For the photometric uncertainties, we generate them using the 5$\sigma$ limits from the LSST Survey Simulator, assuming that observations are sky-noise limited. Further details of TDE simulations in LSST are discussed within \cite{magill_mallorn_2025}.

\begin{figure}
	\includegraphics[width=1\columnwidth]{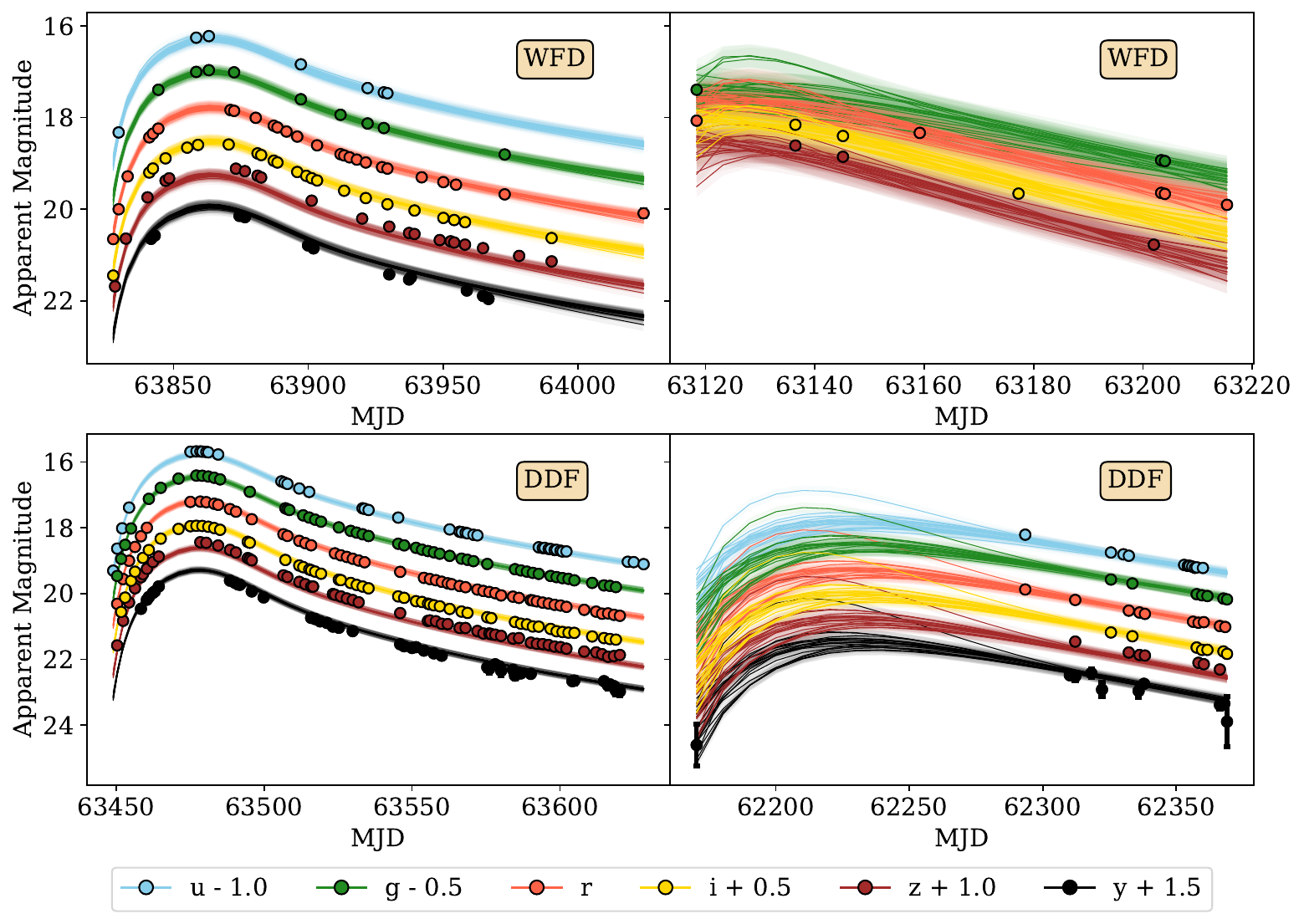}\\
    \caption{Example simulated LSST WFD and DDF light curves at the native redshift of TDE\,2022dbl of $z=0.0284$, alongside their resulting {\tt{MOSFiT}} models. Well sampled examples from each LSST survey are shown on the left, and poorly sampled ones the right. Filters are offset in brightness for clarity.}.
    \label{fig:LSST_sim_LC}
\end{figure}

At the native redshift, each peak is simulated ten times in both the Wide-Fast-Deep (WFD) survey and the Deep Drilling Fields (DDF), yielding forty simulations in total. To allow a controlled comparison of parameter recovery with increased rest-frame UV coverage, we process the $z=0.6$ light curves at the same sky positions and epochs as the native redshift. Due to its lower cadence, at high redshift, too few epochs are recovered with sufficient signal-to-noise in the WFD survey to permit a meaningful light-curve fit within {\tt{MOSFiT}}. Hence for $z=0.6$, we only consider the results of fitting the light curves in the DDF survey.

\begin{figure}
	\includegraphics[width=0.85\columnwidth]{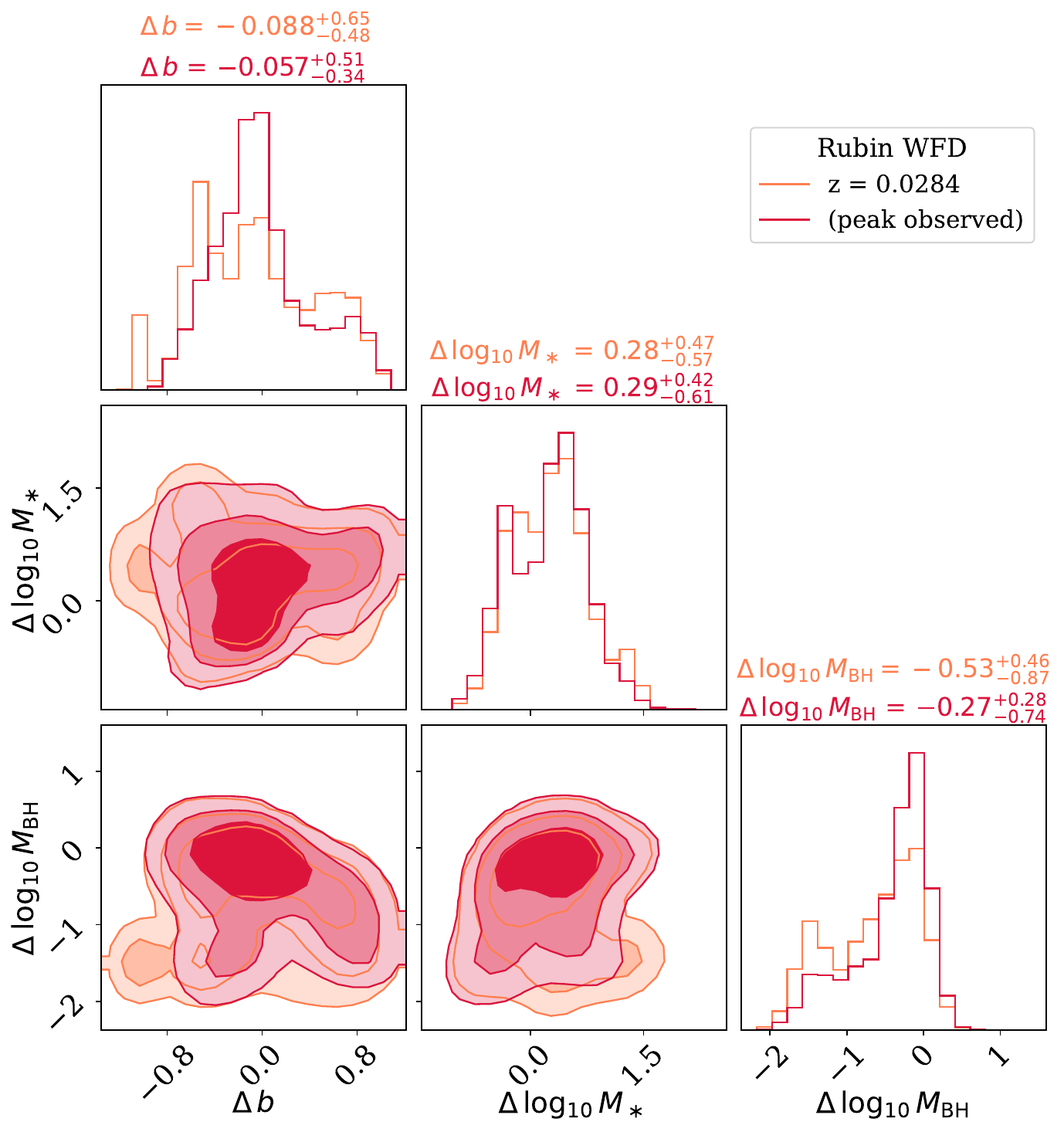}\\
        \includegraphics[width=0.85\columnwidth]{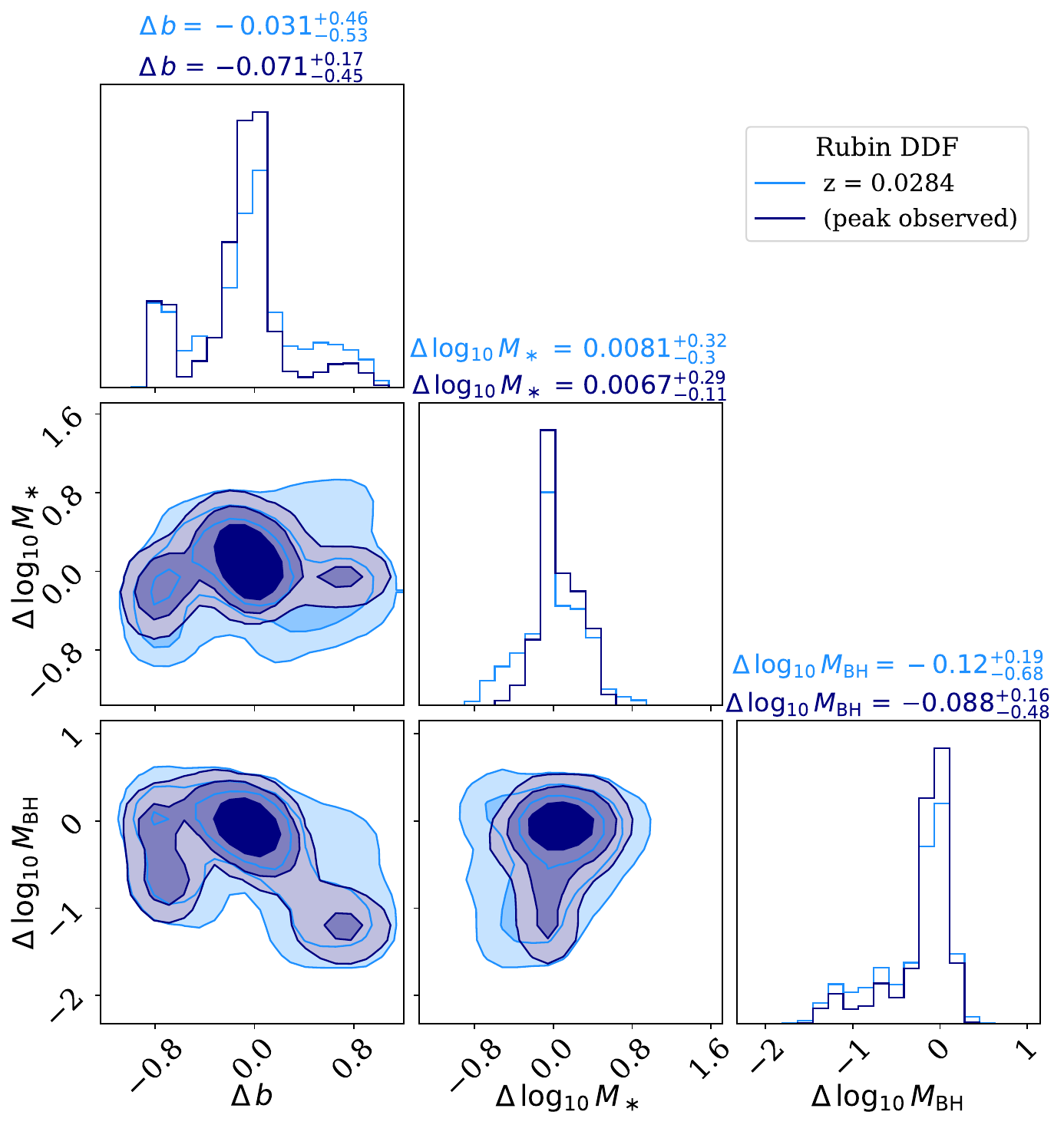}
        \includegraphics[width=0.85
        \columnwidth]{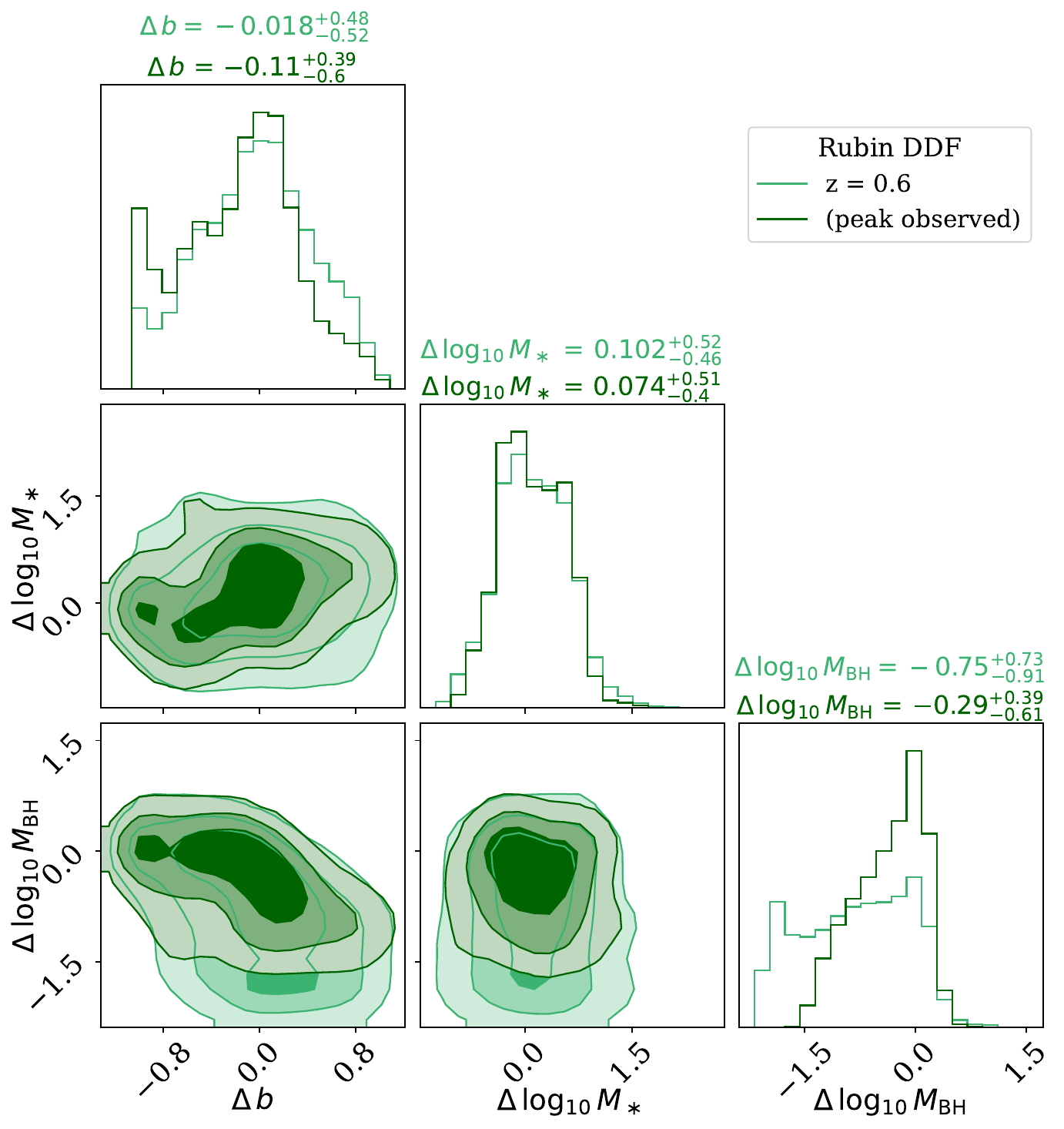}
    \caption{Key parameter recovery rate for simulations in LSST WDFs ({\textit{top}}) and in the DDFs ({\textit{middle}}). We show also parameter recovery in simulations at higher redshift from the DDF fields ({\it{bottom}}). Darker shades indicate residuals when selecting simulations where the peak of the light curve can be recovered.}
    \label{fig:LSST_param_recover}
\end{figure}

Example simulated light curves and their {\tt{MOSFiT}} fits are shown in Figure~\ref{fig:LSST_sim_LC}. All simulated curves are fitted using the broad priors listed in Table~\ref{tab:mosfit_tde_priors}. We visually inspect the resulting fits and discard cases where the model fails to reproduce the light-curve morphology. At $z=0.0284$, we retain 19/20 WFD and 20/20 DDF simulations. At $z=0.6$, only 10/20 DDF simulations meet this criterion. Figure~\ref{fig:LSST_param_recover} shows the posterior offsets relative to the true (input) parameters.

Figure~\ref{fig:LSST_param_recover} demonstrates that the increased cadence of the DDF survey yields substantially better parameter recovery compared to the WFD survey, with a mean offset of $-0.12$\,dex relative to the true $M_{\rm BH}$, compared to $-0.5$\,dex for the WFD simulations. Recovery degrades significantly at higher redshift despite the shift of rest-frame UV emission into the LSST bands, reflecting the trade-off between improved wavelength coverage and reduced signal-to-noise across the light curve. However, when selecting simulations where the peak of the light curve is recovered, or inferred within the light curve data (i.e. detections before and after peak), we see 26--61\% improvement in parameter recovery. 

Across both LSST survey modes, we find that {\tt{MOSFiT}} systematically recovers black hole masses that are lower than the true (input) values, with mean residuals of $\sim\!-0.12$\,dex in the DDF and $\sim\!-0.5$\,dex in the WFD simulations. While this result is based on simulations of a single, well-characterised event and should therefore be viewed as illustrative rather than population-level, a similar tendency towards lower recovered $M_{\rm BH}$ values is also apparent in recent LSST simulations by \citet{french_prospects_2025}. In that work, {\tt{MOSFiT}} fits to synthetic LSST TDE light curves suggest an offset of order $\sim$-0.4\,dex for black holes with $M_{\rm BH}<10^{7}\,M_{\odot}$ (see their Figure~4), although this bias was not explicitly quantified or discussed in the study.

Within the {\tt{MOSFiT}} framework, this might be expected given the cadence and wavelength limitations of the LSST data, where the black hole mass is primarily constrained by the characteristic width of the light curve. Sparse sampling of the rise and peak in the WFD survey may compress the inferred timescale of the flare, which the model interprets as a shorter fallback time and therefore a smaller $M_{\rm BH}$. In addition, the absence of near-UV information forces the SED to be fit only on the Rayleigh--Jeans tail of the assumed underlying black body, leading to systematically cooler inferred temperatures and lower bolometric luminosities. This would favour solutions towards lower accretion rates, thus biasing the posterior towards the lower-mass region of the $(M_{\rm BH}, M_{\star}, \beta)$ degeneracy. This effect is reduced in the DDF simulations, where the higher cadence provides better constraints on the light-curve width and temperature evolution between epochs, but the modest underestimation of $M_{\rm BH}$ persists due to the limited rest-frame UV coverage of LSST without targeted follow-up.

The implications of this systematic bias are non-negligible given the expected LSST TDE yield. Forecasts suggest that the LSST WFD survey will discover several hundred to more than a thousand TDEs per year \citep[e.g.][]{van_velzen_optical_2011,bucar_bricman_rubin_2023}, with the majority detected at $z \lesssim 0.5$. Over the 10-year survey this corresponds to $\sim 3{,}000$--$10{,}000$ TDEs whose inferred $M_{\rm BH}$ values may be affected by the cadence- and SED-driven biases identified here. The DDF fields are expected to yield far fewer events, on the order of a few tens per year \citep{bucar_bricman_rubin_2023}, but with substantially higher-quality light curves that mitigate these biases. Thus, while only a modest fraction of LSST TDEs will benefit from DDF-like sampling, the vast majority discovered in the WFD survey will be impacted by the systematic tendency of {\tt{MOSFiT}} to underpredict $M_{\rm BH}$ when light curves are sparsely sampled or lack rest-frame near-UV coverage. This underscores the important role that photometric follow-up observations will still need to play in the LSST era. Overlapping multi-band surveys, such as ZTF, ATLAS, BlackGEM \citep{groot_blackgem_2024}, alongside surveys designed to fill in gaps in the LSST cadence, such as the La Silla Schmidt Southern Survey \citep{miller_silla_2025}, will be pivotal for improving LSST light curve cadence, whilst expanding wavelength coverage with targeted {\it Swift} data, and later more systematic coverage from the ULTRASAT survey \citep{shvartzvald_ultrasat_2024} will be required to improve rest-frame UV coverage. For estimating black hole masses in large LSST TDE samples, plateau-luminosity scaling relations offer the most observationally economical—and potentially robust—approach \citep{guolo_compact_2025,mummery_tidal_2025} for ensemble measurements of $M_{\rm BH}$. However, the practical recovery of plateau phases from LSST data alone will be increasingly challenging beyond $z \gtrsim 0.1$ in single-visit imaging, and beyond $z \gtrsim 0.3$ even with coadded observations in the DDF fields, reinforcing the need for targeted TDE follow-up in the LSST era.

\section{Conclusions}
\label{sec:conclusions}

We have used three spectroscopically confirmed rpTDEs; TDE\,2020vdq, TDE\,2022dbl, and TDE\,2023adr, to test the robustness of supermassive black hole mass inference from TDE light curves. By independently modelling each flare with a suite of commonly used frameworks (fallback fitting with {\tt{MOSFiT}}, stream–stream collision scalings with {\tt{TDEmass}}, empirical luminosity scaling-relations, and cooling-envelope fits in {\tt{Redback}}, we are able to assess both the internal self-consistency of each framework and the level of agreement across fundamentally different modelling assumptions.

Our main conclusions are as follows:
\begin{itemize}

    \item We find that most methods recover mutually consistent $M_{\rm BH}$ values between flares and are broadly consistent with host-galaxy proxy estimates, with a typical precision of $\sim$0.3–0.5\,dex once model systematics are accounted for. Although we caution that while such agreement may be sufficient for individual objects, recent work suggests that method-dependent biases may limit the ability of some light-curve–based estimators to recover black hole scaling relations in population-level analyses \citep{guolo_compact_2025,mummery_fundamental_2024,mummery_tidal_2025}.

    \item Although $M_{\rm BH}$ appears comparatively robust, the stellar mass $M_{\star}$ and encounter geometry (e.g.\ $\beta$ or the scaled parameter $b$ in {\tt MOSFiT}) remain highly degenerate. In several cases fallback-based fits prefer near-full disruption solutions and low inferred stellar masses despite the rpTDE nature of these systems, indicating that current fallback grids and stellar structure prescriptions are not yet sufficient for reliable inference of stellar properties in repeating partial disruptions.
    
    \item The quality of the light-curve constraints is strongly driven by cadence and wavelength coverage. In particular, near-UV photometry substantially tightens constraints by anchoring the SED near its peak and reducing degeneracies in temperature and bolometric corrections; events lacking near-UV coverage or peak sampling yield noticeably broader posteriors and weaker model discrimination.
    
    \item Simulations of LSST-quality light curves demonstrate that cadence and SED limitations can introduce systematic offsets in model-based inference. Using TDE\,2022dbl as a representative test case, we find that fitting simulated LSST data with {\tt MOSFiT} leads to a systematic underestimation of $M_{\rm BH}$, with mean offsets of $-0.12$\,dex for DDF-like sampling and $-0.5$\,dex for WFD-like sampling. These offsets are primarily driven by sparse constraints on the rise and peak of the light curve and by the absence of rest-frame near-UV information. While this result is based on simulations of a single well-characterised event, a similar tendency towards lower recovered $M_{\rm BH}$ values is also apparent in recent LSST simulations \citep[e.g.][]{french_prospects_2025}, suggesting this effect may be generic and warrants further investigation across a broader range of TDE properties.
    
    \item Given the expected LSST TDE yield, these effects are likely to impact a substantial fraction of the thousands of photometrically selected events. While DDF-like cadence will mitigate biases for a minority of TDEs, robust $M_{\rm BH}$ inference for the broader LSST sample will benefit from complementary follow-up that improves cadence and extends wavelength coverage into the near-UV.
    
\end{itemize}

For handling future large samples of TDE $M_{\rm BH}$ estimates, the most pragmatic approach may be a complementary use of empirical and physically motivated methods. Plateau-luminosity scaling relations offer a comparatively robust and observationally economical route to calibrating $M_{\rm BH}$ at low redshift, where late-time UV/optical plateaus are accessible and can be directly cross-validated against host-galaxy properties. These well-characterised low-$z$ events can then anchor black hole–galaxy scaling relations that are applied at higher redshift, where plateau phases will be increasingly difficult to recover but deep host-galaxy imaging will remain available. At the same time, continued development of light-curve–based models remains essential, as reliably recovering stellar masses and encounter parameters is the only way to test whether the physical mechanisms powering TDE light curve are being correctly captured.

\section*{Acknowledgements}

We thank H. Dykaar, M. Drout and L. Makrygianni for valuable discussions on the {\tt{MOSFiT}} modelling and parameter inference of TDE\,2022dbl.

CRA and MN are supported by the European Research Council (ERC) under the European Union's Horizon 2020 research and innovation program (grant agreement No. 948381). AJS is supported by a DfE studentship. DM acknowledges a studentship funded by the Leverhulme Interdisciplinary Network on Algorithmic Solutions. PR acknowledges support from STFC grant 2742655. SJS acknowledges funding from STFC Grant ST/Y001605/1, a Royal Society Research Professorship and the Hintze Family Charitable Foundation. 

This work made use of data supplied by the UK Swift Science Data Centre at the University of Leicester.

Based on observations obtained with the Samuel Oschin Telescope 48-inch and the 60-inch Telescope at the Palomar Observatory as part of the Zwicky Transient Facility project. ZTF is supported by the National Science Foundation under Grants No. AST-1440341, AST-2034437, and currently Award 2407588. ZTF receives additional funding from the ZTF partnership. Current members include Caltech, USA; Caltech/IPAC, USA; University of Maryland, USA; University of California, Berkeley, USA; University of Wisconsin at Milwaukee, USA; Cornell University, USA; Drexel University, USA; University of North Carolina at Chapel Hill, USA; Institute of Science and Technology, Austria; National Central University, Taiwan, and OKC, University of Stockholm, Sweden. Operations are conducted by Caltech's Optical Observatory (COO), Caltech/IPAC, and the University of Washington at Seattle, USA.

The ZTF forced-photometry service was funded under the Heising-Simons Foundation grant 12540303 (PI: Graham).

This work has made use of data from the Asteroid Terrestrial-impact Last Alert System (ATLAS) project. ATLAS is primarily funded to search for near-Earth asteroids through NASA grants NN12AR55G, 80NSSC18K0284, and 80NSSC18K1575; by-products of the NEO search include images and catalogs from the survey area. The ATLAS science products have been made possible through the contributions of the University of Hawaii Institute for Astronomy, the Queen's University Belfast, and the Space Telescope Science Institute. 

Pan-STARRS is a project of the Institute for Astronomy of the University of Hawaii and is supported by the NASA SSO Near Earth Observation Program under grants 80NSSC18K0971, NNX14AM74G, NNX12AR65G, NNX13AQ47G, NNX08AR22G, and 80NSSC21K1572 and by the State of Hawaii. The Pan-STARRS1 Surveys (PS1) and the PS1 public science archive have been made possible through contributions by the Institute for Astronomy, the University of Hawaii, the Pan-STARRS Project Office, the Max Planck Society and its participating institutes, the Max Planck Institute for Astronomy, Heidelberg, and the Max Planck Institute for Extraterrestrial Physics, Garching, The Johns Hopkins University, Durham University, the University of Edinburgh, the Queen's University Belfast, the Harvard-Smithsonian Center for Astrophysics, the Las Cumbres Observatory Global Telescope Network Incorporated, the National Central University of Taiwan, STScI, NASA under grant NNX08AR22G issued through the Planetary Science Division of the NASA Science Mission Directorate, NSF grant AST-1238877, the University of Maryland, Eotvos Lorand University (ELTE), the Los Alamos National Laboratory, and the Gordon and Betty Moore Foundation.

SED Machine is based upon work supported by the National Science Foundation under grant No. 1106171.

Some of the data presented herein were obtained at Keck Observatory, which is a private 501(c)3 non-profit organization operated as a scientific partnership among the California Institute of Technology, the University of California, and the National Aeronautics and Space Administration. The Observatory was made possible by the generous financial support of the W. M. Keck Foundation.

The authors wish to recognize and acknowledge the very significant cultural role and reverence that the summit of Maunakea has always had within the Native Hawaiian community. We are most fortunate to have the opportunity to conduct observations from this mountain.

\section*{Data Availability}

The data presented in this paper will be provided upon request to the author. Spectra will be uploaded to WiseRep following acceptance. Photometry is listed in online supplementary data tables.
 


\bibliographystyle{mnras}
\bibliography{rpTDE_BHs} 

@article{ivezic_lsst_2019,
	title = {{LSST}: {From} {Science} {Drivers} to {Reference} {Design} and {Anticipated} {Data} {Products}},
	volume = {873},
	issn = {0004-637X},
	shorttitle = {{LSST}},
	url = {https://ui.adsabs.harvard.edu/abs/2019ApJ...873..111I},
	doi = {10.3847/1538-4357/ab042c},
	urldate = {2025-05-19},
	journal = {The Astrophysical Journal},
	author = {{Ivezi{\'c}}, {\v{Z}}eljko and Kahn, Steven M. and Tyson, J. Anthony and Abel, Bob and Acosta, Emily and Allsman, Robyn and Alonso, David and AlSayyad, Yusra and Anderson, Scott F. and Andrew, John and Angel, James Roger P. and Angeli, George Z. and Ansari, Reza and Antilogus, Pierre and Araujo, Constanza and Armstrong, Robert and Arndt, Kirk T. and Astier, Pierre and Aubourg, Éric and Auza, Nicole and Axelrod, Tim S. and Bard, Deborah J. and Barr, Jeff D. and Barrau, Aurelian and Bartlett, James G. and Bauer, Amanda E. and Bauman, Brian J. and Baumont, Sylvain and Bechtol, Ellen and Bechtol, Keith and Becker, Andrew C. and Becla, Jacek and Beldica, Cristina and Bellavia, Steve and Bianco, Federica B. and Biswas, Rahul and Blanc, Guillaume and Blazek, Jonathan and Blandford, Roger D. and Bloom, Josh S. and Bogart, Joanne and Bond, Tim W. and Booth, Michael T. and Borgland, Anders W. and Borne, Kirk and Bosch, James F. and Boutigny, Dominique and Brackett, Craig A. and Bradshaw, Andrew and Brandt, William Nielsen and Brown, Michael E. and Bullock, James S. and Burchat, Patricia and Burke, David L. and Cagnoli, Gianpietro and Calabrese, Daniel and Callahan, Shawn and Callen, Alice L. and Carlin, Jeffrey L. and Carlson, Erin L. and Chandrasekharan, Srinivasan and Charles-Emerson, Glenaver and Chesley, Steve and Cheu, Elliott C. and Chiang, Hsin-Fang and Chiang, James and Chirino, Carol and Chow, Derek and Ciardi, David R. and Claver, Charles F. and Cohen-Tanugi, Johann and Cockrum, Joseph J. and Coles, Rebecca and Connolly, Andrew J. and Cook, Kem H. and Cooray, Asantha and Covey, Kevin R. and Cribbs, Chris and Cui, Wei and Cutri, Roc and Daly, Philip N. and Daniel, Scott F. and Daruich, Felipe and Daubard, Guillaume and Daues, Greg and Dawson, William and Delgado, Francisco and Dellapenna, Alfred and de Peyster, Robert and de Val-Borro, Miguel and Digel, Seth W. and Doherty, Peter and Dubois, Richard and Dubois-Felsmann, Gregory P. and Durech, Josef and Economou, Frossie and Eifler, Tim and Eracleous, Michael and Emmons, Benjamin L. and Fausti Neto, Angelo and Ferguson, Henry and Figueroa, Enrique and Fisher-Levine, Merlin and Focke, Warren and Foss, Michael D. and Frank, James and Freemon, Michael D. and Gangler, Emmanuel and Gawiser, Eric and Geary, John C. and Gee, Perry and Geha, Marla and Gessner, Charles J. B. and Gibson, Robert R. and Gilmore, D. Kirk and Glanzman, Thomas and Glick, William and Goldina, Tatiana and Goldstein, Daniel A. and Goodenow, Iain and Graham, Melissa L. and Gressler, William J. and Gris, Philippe and Guy, Leanne P. and Guyonnet, Augustin and Haller, Gunther and Harris, Ron and Hascall, Patrick A. and Haupt, Justine and Hernandez, Fabio and Herrmann, Sven and Hileman, Edward and Hoblitt, Joshua and Hodgson, John A. and Hogan, Craig and Howard, James D. and Huang, Dajun and Huffer, Michael E. and Ingraham, Patrick and Innes, Walter R. and Jacoby, Suzanne H. and Jain, Bhuvnesh and Jammes, Fabrice and Jee, M. James and Jenness, Tim and Jernigan, Garrett and Jevremović, Darko and Johns, Kenneth and Johnson, Anthony S. and Johnson, Margaret W. G. and Jones, R. Lynne and Juramy-Gilles, Claire and Jurić, Mario and Kalirai, Jason S. and Kallivayalil, Nitya J. and Kalmbach, Bryce and Kantor, Jeffrey P. and Karst, Pierre and Kasliwal, Mansi M. and Kelly, Heather and Kessler, Richard and Kinnison, Veronica and Kirkby, David and Knox, Lloyd and Kotov, Ivan V. and Krabbendam, Victor L. and Krughoff, K. Simon and Kubánek, Petr and Kuczewski, John and Kulkarni, Shri and Ku, John and Kurita, Nadine R. and Lage, Craig S. and Lambert, Ron and Lange, Travis and Langton, J. Brian and Le Guillou, Laurent and Levine, Deborah and Liang, Ming and Lim, Kian-Tat and Lintott, Chris J. and Long, Kevin E. and Lopez, Margaux and Lotz, Paul J. and Lupton, Robert H. and Lust, Nate B. and MacArthur, Lauren A. and Mahabal, Ashish and Mandelbaum, Rachel and Markiewicz, Thomas W. and Marsh, Darren S. and Marshall, Philip J. and Marshall, Stuart and May, Morgan and McKercher, Robert and McQueen, Michelle and Meyers, Joshua and Migliore, Myriam and Miller, Michelle and Mills, David J.},
	month = mar,
	year = {2019},
	note = {Publisher: IOP
ADS Bibcode: 2019ApJ...873..111I},
	pages = {111},
}

@article{ryu_measuring_2020,
	title = {Measuring {Stellar} and {Black} {Hole} {Masses} of {Tidal} {Disruption} {Events}},
	volume = {904},
	issn = {0004-637X},
	url = {https://ui.adsabs.harvard.edu/abs/2020ApJ...904...73R},
	doi = {10.3847/1538-4357/abbf4d},
	urldate = {2025-04-10},
	journal = {The Astrophysical Journal},
	author = {Ryu, Taeho and Krolik, Julian and Piran, Tsvi},
	month = nov,
	year = {2020},
	note = {Publisher: IOP
ADS Bibcode: 2020ApJ...904...73R},
	pages = {73},
}

@article{ryu_tidal_2020-1,
	title = {Tidal {Disruptions} of {Main}-sequence {Stars}. {III}. {Stellar} {Mass} {Dependence} of the {Character} of {Partial} {Disruptions}},
	volume = {904},
	issn = {0004-637X},
	url = {https://ui.adsabs.harvard.edu/abs/2020ApJ...904..100R},
	doi = {10.3847/1538-4357/abb3ce},
	urldate = {2025-04-11},
	journal = {The Astrophysical Journal},
	author = {Ryu, Taeho and Krolik, Julian and Piran, Tsvi and Noble, Scott C.},
	month = dec,
	year = {2020},
	note = {Publisher: IOP
ADS Bibcode: 2020ApJ...904..100R},
	pages = {100},
}

@misc{mummery_tidal_2025,
	title = {Tidal disruption event {Calorimetry}: {Observational} constraints on the physics of {TDE} optical flares},
	shorttitle = {Tidal disruption event {Calorimetry}},
	url = {https://ui.adsabs.harvard.edu/abs/2025arXiv251209143M},
	urldate = {2025-12-11},
	publisher = {arXiv},
	author = {Mummery, Andrew and Metzger, Brian and van Velzen, Sjoert and Guolo, Muryel},
	month = dec,
	year = {2025},
	note = {ADS Bibcode: 2025arXiv251209143M},
}

@article{stein_tdescore_2024,
	title = {tdescore: {An} {Accurate} {Photometric} {Classifier} for {Tidal} {Disruption} {Events}},
	volume = {965},
	issn = {2041-8205},
	shorttitle = {tdescore},
	url = {https://doi.org/10.3847/2041-8213/ad3337},
	doi = {10.3847/2041-8213/ad3337},
	language = {en},
	number = {2},
	urldate = {2026-01-05},
	journal = {The Astrophysical Journal Letters},
	author = {Stein, Robert and Mahabal, Ashish and Reusch, Simeon and Graham, Matthew and Kasliwal, Mansi M. and Kowalski, Marek and Gezari, Suvi and Hammerstein, Erica and Nakoneczny, Szymon J. and Nicholl, Matt and Sollerman, Jesper and van Velzen, Sjoert and Yao, Yuhan and Laher, Russ R. and Rusholme, Ben},
	month = apr,
	year = {2024},
	note = {Publisher: The American Astronomical Society},
	pages = {L14},
}

@article{pasham_potential_2024,
	title = {A {Potential} {Second} {Shutoff} from {AT2018fyk}: {An} {Updated} {Orbital} {Ephemeris} of the {Surviving} {Star} under the {Repeating} {Partial} {Tidal} {Disruption} {Event} {Paradigm}},
	volume = {971},
	issn = {0004-637X},
	shorttitle = {A {Potential} {Second} {Shutoff} from {AT2018fyk}},
	url = {https://ui.adsabs.harvard.edu/abs/2024ApJ...971L..31P},
	doi = {10.3847/2041-8213/ad57b3},
	urldate = {2025-12-19},
	journal = {The Astrophysical Journal},
	author = {Pasham, Dheeraj and Coughlin, E. R. and Guolo, M. and Wevers, T. and Nixon, C. J. and Hinkle, Jason T. and Bandopadhyay, A.},
	month = aug,
	year = {2024},
	note = {Publisher: IOP
ADS Bibcode: 2024ApJ...971L..31P},
	pages = {L31},
}

@article{aleo_anomaly_2024,
	title = {Anomaly {Detection} and {Approximate} {Similarity} {Searches} of {Transients} in {Real}-time {Data} {Streams}},
	volume = {974},
	issn = {0004-637X},
	url = {https://ui.adsabs.harvard.edu/abs/2024ApJ...974..172A},
	doi = {10.3847/1538-4357/ad6869},
	urldate = {2025-12-18},
	journal = {The Astrophysical Journal},
	author = {Aleo, P. D. and Engel, A. W. and Narayan, G. and Angus, C. R. and Malanchev, K. and Auchettl, K. and Baldassare, V. F. and Berres, A. and de Boer, T. J. L. and Boyd, B. M. and Chambers, K. C. and Davis, K. W. and Esquivel, N. and Farias, D. and Foley, R. J. and Gagliano, A. and Gall, C. and Gao, H. and Gomez, S. and Grayling, M. and Jones, D. O. and Lin, C.-C. and Magnier, E. A. and Mandel, K. S. and Matheson, T. and Raimundo, S. I. and Shah, V. G. and Soraisam, M. D. and de Soto, K. M. and Vicencio, S. and Villar, V. A. and Wainscoat, R. J.},
	month = oct,
	year = {2024},
	note = {Publisher: IOP
ADS Bibcode: 2024ApJ...974..172A},
	pages = {172},
}

@article{perley_ztf_2023,
	title = {{ZTF} superluminous supernova candidates},
	volume = {26},
	url = {https://ui.adsabs.harvard.edu/abs/2023TNSAN..26....1P},
	urldate = {2025-12-18},
	journal = {Transient Name Server AstroNote},
	author = {Perley, D. A. and Lunnan, R. and Wise, J. and Gkini, A. and Brennan, S. and Pessi, P. and Schulze, S. and Kangas, T. and Sollerman, J. and Yan, L. and Chen, T. and Gal-Yam, A.},
	month = feb,
	year = {2023},
	note = {ADS Bibcode: 2023TNSAN..26....1P},
	pages = {1},
}

@article{oke_keck_1995,
	title = {The {Keck} {Low}-{Resolution} {Imaging} {Spectrometer}},
	volume = {107},
	issn = {0004-6280},
	url = {https://ui.adsabs.harvard.edu/abs/1995PASP..107..375O},
	doi = {10.1086/133562},
	urldate = {2025-12-18},
	journal = {Publications of the Astronomical Society of the Pacific},
	author = {Oke, J. B. and Cohen, J. G. and Carr, M. and Cromer, J. and Dingizian, A. and Harris, F. H. and Labrecque, S. and Lucinio, R. and Schaal, W. and Epps, H. and Miller, J.},
	month = apr,
	year = {1995},
	note = {Publisher: IOP
ADS Bibcode: 1995PASP..107..375O},
	pages = {375},
}

@article{blagorodnova_sed_2018,
	title = {The {SED} {Machine}: {A} {Robotic} {Spectrograph} for {Fast} {Transient} {Classification}},
	volume = {130},
	issn = {0004-6280},
	shorttitle = {The {SED} {Machine}},
	url = {https://ui.adsabs.harvard.edu/abs/2018PASP..130c5003B},
	doi = {10.1088/1538-3873/aaa53f},
	urldate = {2025-12-18},
	journal = {Publications of the Astronomical Society of the Pacific},
	author = {Blagorodnova, Nadejda and Neill, James D. and Walters, Richard and Kulkarni, Shrinivas R. and Fremling, Christoffer and Ben-Ami, Sagi and Dekany, Richard G. and Fucik, Jason R. and Konidaris, Nick and Nash, Reston and Ngeow, Chow-Choong and Ofek, Eran O. and O' Sullivan, Donal and Quimby, Robert and Ritter, Andreas and Vyhmeister, Karl E.},
	month = mar,
	year = {2018},
	note = {Publisher: IOP
ADS Bibcode: 2018PASP..130c5003B},
	pages = {035003},
}

@article{smartt_pessto_2015,
	title = {{PESSTO}: survey description and products from the first data release by the {Public} {ESO} {Spectroscopic} {Survey} of {Transient} {Objects}},
	volume = {579},
	issn = {0004-6361},
	shorttitle = {{PESSTO}},
	url = {https://ui.adsabs.harvard.edu/abs/2015A&A...579A..40S},
	doi = {10.1051/0004-6361/201425237},
	urldate = {2025-12-18},
	journal = {Astronomy and Astrophysics},
	author = {Smartt, S. J. and Valenti, S. and Fraser, M. and Inserra, C. and Young, D. R. and Sullivan, M. and Pastorello, A. and Benetti, S. and Gal-Yam, A. and Knapic, C. and Molinaro, M. and Smareglia, R. and Smith, K. W. and Taubenberger, S. and Yaron, O. and Anderson, J. P. and Ashall, C. and Balland, C. and Baltay, C. and Barbarino, C. and Bauer, F. E. and Baumont, S. and Bersier, D. and Blagorodnova, N. and Bongard, S. and Botticella, M. T. and Bufano, F. and Bulla, M. and Cappellaro, E. and Campbell, H. and Cellier-Holzem, F. and Chen, T.-W. and Childress, M. J. and Clocchiatti, A. and Contreras, C. and Dall'Ora, M. and Danziger, J. and de Jaeger, T. and De Cia, A. and Della Valle, M. and Dennefeld, M. and Elias-Rosa, N. and Elman, N. and Feindt, U. and Fleury, M. and Gall, E. and Gonzalez-Gaitan, S. and Galbany, L. and Morales Garoffolo, A. and Greggio, L. and Guillou, L. L. and Hachinger, S. and Hadjiyska, E. and Hage, P. E. and Hillebrandt, W. and Hodgkin, S. and Hsiao, E. Y. and James, P. A. and Jerkstrand, A. and Kangas, T. and Kankare, E. and Kotak, R. and Kromer, M. and Kuncarayakti, H. and Leloudas, G. and Lundqvist, P. and Lyman, J. D. and Hook, I. M. and Maguire, K. and Manulis, I. and Margheim, S. J. and Mattila, S. and Maund, J. R. and Mazzali, P. A. and McCrum, M. and McKinnon, R. and Moreno-Raya, M. E. and Nicholl, M. and Nugent, P. and Pain, R. and Pignata, G. and Phillips, M. M. and Polshaw, J. and Pumo, M. L. and Rabinowitz, D. and Reilly, E. and Romero-Cañizales, C. and Scalzo, R. and Schmidt, B. and Schulze, S. and Sim, S. and Sollerman, J. and Taddia, F. and Tartaglia, L. and Terreran, G. and Tomasella, L. and Turatto, M. and Walker, E. and Walton, N. A. and Wyrzykowski, L. and Yuan, F. and Zampieri, L.},
	month = jul,
	year = {2015},
	note = {Publisher: EDP
ADS Bibcode: 2015A\&A...579A..40S},
	pages = {A40},
}

@article{angus_fast-rising_2022,
	title = {A fast-rising tidal disruption event from a candidate intermediate-mass black hole},
	volume = {6},
	copyright = {2022 The Author(s), under exclusive licence to Springer Nature Limited},
	issn = {2397-3366},
	url = {https://www.nature.com/articles/s41550-022-01811-y},
	doi = {10.1038/s41550-022-01811-y},
	language = {en},
	number = {12},
	urldate = {2025-12-18},
	journal = {Nature Astronomy},
	author = {Angus, C. R. and Baldassare, V. F. and Mockler, B. and Foley, R. J. and Ramirez-Ruiz, E. and Raimundo, S. I. and French, K. D. and Auchettl, K. and Pfister, H. and Gall, C. and Hjorth, J. and Drout, M. R. and Alexander, K. D. and Dimitriadis, G. and Hung, T. and Jones, D. O. and Rest, A. and Siebert, M. R. and Taggart, K. and Terreran, G. and Tinyanont, S. and Carroll, C. M. and DeMarchi, L. and Earl, N. and Gagliano, A. and Izzo, L. and Villar, V. A. and Zenati, Y. and Arendse, N. and Cold, C. and de Boer, T. J. L. and Chambers, K. C. and Coulter, D. A. and Khetan, N. and Lin, C. C. and Magnier, E. A. and Rojas-Bravo, C. and Wainscoat, R. J. and Wojtak, R.},
	month = dec,
	year = {2022},
	note = {Publisher: Nature Publishing Group},
	pages = {1452--1463},
}

@article{krolik_tidal_2020,
	title = {Tidal {Disruptions} of {Main}-sequence {Stars}. {V}. {The} {Varieties} of {Disruptions}},
	volume = {904},
	issn = {0004-637X},
	url = {https://ui.adsabs.harvard.edu/abs/2020ApJ...904...68K},
	doi = {10.3847/1538-4357/abc0f6},
	urldate = {2025-04-11},
	journal = {The Astrophysical Journal},
	author = {Krolik, Julian and Piran, Tsvi and Ryu, Taeho},
	month = nov,
	year = {2020},
	note = {Publisher: IOP
ADS Bibcode: 2020ApJ...904...68K},
	pages = {68},
}

@misc{french_prospects_2025,
	title = {Prospects for {Measuring} {Black} {Hole} {Masses} using {TDEs} with the {Vera} {C}. {Rubin} {Observatory}},
	url = {https://ui.adsabs.harvard.edu/abs/2025arXiv251213409F},
	urldate = {2025-12-16},
	publisher = {arXiv},
	author = {French, K. Decker and Mockler, Brenna and Earl, Nicholas and Murphey, Tanner},
	month = dec,
	year = {2025},
	note = {ADS Bibcode: 2025arXiv251213409F},
}

@article{speagle_dynesty_2020,
	title = {{DYNESTY}: a dynamic nested sampling package for estimating {Bayesian} posteriors and evidences},
	volume = {493},
	issn = {0035-8711},
	shorttitle = {{DYNESTY}},
	url = {https://ui.adsabs.harvard.edu/abs/2020MNRAS.493.3132S},
	doi = {10.1093/mnras/staa278},
	urldate = {2025-04-11},
	journal = {Monthly Notices of the Royal Astronomical Society},
	author = {Speagle, Joshua S.},
	month = apr,
	year = {2020},
	note = {Publisher: OUP
ADS Bibcode: 2020MNRAS.493.3132S},
	pages = {3132--3158},
}

@article{sarin_redback_2024,
	title = {{REDBACK}: a {Bayesian} inference software package for electromagnetic transients},
	volume = {531},
	issn = {0035-8711},
	shorttitle = {{REDBACK}},
	url = {https://ui.adsabs.harvard.edu/abs/2024MNRAS.531.1203S},
	doi = {10.1093/mnras/stae1238},
	urldate = {2025-04-11},
	journal = {Monthly Notices of the Royal Astronomical Society},
	author = {Sarin, Nikhil and Hübner, Moritz and Omand, Conor M. B. and Setzer, Christian N. and Schulze, Steve and Adhikari, Naresh and Sagués-Carracedo, Ana and Galaudage, Shanika and Wallace, Wendy F. and Lamb, Gavin P. and Lin, En-Tzu},
	month = jun,
	year = {2024},
	note = {Publisher: OUP
ADS Bibcode: 2024MNRAS.531.1203S},
	pages = {1203--1227},
}

@article{ramsden_bulge_2022,
	title = {The bulge masses of {TDE} host galaxies and their scaling with black hole mass},
	volume = {515},
	issn = {0035-8711},
	url = {https://ui.adsabs.harvard.edu/abs/2022MNRAS.515.1146R},
	doi = {10.1093/mnras/stac1810},
	urldate = {2025-04-11},
	journal = {Monthly Notices of the Royal Astronomical Society},
	author = {Ramsden, Paige and Lanning, Daniel and Nicholl, Matt and McGee, Sean L.},
	month = sep,
	year = {2022},
	note = {Publisher: OUP
ADS Bibcode: 2022MNRAS.515.1146R},
	pages = {1146--1157},
}

@article{mummery_optical_2025,
	title = {The optical, {UV}-plateau, and {X}-ray tidal disruption event luminosity functions reproduced from first principles},
	volume = {541},
	issn = {0035-8711},
	url = {https://ui.adsabs.harvard.edu/abs/2025MNRAS.541..429M},
	doi = {10.1093/mnras/staf938},
	urldate = {2025-11-04},
	journal = {Monthly Notices of the Royal Astronomical Society},
	author = {Mummery, Andrew and van Velzen, Sjoert},
	month = jul,
	year = {2025},
	note = {Publisher: OUP
ADS Bibcode: 2025MNRAS.541..429M},
	pages = {429--445},
}

@article{malyali_rebrightening_2023,
	title = {The rebrightening of a {ROSAT}-selected tidal disruption event: repeated weak partial disruption flares from a quiescent galaxy?},
	volume = {520},
	issn = {0035-8711},
	shorttitle = {The rebrightening of a {ROSAT}-selected tidal disruption event},
	url = {https://ui.adsabs.harvard.edu/abs/2023MNRAS.520.3549M},
	doi = {10.1093/mnras/stad022},
	urldate = {2025-04-11},
	journal = {Monthly Notices of the Royal Astronomical Society},
	author = {Malyali, A. and Liu, Z. and Rau, A. and Grotova, I. and Merloni, A. and Goodwin, A. J. and Anderson, G. E. and Miller-Jones, J. C. A. and Kawka, A. and Arcodia, R. and Buchner, J. and Nandra, K. and Homan, D. and Krumpe, M.},
	month = apr,
	year = {2023},
	note = {Publisher: OUP
ADS Bibcode: 2023MNRAS.520.3549M},
	pages = {3549--3559},
}

@misc{magill_mallorn_2025,
	title = {{MALLORN}: {Many} {Artificial} {LSST} {Lightcurves} based on {Observations} of {Real} {Nuclear} transients},
	shorttitle = {{MALLORN}},
	url = {https://ui.adsabs.harvard.edu/abs/2025arXiv251204946M},
	urldate = {2025-12-05},
	publisher = {arXiv},
	author = {Magill, Dylan and Nicholl, Matt and Anilkumar, Vysakh and van Velzen, Sjoert and Sheng, Xinyue and Son Mai, Thai and Viet Tran, Hung and Phu Doan, Ngoc and Moore, Thomas and Srivastav, Shubham and Young, David R. and Angus, Charlotte R. and Weston, Joshua},
	month = dec,
	year = {2025},
	note = {ADS Bibcode: 2025arXiv251204946M},
}

@misc{guolo_compact_2025,
	title = {Compact {Accretion} {Disks} in the {Aftermath} of {Tidal} {Disruption} {Events}: {Parameter} {Inference} from {Joint} {X}-ray {Spectra} and {UV}/{Optical} {Photometry} {Fitting}},
	shorttitle = {Compact {Accretion} {Disks} in the {Aftermath} of {Tidal} {Disruption} {Events}},
	url = {https://ui.adsabs.harvard.edu/abs/2025arXiv251026774G},
	doi = {10.48550/arXiv.2510.26774},
	urldate = {2025-11-03},
	publisher = {arXiv},
	author = {Guolo, M. and Mummery, A. and van Velzen, S. and Gezari, S. and Nicholl, M. and Yao, Y. and Karmen, M. and Ajay, Y. and Wevers, T. and LeBaron, N. and Chornock, R.},
	month = oct,
	year = {2025},
	note = {ADS Bibcode: 2025arXiv251026774G},
}

@article{bucar_bricman_rubin_2023,
	title = {Rubin {Observatory}'s {Survey} {Strategy} {Performance} for {Tidal} {Disruption} {Events}},
	volume = {268},
	issn = {0067-0049},
	url = {https://ui.adsabs.harvard.edu/abs/2023ApJS..268...13B},
	doi = {10.3847/1538-4365/ace1e7},
	urldate = {2025-05-19},
	journal = {The Astrophysical Journal Supplement Series},
	author = {Bučar Bricman, K. and van Velzen, S. and Nicholl, M. and Gomboc, A.},
	month = sep,
	year = {2023},
	note = {Publisher: IOP
ADS Bibcode: 2023ApJS..268...13B},
	pages = {13},
}

@article{schlafly_measuring_2011,
	title = {Measuring {Reddening} with {Sloan} {Digital} {Sky} {Survey} {Stellar} {Spectra} and {Recalibrating} {SFD}},
	volume = {737},
	issn = {0004-637X},
	url = {https://ui.adsabs.harvard.edu/abs/2011ApJ...737..103S},
	doi = {10.1088/0004-637X/737/2/103},
	urldate = {2025-12-01},
	journal = {The Astrophysical Journal},
	author = {Schlafly, Edward F. and Finkbeiner, Douglas P.},
	month = aug,
	year = {2011},
	note = {Publisher: IOP
ADS Bibcode: 2011ApJ...737..103S},
	pages = {103},
}

@misc{quintin_lost_2025,
	title = {Lost and {Found} - {A} gallery of overlooked optical nuclear transients from the {ZTF} archive},
	url = {https://ui.adsabs.harvard.edu/abs/2025arXiv251119016Q},
	doi = {10.48550/arXiv.2511.19016},
	urldate = {2025-12-01},
	publisher = {arXiv},
	author = {Quintin, E. and Russeil, E. and Llamas Lanza, M. and Karpov, S. and Ishida, E. E. O. and Peloton, J. and Pruzhinskaya, M. V. and Möller, A. and Giustini, M. and Miniutti, G. and Saxton, R. S. and Sánchez-Sáez, P. and Zheltoukhov, S. and Dodin, A. and Belinski, A.},
	month = nov,
	year = {2025},
	note = {ADS Bibcode: 2025arXiv251119016Q},
}

@article{shvartzvald_ultrasat_2024,
	title = {{ULTRASAT}: {A} {Wide}-field {Time}-domain {UV} {Space} {Telescope}},
	volume = {964},
	issn = {0004-637X},
	shorttitle = {{ULTRASAT}},
	url = {https://ui.adsabs.harvard.edu/abs/2024ApJ...964...74S},
	doi = {10.3847/1538-4357/ad2704},
	urldate = {2025-11-28},
	journal = {The Astrophysical Journal},
	author = {Shvartzvald, Y. and Waxman, E. and Gal-Yam, A. and Ofek, E. O. and Ben-Ami, S. and Berge, D. and Kowalski, M. and Bühler, R. and Worm, S. and Rhoads, J. E. and Arcavi, I. and Maoz, D. and Polishook, D. and Stone, N. and Trakhtenbrot, B. and Ackermann, M. and Aharonson, O. and Birnholtz, O. and Chelouche, D. and Guetta, D. and Hallakoun, N. and Horesh, A. and Kushnir, D. and Mazeh, T. and Nordin, J. and Ofir, A. and Ohm, S. and Parsons, D. and Pe'er, A. and Perets, H. B. and Perdelwitz, V. and Poznanski, D. and Sadeh, I. and Sagiv, I. and Shahaf, S. and Soumagnac, M. and Tal-Or, L. and Santen, J. Van and Zackay, B. and Guttman, O. and Rekhi, P. and Townsend, A. and Weinstein, A. and Wold, I.},
	month = mar,
	year = {2024},
	note = {Publisher: IOP
ADS Bibcode: 2024ApJ...964...74S},
	pages = {74},
}

@article{miller_silla_2025,
	title = {The {La} {Silla} {Schmidt} {Southern} {Survey}},
	volume = {137},
	issn = {0004-6280},
	url = {https://ui.adsabs.harvard.edu/abs/2025PASP..137i4204M},
	doi = {10.1088/1538-3873/ae02c5},
	urldate = {2025-11-28},
	journal = {Publications of the Astronomical Society of the Pacific},
	author = {Miller, Adam A. and Abrams, Natasha S. and Aldering, Greg and Anand, Shreya and Angus, Charlotte R. and Arcavi, Iair and Baltay, Charles and Bauer, Franz E. and Brethauer, Daniel and Bloom, Joshua S. and Bommireddy, Hemanth and Catelan, Márcio and Chornock, Ryan and Clark, Peter and Collett, Thomas E. and Dimitriadis, Georgios and Faris, Sara and Förster, Francisco and Franckowiak, Anna and Frohmaier, Christopher and Galbany, Lluís and Galleguillos, Renato B. and Goobar, Ariel and Graur, Or and Gutiérrez, Claudia P. and Hall, Saarah and Hammerstein, Erica and Herner, Kenneth R. and Hook, Isobel M. and Huston, Macy J. and Johansson, Joel and Kilpatrick, Charles D. and Kim, Alex G. and Knop, Robert A. and Kowalski, Marek P. and Kwok, Lindsey A. and LeBaron, Natalie and Lin, Kenneth W. and Liu, Chang and Lu, Jessica R. and Lu, Wenbin and Lunnan, Ragnhild and Maguire, Kate and Makrygianni, Lydia and Margutti, Raffaella and Maoz, Dan and Veres, Patrik Milán and Moore, Thomas and Nayana, A. J. and Nicholl, Matt and Nordin, Jakob and Oates, S. R. and Pignata, Giuliano and Polin, Abigail and Poznanski, Dovi and Prieto, Jose L. and Rabinowitz, David L. and Rehemtulla, Nabeel and Rigault, Mickael and Ryczanowski, Dan and Sarin, Nikhil and Schulze, Steve and Shah, Ved G. and Sheng, Xinyue and Shilling, Samuel P. R. and Simmons, Brooke D. and Singh, Avinash and Smith, Graham P. and Smith, Mathew and Sollerman, Jesper and Soumagnac, Maayane T. and Stubbs, Christopher W. and Sullivan, Mark and Suresh, Aswin and Trakhtenbrot, Benny and Ward, Charlotte and Wiston, Eli and Xiong, Helen and Yao, Yuhan and Nugent, Peter E.},
	month = sep,
	year = {2025},
	note = {Publisher: IOP
ADS Bibcode: 2025PASP..137i4204M},
	pages = {094204},
}

@article{groot_blackgem_2024,
	title = {The {BlackGEM} {Telescope} {Array}. {I}. {Overview}},
	volume = {136},
	issn = {0004-6280},
	url = {https://ui.adsabs.harvard.edu/abs/2024PASP..136k5003G},
	doi = {10.1088/1538-3873/ad8b6a},
	urldate = {2025-11-28},
	journal = {Publications of the Astronomical Society of the Pacific},
	author = {Groot, P. J. and Bloemen, S. and Vreeswijk, P. M. and van Roestel, J. C. J. and Jonker, P. G. and Nelemans, G. and Klein-Wolt, M. and Lepoole, R. and Pieterse, D. L. A. and Rodenhuis, M. and Boland, W. and Haverkorn, M. and Aerts, C. and Bakker, R. and Balster, H. and Bekema, M. and Dijkstra, E. and Dolron, P. and Elswijk, E. and van Elteren, A. and Engels, A. and Fokker, M. and de Haan, M. and Hahn, F. and ter Horst, R. and Lesman, D. and Kragt, J. and Morren, J. and Nillissen, H. and Pessemier, W. and Raskin, G. and de Rijke, A. and Scheers, L. H. A. and Schuil, M. and Timmer, S. T. and Antunes Amaral, L. and Arancibia-Rojas, E. and Arcavi, I. and Blagorodnova, N. and Biswas, S. and Breton, R. P. and Dawson, H. and Dayal, P. and De Wet, S. and Duffy, C. and Faris, S. and Fausnaugh, M. and Gal-Yam, A. and Geier, S. and Horesh, A. and Johnston, C. and Katusiime, G. and Kelley, C. and Kosakowski, A. and Kupfer, T. and Leloudas, G. and Levan, A. and Modiano, D. and Mogawana, O. and Munday, J. and Paice, J. and Patat, F. and Pelisoli, I. and Ramsay, G. and Ranaivomanana, P. T. and Ruiz-Carmona, R. and Schaffenroth, V. and Scaringi, S. and Stoppa, F. and Street, R. and Tranin, H. and Uzundag, M. and Valenti, S. and Veresvarska, M. and Vuc̆ković, M. and Wichern, H. C. I. and Wijers, R. A. M. J. and Wijnands, R. A. D. and Zimmerman, E.},
	month = nov,
	year = {2024},
	note = {Publisher: IOP
ADS Bibcode: 2024PASP..136k5003G},
	pages = {115003},
}

@misc{peter_yoachim_lsstrubin_scheduler_2025,
	title = {lsst/rubin\_scheduler: v3.20.1},
	copyright = {GNU General Public License v3.0 only},
	shorttitle = {lsst/rubin\_scheduler},
	url = {https://zenodo.org/doi/10.5281/zenodo.17611865},
	urldate = {2025-11-27},
	publisher = {Zenodo},
	author = {Peter Yoachim and Lynne Jones and Eric H. Neilsen, Jr and Sean MacBride and Keith Bechtol and Matthew R. Becker and Ross and Humna},
	month = nov,
	year = {2025},
	doi = {10.5281/ZENODO.17611865},
}

@misc{peter_yoachim_lsstrubin_sim_2025,
	title = {lsst/rubin\_sim: v2.6.0},
	copyright = {GNU General Public License v3.0 only},
	shorttitle = {lsst/rubin\_sim},
	url = {https://zenodo.org/doi/10.5281/zenodo.17505313},
	urldate = {2025-11-27},
	publisher = {Zenodo},
	author = {Peter Yoachim and Lynne Jones and Eric H. Neilsen, Jr and Tiago and Boris Leistedt and John Parejko and Eric Bellm and Rachel Street and Jeff Carlin and Humna and Matthew R. Becker and pgris and erykoff and reneehlozek and Loredana Prisinzano and Giovanni A. Gollotti and Erik Dennihy and Jonathan Sick and lmptc and Ross and Leanne Guy and Roberto J. Assef and Natasha Abrams and LI and Katja Bricman and Johan Bregeon and Kian-Tat Lim and Michael Kelley and Igor Andreoni},
	month = nov,
	year = {2025},
	doi = {10.5281/ZENODO.17505313},
}

@misc{pursiainen_muse_2025,
	title = {{MUSE} {IFU} observations of galaxies hosting of {Tidal} {Disruption} {Events}},
	url = {https://ui.adsabs.harvard.edu/abs/2025arXiv250712520P},
	doi = {10.48550/arXiv.2507.12520},
	urldate = {2025-11-26},
	publisher = {arXiv},
	author = {Pursiainen, M. and Leloudas, G. and Lyman, J. and Byrne, C. M. and Charalampopoulos, P. and Ramsden, P. and Kim, S. and Schulze, S. and Anderson, J. P. and Bauer, F. E. and Dai, L. and Galbany, L. and Kuncarayakti, H. and Nicholl, M. and Pessi, T. and Prieto, J. L. and Sanchez, S. F.},
	month = jul,
	year = {2025},
	note = {ADS Bibcode: 2025arXiv250712520P},
}

@article{gebhardt_relationship_2000,
	title = {A {Relationship} between {Nuclear} {Black} {Hole} {Mass} and {Galaxy} {Velocity} {Dispersion}},
	volume = {539},
	issn = {0004-637X},
	url = {https://ui.adsabs.harvard.edu/abs/2000ApJ...539L..13G},
	doi = {10.1086/312840},
	urldate = {2025-11-25},
	journal = {The Astrophysical Journal},
	author = {Gebhardt, Karl and Bender, Ralf and Bower, Gary and Dressler, Alan and Faber, S. M. and Filippenko, Alexei V. and Green, Richard and Grillmair, Carl and Ho, Luis C. and Kormendy, John and Lauer, Tod R. and Magorrian, John and Pinkney, Jason and Richstone, Douglas and Tremaine, Scott},
	month = aug,
	year = {2000},
	note = {Publisher: IOP
ADS Bibcode: 2000ApJ...539L..13G},
	pages = {L13--L16},
}

@article{di_matteo_energy_2005,
	title = {Energy input from quasars regulates the growth and activity of black holes and their host galaxies},
	volume = {433},
	issn = {0028-0836},
	url = {https://ui.adsabs.harvard.edu/abs/2005Natur.433..604D},
	doi = {10.1038/nature03335},
	urldate = {2025-11-25},
	journal = {Nature},
	author = {Di Matteo, Tiziana and Springel, Volker and Hernquist, Lars},
	month = feb,
	year = {2005},
	note = {ADS Bibcode: 2005Natur.433..604D},
	pages = {604--607},
}

@article{kauffmann_unified_2000,
	title = {A unified model for the evolution of galaxies and quasars},
	volume = {311},
	issn = {0035-8711},
	url = {https://ui.adsabs.harvard.edu/abs/2000MNRAS.311..576K},
	doi = {10.1046/j.1365-8711.2000.03077.x},
	urldate = {2025-11-25},
	journal = {Monthly Notices of the Royal Astronomical Society},
	author = {Kauffmann, Guinevere and Haehnelt, Martin},
	month = jan,
	year = {2000},
	note = {Publisher: OUP
ADS Bibcode: 2000MNRAS.311..576K},
	pages = {576--588},
}

@article{fabian_observational_2012,
	title = {Observational {Evidence} of {Active} {Galactic} {Nuclei} {Feedback}},
	volume = {50},
	issn = {0066-4146},
	url = {https://ui.adsabs.harvard.edu/abs/2012ARA&A..50..455F},
	doi = {10.1146/annurev-astro-081811-125521},
	urldate = {2025-11-25},
	journal = {Annual Review of Astronomy and Astrophysics},
	author = {Fabian, A. C.},
	month = sep,
	year = {2012},
	note = {ADS Bibcode: 2012ARA\&A..50..455F},
	pages = {455--489},
}

@misc{silk_quasars_1998,
	title = {Quasars and galaxy formation},
	url = {https://ui.adsabs.harvard.edu/abs/1998A&A...331L...1S},
	doi = {10.48550/arXiv.astro-ph/9801013},
	urldate = {2025-11-25},
	publisher = {arXiv},
	author = {Silk, Joseph and Rees, Martin J.},
	month = mar,
	year = {1998},
	note = {ISSN: 0004-6361
Volume: 331
ADS Bibcode: 1998A\&A...331L...1S},
}

@article{graur_dependence_2018,
	title = {A {Dependence} of the {Tidal} {Disruption} {Event} {Rate} on {Global} {Stellar} {Surface} {Mass} {Density} and {Stellar} {Velocity} {Dispersion}},
	volume = {853},
	issn = {0004-637X, 1538-4357},
	url = {https://iopscience.iop.org/article/10.3847/1538-4357/aaa3fd},
	doi = {10.3847/1538-4357/aaa3fd},
	number = {1},
	urldate = {2025-11-25},
	journal = {The Astrophysical Journal},
	author = {Graur, Or and French, K. Decker and Zahid, H. Jabran and Guillochon, James and Mandel, Kaisey S. and Auchettl, Katie and Zabludoff, Ann I.},
	month = jan,
	year = {2018},
	pages = {39},
}

@article{wevers_black_2019,
	title = {Black hole masses of tidal disruption event host galaxies {II}},
	volume = {487},
	copyright = {https://academic.oup.com/journals/pages/open\_access/funder\_policies/chorus/standard\_publication\_model},
	issn = {0035-8711, 1365-2966},
	url = {https://academic.oup.com/mnras/article/487/3/4136/5513460},
	doi = {10.1093/mnras/stz1602},
	language = {en},
	number = {3},
	urldate = {2025-11-25},
	journal = {Monthly Notices of the Royal Astronomical Society},
	author = {Wevers, Thomas and Stone, Nicholas C and van Velzen, Sjoert and Jonker, Peter G and Hung, Tiara and Auchettl, Katie and Gezari, Suvi and Onori, Francesca and Mata Sánchez, Daniel and Kostrzewa-Rutkowska, Zuzanna and Casares, Jorge},
	month = aug,
	year = {2019},
	pages = {4136--4152},
}

@article{french_tidal_2016,
	title = {Tidal {Disruption} {Events} {Prefer} {Unusual} {Host} {Galaxies}},
	volume = {818},
	issn = {0004-637X},
	url = {https://ui.adsabs.harvard.edu/abs/2016ApJ...818L..21F},
	doi = {10.3847/2041-8205/818/1/L21},
	urldate = {2025-11-25},
	journal = {The Astrophysical Journal},
	author = {French, K. Decker and Arcavi, Iair and Zabludoff, Ann},
	month = feb,
	year = {2016},
	note = {Publisher: IOP
ADS Bibcode: 2016ApJ...818L..21F},
	pages = {L21},
}

@article{lower_how_2020,
	title = {How {Well} {Can} {We} {Measure} the {Stellar} {Mass} of a {Galaxy}: {The} {Impact} of the {Assumed} {Star} {Formation} {History} {Model} in {SED} {Fitting}},
	volume = {904},
	issn = {0004-637X},
	shorttitle = {How {Well} {Can} {We} {Measure} the {Stellar} {Mass} of a {Galaxy}},
	url = {https://ui.adsabs.harvard.edu/abs/2020ApJ...904...33L},
	doi = {10.3847/1538-4357/abbfa7},
	urldate = {2025-11-24},
	journal = {The Astrophysical Journal},
	author = {Lower, Sidney and Narayanan, Desika and Leja, Joel and Johnson, Benjamin D. and Conroy, Charlie and Davé, Romeel},
	month = nov,
	year = {2020},
	note = {Publisher: IOP
ADS Bibcode: 2020ApJ...904...33L},
	pages = {33},
}

@article{reines_relations_2015,
	title = {Relations between {Central} {Black} {Hole} {Mass} and {Total} {Galaxy} {Stellar} {Mass} in the {Local} {Universe}},
	volume = {813},
	issn = {0004-637X},
	url = {https://ui.adsabs.harvard.edu/abs/2015ApJ...813...82R},
	doi = {10.1088/0004-637X/813/2/82},
	urldate = {2025-11-24},
	journal = {The Astrophysical Journal},
	author = {Reines, Amy E. and Volonteri, Marta},
	month = nov,
	year = {2015},
	note = {Publisher: IOP
ADS Bibcode: 2015ApJ...813...82R},
	pages = {82},
}

@misc{honscheid_dark_2008,
	title = {The {Dark} {Energy} {Camera} ({DECam})},
	url = {https://ui.adsabs.harvard.edu/abs/2008arXiv0810.3600H},
	doi = {10.48550/arXiv.0810.3600},
	urldate = {2025-11-24},
	publisher = {arXiv},
	author = {Honscheid, K. and DePoy, D. L.},
	month = oct,
	year = {2008},
	note = {ADS Bibcode: 2008arXiv0810.3600H},
}

@article{leja_deriving_2017,
	title = {Deriving {Physical} {Properties} from {Broadband} {Photometry} with {Prospector}: {Description} of the {Model} and a {Demonstration} of its {Accuracy} {Using} 129 {Galaxies} in the {Local} {Universe}},
	volume = {837},
	issn = {0004-637X},
	shorttitle = {Deriving {Physical} {Properties} from {Broadband} {Photometry} with {Prospector}},
	url = {https://ui.adsabs.harvard.edu/abs/2017ApJ...837..170L},
	doi = {10.3847/1538-4357/aa5ffe},
	urldate = {2025-11-20},
	journal = {The Astrophysical Journal},
	author = {Leja, Joel and Johnson, Benjamin D. and Conroy, Charlie and van Dokkum, Pieter G. and Byler, Nell},
	month = mar,
	year = {2017},
	note = {Publisher: IOP
ADS Bibcode: 2017ApJ...837..170L},
	pages = {170},
}

@article{johnson_stellar_2021,
	title = {Stellar {Population} {Inference} with {Prospector}},
	volume = {254},
	issn = {0067-0049},
	url = {https://ui.adsabs.harvard.edu/abs/2021ApJS..254...22J},
	doi = {10.3847/1538-4365/abef67},
	urldate = {2025-11-20},
	journal = {The Astrophysical Journal Supplement Series},
	author = {Johnson, Benjamin D. and Leja, Joel and Conroy, Charlie and Speagle, Joshua S.},
	month = jun,
	year = {2021},
	note = {Publisher: IOP
ADS Bibcode: 2021ApJS..254...22J},
	pages = {22},
}

@article{martin_galaxy_2005,
	title = {The \textit{{Galaxy} {Evolution} {Explorer}} : {A} {Space} {Ultraviolet} {Survey} {Mission}},
	volume = {619},
	issn = {0004-637X, 1538-4357},
	shorttitle = {The \textit{{Galaxy} {Evolution} {Explorer}}},
	url = {https://iopscience.iop.org/article/10.1086/426387},
	doi = {10.1086/426387},
	language = {en},
	number = {1},
	urldate = {2025-11-07},
	journal = {The Astrophysical Journal},
	author = {Martin, D. Christopher and Fanson, James and Schiminovich, David and Morrissey, Patrick and Friedman, Peter G. and Barlow, Tom A. and Conrow, Tim and Grange, Robert and Jelinsky, Patrick N. and Milliard, Bruno and Siegmund, Oswald H. W. and Bianchi, Luciana and Byun, Yong-Ik and Donas, Jose and Forster, Karl and Heckman, Timothy M. and Lee, Young-Wook and Madore, Barry F. and Malina, Roger F. and Neff, Susan G. and Rich, R. Michael and Small, Todd and Surber, Frank and Szalay, Alex S. and Welsh, Barry and Wyder, Ted K.},
	month = jan,
	year = {2005},
	pages = {L1--L6},
}

@article{schlafly_unwise_2019,
	title = {The {unWISE} {Catalog}: {Two} {Billion} {Infrared} {Sources} from {Five} {Years} of \textit{{WISE}} {Imaging}},
	volume = {240},
	issn = {1538-4365},
	shorttitle = {The {unWISE} {Catalog}},
	url = {https://iopscience.iop.org/article/10.3847/1538-4365/aafbea},
	doi = {10.3847/1538-4365/aafbea},
	number = {2},
	urldate = {2025-11-07},
	journal = {The Astrophysical Journal Supplement Series},
	author = {Schlafly, Edward F. and Meisner, Aaron M. and Green, Gregory M.},
	month = feb,
	year = {2019},
	pages = {30},
}

@article{skrutskie_two_2006,
	title = {The {Two} {Micron} {All} {Sky} {Survey} ({2MASS})},
	volume = {131},
	issn = {0004-6256, 1538-3881},
	url = {https://iopscience.iop.org/article/10.1086/498708},
	doi = {10.1086/498708},
	language = {en},
	number = {2},
	urldate = {2025-11-07},
	journal = {The Astronomical Journal},
	author = {Skrutskie, M. F. and Cutri, R. M. and Stiening, R. and Weinberg, M. D. and Schneider, S. and Carpenter, J. M. and Beichman, C. and Capps, R. and Chester, T. and Elias, J. and Huchra, J. and Liebert, J. and Lonsdale, C. and Monet, D. G. and Price, S. and Seitzer, P. and Jarrett, T. and Kirkpatrick, J. D. and Gizis, J. E. and Howard, E. and Evans, T. and Fowler, J. and Fullmer, L. and Hurt, R. and Light, R. and Kopan, E. L. and Marsh, K. A. and McCallon, H. L. and Tam, R. and Van Dyk, S. and Wheelock, S.},
	month = feb,
	year = {2006},
	pages = {1163--1183},
}

@article{ramsden_evidence_2025,
	title = {Evidence for a steeper {SMBH}–bulge mass relationship extended to low masses using {TDE} host galaxies},
	volume = {541},
	issn = {0035-8711},
	url = {https://ui.adsabs.harvard.edu/abs/2025MNRAS.541.1218R},
	doi = {10.1093/mnras/staf1059},
	urldate = {2025-11-07},
	journal = {Monthly Notices of the Royal Astronomical Society},
	author = {Ramsden, Paige and Nicholl, Matt and McGee, Sean L. and Mummery, Andrew},
	month = aug,
	year = {2025},
	note = {Publisher: OUP
ADS Bibcode: 2025MNRAS.541.1218R},
	pages = {1218--1230},
}

@article{wevers_black_2017,
	title = {Black hole masses of tidal disruption event host galaxies},
	volume = {471},
	issn = {0035-8711},
	url = {https://ui.adsabs.harvard.edu/abs/2017MNRAS.471.1694W},
	doi = {10.1093/mnras/stx1703},
	urldate = {2025-11-07},
	journal = {Monthly Notices of the Royal Astronomical Society},
	author = {Wevers, Thomas and van Velzen, Sjoert and Jonker, Peter G. and Stone, Nicholas C. and Hung, Tiara and Onori, Francesca and Gezari, Suvi and Blagorodnova, Nadejda},
	month = oct,
	year = {2017},
	note = {Publisher: OUP
ADS Bibcode: 2017MNRAS.471.1694W},
	pages = {1694--1708},
}

@article{ashton_bilby_2019,
	title = {Bilby: {A} {User}-friendly {Bayesian} {Inference} {Library} for {Gravitational}-wave {Astronomy}},
	volume = {241},
	issn = {0067-0049},
	shorttitle = {Bilby},
	url = {https://doi.org/10.3847/1538-4365/ab06fc},
	doi = {10.3847/1538-4365/ab06fc},
	language = {en},
	number = {2},
	urldate = {2025-11-06},
	journal = {The Astrophysical Journal Supplement Series},
	author = {Ashton, Gregory and Hübner, Moritz and Lasky, Paul D. and Talbot, Colm and Ackley, Kendall and Biscoveanu, Sylvia and Chu, Qi and Divakarla, Atul and Easter, Paul J. and Goncharov, Boris and Vivanco, Francisco Hernandez and Harms, Jan and Lower, Marcus E. and Meadors, Grant D. and Melchor, Denyz and Payne, Ethan and Pitkin, Matthew D. and Powell, Jade and Sarin, Nikhil and Smith, Rory J. E. and Thrane, Eric},
	month = apr,
	year = {2019},
	note = {Publisher: The American Astronomical Society},
	pages = {27},
}

@article{yao_tidal_2022,
	title = {The {Tidal} {Disruption} {Event} {AT2021ehb}: {Evidence} of {Relativistic} {Disk} {Reflection}, and {Rapid} {Evolution} of the {Disk}–{Corona} {System}},
	volume = {937},
	issn = {0004-637X},
	shorttitle = {The {Tidal} {Disruption} {Event} {AT2021ehb}},
	url = {https://doi.org/10.3847/1538-4357/ac898a},
	doi = {10.3847/1538-4357/ac898a},
	language = {en},
	number = {1},
	urldate = {2025-11-06},
	journal = {The Astrophysical Journal},
	author = {Yao, Yuhan and Lu, Wenbin and Guolo, Muryel and Pasham, Dheeraj R. and Gezari, Suvi and Gilfanov, Marat and Gendreau, Keith C. and Harrison, Fiona and Cenko, S. Bradley and Kulkarni, S. R. and Miller, Jon M. and Walton, Dominic J. and García, Javier A. and Velzen, Sjoert van and Alexander, Kate D. and Miller-Jones, James C. A. and Nicholl, Matt and Hammerstein, Erica and Medvedev, Pavel and Stern, Daniel and Ravi, Vikram and Sunyaev, R. and Bloom, Joshua S. and Graham, Matthew J. and Kool, Erik C. and Mahabal, Ashish A. and Masci, Frank J. and Purdum, Josiah and Rusholme, Ben and Sharma, Yashvi and Smith, Roger and Sollerman, Jesper},
	month = sep,
	year = {2022},
	note = {Publisher: The American Astronomical Society},
	pages = {8},
}

@misc{wise_at2019cmw_2025,
	title = {{AT2019cmw}: {A} highly luminous, cooling featureless {TDE} candidate from the disruption of a high mass star in an early-type galaxy},
	shorttitle = {{AT2019cmw}},
	url = {https://ui.adsabs.harvard.edu/abs/2025arXiv250707380W},
	doi = {10.48550/arXiv.2507.07380},
	urldate = {2025-11-06},
	publisher = {arXiv},
	author = {Wise, Jacob and Perley, Daniel and Sarin, Nikhil and Matsumoto, Tatsuya and Hinds, K-Ryan and Yao, Yuhan and Sollerman, Jesper and Schulze, Steve and Bochenek, Aleksandra and Coughlin, Michael W. and De, Kishalay and Dekany, Richard and Frederick, Sara and Fremling, Christoffer and Gezari, Suvi and Graham, Matthew J. and Ho, Anna Y. Q. and Kulkarni, Shrinivas and Laher, Russ R. and Omand, Conor and Pletskova, Natalya and Sharma, Yashvi and Taggart, Kirsty and Ward, Charlotte and Wold, Avery and Yan, Lin},
	month = jul,
	year = {2025},
	note = {ADS Bibcode: 2025arXiv250707380W},
}

@article{mummery_fundamental_2024,
	title = {Fundamental scaling relationships revealed in the optical light curves of tidal disruption events},
	volume = {527},
	issn = {0035-8711},
	url = {https://ui.adsabs.harvard.edu/abs/2024MNRAS.527.2452M},
	doi = {10.1093/mnras/stad3001},
	urldate = {2025-11-04},
	journal = {Monthly Notices of the Royal Astronomical Society},
	author = {Mummery, Andrew and van Velzen, Sjoert and Nathan, Edward and Ingram, Adam and Hammerstein, Erica and Fraser-Taliente, Ludovic and Balbus, Steven},
	month = jan,
	year = {2024},
	note = {Publisher: OUP
ADS Bibcode: 2024MNRAS.527.2452M},
	pages = {2452--2489},
}

@article{mummery_spectral_2020,
	title = {The spectral evolution of disc dominated tidal disruption events},
	volume = {492},
	issn = {0035-8711},
	url = {https://ui.adsabs.harvard.edu/abs/2020MNRAS.492.5655M},
	doi = {10.1093/mnras/staa192},
	urldate = {2025-11-04},
	journal = {Monthly Notices of the Royal Astronomical Society},
	author = {Mummery, Andrew and Balbus, Steven A.},
	month = mar,
	year = {2020},
	note = {Publisher: OUP
ADS Bibcode: 2020MNRAS.492.5655M},
	pages = {5655--5674},
}

@article{gezari_x-ray_2017,
	title = {X-{Ray} {Brightening} and {UV} {Fading} of {Tidal} {Disruption} {Event} {ASASSN}-15oi},
	volume = {851},
	issn = {0004-637X},
	url = {https://ui.adsabs.harvard.edu/abs/2017ApJ...851L..47G},
	doi = {10.3847/2041-8213/aaa0c2},
	urldate = {2025-11-04},
	journal = {The Astrophysical Journal},
	author = {Gezari, S. and Cenko, S. B. and Arcavi, I.},
	month = dec,
	year = {2017},
	note = {Publisher: IOP
ADS Bibcode: 2017ApJ...851L..47G},
	pages = {L47},
}

@article{charalampopoulos_at_2023,
	title = {{AT} 2020wey and the class of faint and fast tidal disruption events},
	volume = {673},
	issn = {0004-6361},
	url = {https://ui.adsabs.harvard.edu/abs/2023A&A...673A..95C},
	doi = {10.1051/0004-6361/202245065},
	urldate = {2025-11-04},
	journal = {Astronomy and Astrophysics},
	author = {Charalampopoulos, P. and Pursiainen, M. and Leloudas, G. and Arcavi, I. and Newsome, M. and Schulze, S. and Burke, J. and Nicholl, M.},
	month = may,
	year = {2023},
	note = {Publisher: EDP
ADS Bibcode: 2023A\&A...673A..95C},
	pages = {A95},
}

@article{brown_long_2017,
	title = {The {Long} {Term} {Evolution} of {ASASSN}-14li},
	volume = {466},
	issn = {0035-8711},
	url = {https://ui.adsabs.harvard.edu/abs/2017MNRAS.466.4904B},
	doi = {10.1093/mnras/stx033},
	urldate = {2025-11-04},
	journal = {Monthly Notices of the Royal Astronomical Society},
	author = {Brown, J. S. and Holoien, T. W.-S. and Auchettl, K. and Stanek, K. Z. and Kochanek, C. S. and Shappee, B. J. and Prieto, J. L. and Grupe, D.},
	month = apr,
	year = {2017},
	note = {Publisher: OUP
ADS Bibcode: 2017MNRAS.466.4904B},
	pages = {4904--4916},
}

@article{gezari_ps1-10jh_2015,
	title = {{PS1}-10jh {Continues} to {Follow} the {Fallback} {Accretion} {Rate} of a {Tidally} {Disrupted} {Star}},
	volume = {815},
	issn = {0004-637X},
	url = {https://ui.adsabs.harvard.edu/abs/2015ApJ...815L...5G},
	doi = {10.1088/2041-8205/815/1/L5},
	urldate = {2025-11-04},
	journal = {The Astrophysical Journal},
	author = {Gezari, S. and Chornock, R. and Lawrence, A. and Rest, A. and Jones, D. O. and Berger, E. and Challis, P. M. and Narayan, G.},
	month = dec,
	year = {2015},
	note = {Publisher: IOP
ADS Bibcode: 2015ApJ...815L...5G},
	pages = {L5},
}

@article{nixon_partial_2021,
	title = {Partial, {Zombie}, and {Full} {Tidal} {Disruption} of {Stars} by {Supermassive} {Black} {Holes}},
	volume = {922},
	issn = {0004-637X},
	url = {https://ui.adsabs.harvard.edu/abs/2021ApJ...922..168N},
	doi = {10.3847/1538-4357/ac1bb8},
	urldate = {2025-11-03},
	journal = {The Astrophysical Journal},
	author = {Nixon, C. J. and Coughlin, Eric R. and Miles, Patrick R.},
	month = dec,
	year = {2021},
	note = {Publisher: IOP
ADS Bibcode: 2021ApJ...922..168N},
	pages = {168},
}

@inproceedings{breeveld_updated_2011,
	address = {eprint: arXiv:1102.4717},
	title = {An {Updated} {Ultraviolet} {Calibration} for the {Swift}/{UVOT}},
	volume = {1358},
	url = {https://ui.adsabs.harvard.edu/abs/2011AIPC.1358..373B},
	doi = {10.1063/1.3621807},
	urldate = {2025-08-02},
	publisher = {AIP},
	author = {Breeveld, A. A. and Landsman, W. and Holland, S. T. and Roming, P. and Kuin, N. P. M. and Page, M. J.},
	month = aug,
	year = {2011},
	note = {ADS Bibcode: 2011AIPC.1358..373B},
	pages = {373--376},
}

@article{dalen_epessto_2024,
	title = {{ePESSTO}+ {Transient} {Classification} {Report} for 2024-04-10},
	volume = {2024-1012},
	url = {https://ui.adsabs.harvard.edu/abs/2024TNSCR1012....1D},
	urldate = {2025-07-28},
	journal = {Transient Name Server Classification Report},
	author = {Dalen, J. V. and Hoof, A. V. and Shlentsova, A. and Fraser, M. and Yaron, O.},
	month = apr,
	year = {2024},
	note = {ADS Bibcode: 2024TNSCR1012....1D},
	pages = {1},
}

@article{smith_design_2020,
	title = {Design and {Operation} of the {ATLAS} {Transient} {Science} {Server}},
	volume = {132},
	issn = {0004-6280},
	url = {https://ui.adsabs.harvard.edu/abs/2020PASP..132h5002S},
	doi = {10.1088/1538-3873/ab936e},
	urldate = {2025-04-13},
	journal = {Publications of the Astronomical Society of the Pacific},
	author = {Smith, K. W. and Smartt, S. J. and Young, D. R. and Tonry, J. L. and Denneau, L. and Flewelling, H. and Heinze, A. N. and Weiland, H. J. and Stalder, B. and Rest, A. and Stubbs, C. W. and Anderson, J. P. and Chen, T. -W. and Clark, P. and Do, A. and Förster, F. and Fulton, M. and Gillanders, J. and McBrien, O. R. and O'Neill, D. and Srivastav, S. and Wright, D. E.},
	month = aug,
	year = {2020},
	note = {Publisher: IOP
ADS Bibcode: 2020PASP..132h5002S},
	pages = {085002},
}

@article{shingles_release_2021,
	title = {Release of the {ATLAS} {Forced} {Photometry} server for public use},
	volume = {7},
	url = {https://ui.adsabs.harvard.edu/abs/2021TNSAN...7....1S},
	urldate = {2025-04-13},
	journal = {Transient Name Server AstroNote},
	author = {Shingles, L. and Smith, K. W. and Young, D. R. and Smartt, S. J. and Tonry, J. and Denneau, L. and Heinze, A. and Weiland, H. and Flewelling, H. and Stalder, B. and Clocchiatti, A. and Förster, F. and Pignata, G. and Rest, A. and Anderson, J. and Stubbs, C. and Erasmus, N.},
	month = jan,
	year = {2021},
	note = {ADS Bibcode: 2021TNSAN...7....1S},
	pages = {1--7},
}

@article{roming_swift_2005,
	title = {The {Swift} {Ultra}-{Violet}/{Optical} {Telescope}},
	volume = {120},
	issn = {0038-6308},
	url = {https://ui.adsabs.harvard.edu/abs/2005SSRv..120...95R},
	doi = {10.1007/s11214-005-5095-4},
	urldate = {2025-04-13},
	journal = {Space Science Reviews},
	author = {Roming, Peter W. A. and Kennedy, Thomas E. and Mason, Keith O. and Nousek, John A. and Ahr, Lindy and Bingham, Richard E. and Broos, Patrick S. and Carter, Mary J. and Hancock, Barry K. and Huckle, Howard E. and Hunsberger, S. D. and Kawakami, Hajime and Killough, Ronnie and Koch, T. Scott and McLelland, Michael K. and Smith, Kelly and Smith, Philip J. and Soto, Juan Carlos and Boyd, Patricia T. and Breeveld, Alice A. and Holland, Stephen T. and Ivanushkina, Mariya and Pryzby, Michael S. and Still, Martin D. and Stock, Joseph},
	month = oct,
	year = {2005},
	note = {ADS Bibcode: 2005SSRv..120...95R},
	pages = {95--142},
}

@article{gehrels_swift_2004,
	title = {The {Swift} {Gamma}-{Ray} {Burst} {Mission}},
	volume = {611},
	issn = {0004-637X},
	url = {https://ui.adsabs.harvard.edu/abs/2004ApJ...611.1005G},
	doi = {10.1086/422091},
	urldate = {2025-04-13},
	journal = {The Astrophysical Journal},
	author = {Gehrels, N. and Chincarini, G. and Giommi, P. and Mason, K. O. and Nousek, J. A. and Wells, A. A. and White, N. E. and Barthelmy, S. D. and Burrows, D. N. and Cominsky, L. R. and Hurley, K. C. and Marshall, F. E. and Mészáros, P. and Roming, P. W. A. and Angelini, L. and Barbier, L. M. and Belloni, T. and Campana, S. and Caraveo, P. A. and Chester, M. M. and Citterio, O. and Cline, T. L. and Cropper, M. S. and Cummings, J. R. and Dean, A. J. and Feigelson, E. D. and Fenimore, E. E. and Frail, D. A. and Fruchter, A. S. and Garmire, G. P. and Gendreau, K. and Ghisellini, G. and Greiner, J. and Hill, J. E. and Hunsberger, S. D. and Krimm, H. A. and Kulkarni, S. R. and Kumar, P. and Lebrun, F. and Lloyd-Ronning, N. M. and Markwardt, C. B. and Mattson, B. J. and Mushotzky, R. F. and Norris, J. P. and Osborne, J. and Paczynski, B. and Palmer, D. M. and Park, H. -S. and Parsons, A. M. and Paul, J. and Rees, M. J. and Reynolds, C. S. and Rhoads, J. E. and Sasseen, T. P. and Schaefer, B. E. and Short, A. T. and Smale, A. P. and Smith, I. A. and Stella, L. and Tagliaferri, G. and Takahashi, T. and Tashiro, M. and Townsley, L. K. and Tueller, J. and Turner, M. J. L. and Vietri, M. and Voges, W. and Ward, M. J. and Willingale, R. and Zerbi, F. M. and Zhang, W. W.},
	month = aug,
	year = {2004},
	note = {Publisher: IOP
ADS Bibcode: 2004ApJ...611.1005G},
	pages = {1005--1020},
}

@misc{masci_new_2023,
	title = {A {New} {Forced} {Photometry} {Service} for the {Zwicky} {Transient} {Facility}},
	url = {https://ui.adsabs.harvard.edu/abs/2023arXiv230516279M},
	doi = {10.48550/arXiv.2305.16279},
	urldate = {2025-07-17},
	publisher = {arXiv},
	author = {Masci, Frank J. and Laher, Russ R. and Rusholme, Benjamin and Shupe, David and Paladini, Roberta and Groom, Steve and Wold, Avery and Miller, Adam A. and Drake, Andrew},
	month = may,
	year = {2023},
	note = {ADS Bibcode: 2023arXiv230516279M},
}

@article{huber_pan-starrs_2015,
	title = {The {Pan}-{STARRS} {Survey} for {Transients} ({PSST}) - first announcement and public release},
	volume = {7153},
	url = {https://ui.adsabs.harvard.edu/abs/2015ATel.7153....1H},
	urldate = {2025-07-16},
	journal = {The Astronomer's Telegram},
	author = {Huber, M. and Chambers, K. C. and Flewelling, H. and Willman, M. and Primak, N. and Schultz, A. and Gibson, B. and Magnier, E. and Waters, C. and Tonry, J. and Wainscoat, R. J. and Smith, K. W. and Wright, D. and Smartt, S. J. and Foley, R. J. and Jha, S. W. and Rest, A. and Scolnic, D.},
	month = feb,
	year = {2015},
	note = {ADS Bibcode: 2015ATel.7153....1H},
	pages = {1},
}

@article{guo_reverberation_2025,
	title = {Reverberation {Evidence} for {Stream} {Collision} and {Delayed} {Disk} {Formation} in {Tidal} {Disruption} {Events}},
	volume = {979},
	issn = {0004-637X},
	url = {https://dx.doi.org/10.3847/1538-4357/ada274},
	doi = {10.3847/1538-4357/ada274},
	language = {en},
	number = {2},
	urldate = {2025-07-16},
	journal = {The Astrophysical Journal},
	author = {Guo, Hengxiao and Sun, Jingbo and Li, Shuangliang and Jiang, Yan-Fei and Wang, Tinggui and Bu, Defu and Jiang, Ning and Wang, Yanan and Yao, Yuhan and Shen, Rongfeng and Gu, Minfeng and Sun, Mouyuan},
	month = jan,
	year = {2025},
	note = {Publisher: The American Astronomical Society},
	pages = {235},
}

@article{zhong_modeling_2025,
	title = {Modeling the {UV}/{Optical} {Light} {Curve} of {Re}-brightening {Tidal} {Disruption} {Events}},
	volume = {983},
	issn = {0004-637X},
	url = {https://dx.doi.org/10.3847/1538-4357/adc005},
	doi = {10.3847/1538-4357/adc005},
	language = {en},
	number = {2},
	urldate = {2025-07-16},
	journal = {The Astrophysical Journal},
	author = {Zhong, Shiyan},
	month = apr,
	year = {2025},
	note = {Publisher: The American Astronomical Society},
	pages = {131},
}

@article{somalwar_first_2025,
	title = {The {First} {Systematically} {Identified} {Repeating} {Partial} {Tidal} {Disruption} {Event}},
	volume = {985},
	issn = {0004-637X},
	url = {https://ui.adsabs.harvard.edu/abs/2025ApJ...985..175S},
	doi = {10.3847/1538-4357/adcc19},
	urldate = {2025-07-16},
	journal = {The Astrophysical Journal},
	author = {Somalwar, Jean J. and Ravi, Vikram and Yao, Yuhan and Guolo, Muryel and Graham, Matthew and Hammerstein, Erica and Lu, Wenbin and Nicholl, Matt and Sharma, Yashvi and Stein, Robert and van Velzen, Sjoert and Bellm, Eric C. and Coughlin, Michael W. and Groom, Steven L. and Masci, Frank J. and Riddle, Reed},
	month = jun,
	year = {2025},
	note = {Publisher: IOP
ADS Bibcode: 2025ApJ...985..175S},
	pages = {175},
}

@article{bandopadhyay_repeating_2024,
	title = {Repeating {Nuclear} {Transients} from {Repeating} {Partial} {Tidal} {Disruption} {Events}: {Reproducing} {ASASSN}-14ko and {AT2020vdq}},
	volume = {974},
	issn = {0004-637X},
	shorttitle = {Repeating {Nuclear} {Transients} from {Repeating} {Partial} {Tidal} {Disruption} {Events}},
	url = {https://ui.adsabs.harvard.edu/abs/2024ApJ...974...80B},
	doi = {10.3847/1538-4357/ad6a5a},
	urldate = {2025-04-11},
	journal = {The Astrophysical Journal},
	author = {Bandopadhyay, Ananya and Coughlin, Eric R. and Nixon, C. J. and Pasham, Dheeraj R.},
	month = oct,
	year = {2024},
	note = {Publisher: IOP
ADS Bibcode: 2024ApJ...974...80B},
	pages = {80},
}

@article{holoien_asassn-15oi_2016,
	title = {{ASASSN}-15oi: a rapidly evolving, luminous tidal disruption event at 216 {Mpc}},
	volume = {463},
	issn = {0035-8711},
	shorttitle = {{ASASSN}-15oi},
	url = {https://ui.adsabs.harvard.edu/abs/2016MNRAS.463.3813H},
	doi = {10.1093/mnras/stw2272},
	urldate = {2025-07-16},
	journal = {Monthly Notices of the Royal Astronomical Society},
	author = {Holoien, T. W. -S. and Kochanek, C. S. and Prieto, J. L. and Grupe, D. and Chen, Ping and Godoy-Rivera, D. and Stanek, K. Z. and Shappee, B. J. and Dong, Subo and Brown, J. S. and Basu, U. and Beacom, J. F. and Bersier, D. and Brimacombe, J. and Carlson, E. K. and Falco, E. and Johnston, E. and Madore, B. F. and Pojmanski, G. and Seibert, M.},
	month = dec,
	year = {2016},
	note = {Publisher: OUP
ADS Bibcode: 2016MNRAS.463.3813H},
	pages = {3813--3828},
}

@article{sun_at2021aeuk_2025,
	title = {{AT2021aeuk}: {A} {Repeating} {Partial} {Tidal} {Disruption} {Event} {Candidate} in a {Narrow}-line {Seyfert} 1 {Galaxy}},
	volume = {982},
	issn = {0004-637X},
	shorttitle = {{AT2021aeuk}},
	url = {https://ui.adsabs.harvard.edu/abs/2025ApJ...982..150S},
	doi = {10.3847/1538-4357/adb724},
	urldate = {2025-04-11},
	journal = {The Astrophysical Journal},
	author = {Sun, Jingbo and Guo, Hengxiao and Gu, Minfeng and Li, Ya-Ping and Chen, Yongjun and González-Buitrago, D. and Wang, Jian-Guo and Li, Sha-Sha and Feng, Hai-Cheng and Xiong, Dingrong and Wang, Yanan and Yuan, Qi and Jin, Jun-jie and Zhang, Wenda and Deng, Hongping and Zhang, Minghao},
	month = apr,
	year = {2025},
	note = {Publisher: IOP
ADS Bibcode: 2025ApJ...982..150S},
	pages = {150},
}

@article{kiroglu_partial_2023,
	title = {Partial {Tidal} {Disruptions} of {Main}-sequence {Stars} by {Intermediate}-mass {Black} {Holes}},
	volume = {948},
	issn = {0004-637X},
	url = {https://ui.adsabs.harvard.edu/abs/2023ApJ...948...89K},
	doi = {10.3847/1538-4357/acc24c},
	urldate = {2025-07-14},
	journal = {The Astrophysical Journal},
	author = {Kıroğlu, Fulya and Lombardi, James C. and Kremer, Kyle and Fragione, Giacomo and Fogarty, Shane and Rasio, Frederic A.},
	month = may,
	year = {2023},
	note = {Publisher: IOP
ADS Bibcode: 2023ApJ...948...89K},
	pages = {89},
}

@article{liu_tidal_2023,
	title = {Tidal {Disruption} {Events} from {Eccentric} {Orbits} and {Lessons} {Learned} from the {Noteworthy} {ASASSN}-14ko},
	volume = {944},
	issn = {0004-637X},
	url = {https://ui.adsabs.harvard.edu/abs/2023ApJ...944..184L},
	doi = {10.3847/1538-4357/acafe1},
	urldate = {2025-07-14},
	journal = {The Astrophysical Journal},
	author = {Liu, Chang and Mockler, Brenna and Ramirez-Ruiz, Enrico and Yarza, Ricardo and Law-Smith, Jamie A. P. and Naoz, Smadar and Melchor, Denyz and Rose, Sanaea},
	month = feb,
	year = {2023},
	note = {Publisher: IOP
ADS Bibcode: 2023ApJ...944..184L},
	pages = {184},
}

@article{liu_deciphering_2023,
	title = {Deciphering the extreme {X}-ray variability of the nuclear transient {eRASSt} {J045650}.3−203750. {A} likely repeating partial tidal disruption event},
	volume = {669},
	issn = {0004-6361},
	url = {https://ui.adsabs.harvard.edu/abs/2023A&A...669A..75L},
	doi = {10.1051/0004-6361/202244805},
	urldate = {2025-07-14},
	journal = {Astronomy and Astrophysics},
	author = {Liu, Z. and Malyali, A. and Krumpe, M. and Homan, D. and Goodwin, A. J. and Grotova, I. and Kawka, A. and Rau, A. and Merloni, A. and Anderson, G. E. and Miller-Jones, J. C. A. and Markowitz, A. G. and Ciroi, S. and Di Mille, F. and Schramm, M. and Tang, S. and Buckley, D. A. H. and Gromadzki, M. and Jin, C. and Buchner, J.},
	month = jan,
	year = {2023},
	note = {Publisher: EDP
ADS Bibcode: 2023A\&A...669A..75L},
	pages = {A75},
}

@article{evans_monthly_2023,
	title = {Monthly quasi-periodic eruptions from repeated stellar disruption by a massive black hole},
	volume = {7},
	issn = {2397-3366},
	url = {https://ui.adsabs.harvard.edu/abs/2023NatAs...7.1368E},
	doi = {10.1038/s41550-023-02073-y},
	urldate = {2025-07-14},
	journal = {Nature Astronomy},
	author = {Evans, P. A. and Nixon, C. J. and Campana, S. and Charalampopoulos, P. and Perley, D. A. and Breeveld, A. A. and Page, K. L. and Oates, S. R. and Eyles-Ferris, R. A. J. and Malesani, D. B. and Izzo, L. and Goad, M. R. and O'Brien, P. T. and Osborne, J. P. and Sbarufatti, B.},
	month = nov,
	year = {2023},
	note = {ADS Bibcode: 2023NatAs...7.1368E},
	pages = {1368--1375},
}

@article{guolo_x-ray_2024,
	title = {X-ray eruptions every 22 days from the nucleus of a nearby galaxy},
	volume = {8},
	issn = {2397-3366},
	url = {https://ui.adsabs.harvard.edu/abs/2024NatAs...8..347G},
	doi = {10.1038/s41550-023-02178-4},
	urldate = {2025-07-14},
	journal = {Nature Astronomy},
	author = {Guolo, Muryel and Pasham, Dheeraj R. and Zajaček, Michal and Coughlin, Eric R. and Gezari, Suvi and Suková, Petra and Wevers, Thomas and Witzany, Vojtěch and Tombesi, Francesco and van Velzen, Sjoert and Alexander, Kate D. and Yao, Yuhan and Arcodia, Riccardo and Karas, Vladimír and Miller-Jones, James C. A. and Remillard, Ronald and Gendreau, Keith and Ferrara, Elizabeth C.},
	month = mar,
	year = {2024},
	note = {ADS Bibcode: 2024NatAs...8..347G},
	pages = {347--358},
}

@article{hampel_new_2022,
	title = {A {New} {X}-{Ray} {Tidal} {Disruption} {Event} {Candidate} with {Fast} {Variability}},
	volume = {22},
	issn = {1674-4527},
	url = {https://ui.adsabs.harvard.edu/abs/2022RAA....22e5004H},
	doi = {10.1088/1674-4527/ac5800},
	urldate = {2025-07-14},
	journal = {Research in Astronomy and Astrophysics},
	author = {Hampel, J. and Komossa, S. and Greiner, J. and Reiprich, T. H. and Freyberg, M. and Erben, T.},
	month = may,
	year = {2022},
	note = {Publisher: IOP
ADS Bibcode: 2022RAA....22e5004H},
	pages = {055004},
}

@article{bao_gleeoks_2024,
	title = {Gleeok's {Fire}-breathing: {Triple} {Flares} of {AT} 2021aeuk within {Five} {Years} from the {Active} {Galaxy} {SDSS} {J161259}.83+421940.3},
	volume = {977},
	issn = {0004-637X},
	shorttitle = {Gleeok's {Fire}-breathing},
	url = {https://ui.adsabs.harvard.edu/abs/2024ApJ...977..279B},
	doi = {10.3847/1538-4357/ad9246},
	urldate = {2025-07-14},
	journal = {The Astrophysical Journal},
	author = {Bao, Dong-Wei and Guo, Wei-Jian and Zhang, Zhi-Xiang and Cheng, Cheng and Yao, Zhu-Heng and Li, Yan-Rong and Yuan, Ye-Fei and Xue, Sui-Jian and Wang, Jian-Min and Tsai, Chao-Wei and Zou, Hu and Chen, Yong-Jie and Li, Wenxiong and Zhong, Shiyan and Chen, Zhi-Qiang},
	month = dec,
	year = {2024},
	note = {Publisher: IOP
ADS Bibcode: 2024ApJ...977..279B},
	pages = {279},
}

@article{payne_asassn-14ko_2021,
	title = {{ASASSN}-14ko is a {Periodic} {Nuclear} {Transient} in {ESO} 253-{G003}},
	volume = {910},
	issn = {0004-637X},
	url = {https://ui.adsabs.harvard.edu/abs/2021ApJ...910..125P},
	doi = {10.3847/1538-4357/abe38d},
	urldate = {2025-07-14},
	journal = {The Astrophysical Journal},
	author = {Payne, Anna V. and Shappee, Benjamin J. and Hinkle, Jason T. and Vallely, Patrick J. and Kochanek, Christopher S. and Holoien, Thomas W. -S. and Auchettl, Katie and Stanek, K. Z. and Thompson, Todd A. and Neustadt, Jack M. M. and Tucker, Michael A. and Armstrong, James D. and Brimacombe, Joseph and Cacella, Paulo and Cornect, Robert and Denneau, Larry and Fausnaugh, Michael M. and Flewelling, Heather and Grupe, Dirk and Heinze, A. N. and Lopez, Laura A. and Monard, Berto and Prieto, Jose L. and Schneider, Adam C. and Sheppard, Scott S. and Tonry, John L. and Weiland, Henry},
	month = apr,
	year = {2021},
	note = {Publisher: IOP
ADS Bibcode: 2021ApJ...910..125P},
	pages = {125},
}

@misc{langis_repeating_2025,
	title = {Repeating {Flares}, {X}-ray {Outbursts} and {Delayed} {Infrared} {Emission}: {A} {Comprehensive} {Compilation} of {Optical} {Tidal} {Disruption} {Events}},
	shorttitle = {Repeating {Flares}, {X}-ray {Outbursts} and {Delayed} {Infrared} {Emission}},
	url = {https://ui.adsabs.harvard.edu/abs/2025arXiv250605476L},
	doi = {10.48550/arXiv.2506.05476},
	urldate = {2025-07-14},
	publisher = {arXiv},
	author = {Langis, D. A. and Liodakis, I. and Koljonen, K. I. I. and Paggi, A. and Globus, N. and Wyrzykowski, L. and Mikołajczyk, P. J. and Kotysz, K. and Zieliński, P. and Ihanec, N. and Ding, J. and Morshed, D. and Torres, Z.},
	month = jun,
	year = {2025},
	note = {ADS Bibcode: 2025arXiv250605476L},
}

@article{llamas_lanza_identification_2024,
	title = {Identification of {AT} 2023adr as a candidate repeated partial {TDE}},
	volume = {178},
	url = {https://ui.adsabs.harvard.edu/abs/2024TNSAN.178....1L},
	urldate = {2025-07-14},
	journal = {Transient Name Server AstroNote},
	author = {Llamas Lanza, M. and Quintin, E. and Russeil, E. and Ishida, E. and Peloton, J. and Karpov, S. and Pruzhinskaya, M. V.},
	month = jul,
	year = {2024},
	note = {ADS Bibcode: 2024TNSAN.178....1L},
	pages = {1},
}

@article{guillochon_mosfit_2018,
	title = {{MOSFiT}: {Modular} {Open} {Source} {Fitter} for {Transients}},
	volume = {236},
	issn = {0067-0049},
	shorttitle = {{MOSFiT}},
	url = {https://ui.adsabs.harvard.edu/abs/2018ApJS..236....6G},
	doi = {10.3847/1538-4365/aab761},
	urldate = {2025-04-11},
	journal = {The Astrophysical Journal Supplement Series},
	author = {Guillochon, James and Nicholl, Matt and Villar, V. Ashley and Mockler, Brenna and Narayan, Gautham and Mandel, Kaisey S. and Berger, Edo and Williams, Peter K. G.},
	month = may,
	year = {2018},
	note = {Publisher: IOP
ADS Bibcode: 2018ApJS..236....6G},
	pages = {6},
}

@article{goodwin_radio-emitting_2023,
	title = {A radio-emitting outflow produced by the tidal disruption event {AT2020vwl}},
	volume = {522},
	issn = {0035-8711},
	url = {https://ui.adsabs.harvard.edu/abs/2023MNRAS.522.5084G},
	doi = {10.1093/mnras/stad1258},
	urldate = {2025-04-11},
	journal = {Monthly Notices of the Royal Astronomical Society},
	author = {Goodwin, A. J. and Alexander, K. D. and Miller-Jones, J. C. A. and Bietenholz, M. F. and van Velzen, S. and Anderson, G. E. and Berger, E. and Cendes, Y. and Chornock, R. and Coppejans, D. L. and Eftekhari, T. and Gezari, S. and Laskar, T. and Ramirez-Ruiz, E. and Saxton, R.},
	month = jul,
	year = {2023},
	note = {Publisher: OUP
ADS Bibcode: 2023MNRAS.522.5084G},
	pages = {5084--5097},
}

@article{arcavi_continuum_2014,
	title = {A {Continuum} of {H}- to {He}-rich {Tidal} {Disruption} {Candidates} {With} a {Preference} for {E}+{A} {Galaxies}},
	volume = {793},
	issn = {0004-637X},
	url = {https://ui.adsabs.harvard.edu/abs/2014ApJ...793...38A},
	doi = {10.1088/0004-637X/793/1/38},
	urldate = {2025-04-11},
	journal = {The Astrophysical Journal},
	author = {Arcavi, Iair and Gal-Yam, Avishay and Sullivan, Mark and Pan, Yen-Chen and Cenko, S. Bradley and Horesh, Assaf and Ofek, Eran O. and De Cia, Annalisa and Yan, Lin and Yang, Chen-Wei and Howell, D. A. and Tal, David and Kulkarni, Shrinivas R. and Tendulkar, Shriharsh P. and Tang, Sumin and Xu, Dong and Sternberg, Assaf and Cohen, Judith G. and Bloom, Joshua S. and Nugent, Peter E. and Kasliwal, Mansi M. and Perley, Daniel A. and Quimby, Robert M. and Miller, Adam A. and Theissen, Christopher A. and Laher, Russ R.},
	month = sep,
	year = {2014},
	note = {Publisher: IOP
ADS Bibcode: 2014ApJ...793...38A},
	pages = {38},
}

@article{makrygianni_double_2025,
	title = {The {Double} {Tidal} {Disruption} {Event} {AT} 2022dbl {Implies} that at {Least} {Some} “{Standard}” {Optical} {Tidal} {Disruption} {Events} {Are} {Partial} {Disruptions}},
	volume = {987},
	issn = {2041-8205},
	url = {https://dx.doi.org/10.3847/2041-8213/ade155},
	doi = {10.3847/2041-8213/ade155},
	language = {en},
	number = {1},
	urldate = {2025-07-14},
	journal = {The Astrophysical Journal Letters},
	author = {Makrygianni, Lydia and Arcavi, Iair and Newsome, Megan and Bandopadhyay, Ananya and Coughlin, Eric R. and Linial, Itai and Mockler, Brenna and Quataert, Eliot and Nixon, Chris and Godson, Benjamin and Pursiainen, Miika and Leloudas, Giorgos and French, K. Decker and Zitrin, Adi and Faris, Sara and Lam, Marco C. and Horesh, Assaf and Sfaradi, Itai and Fausnaugh, Michael and Nakar, Ehud and Ackley, Kendall and Andrews, Moira and Charalampopoulos, Panos and Davies, Benjamin D. R. and Dgany, Yael and Dyer, Martin J. and Farah, Joseph and Fender, Rob and Green, David A. and Howell, D. Andrew and Killestein, Thomas and Koivisto, Niilo and Lyman, Joseph and McCully, Curtis and Mitchell, Morgan A. and Padilla Gonzalez, Estefania and Rhodes, Lauren and Sahu, Anwesha and Terreran, Giacomo and Warwick, Ben},
	month = jul,
	year = {2025},
	note = {Publisher: The American Astronomical Society},
	pages = {L20},
}

@article{soumagnac_most_2024,
	title = {The {MOST} {Hosts} {Survey}: {Spectroscopic} {Observation} of the {Host} {Galaxies} of ∼40,000 {Transients} {Using} {DESI}},
	volume = {275},
	issn = {0067-0049},
	shorttitle = {The {MOST} {Hosts} {Survey}},
	url = {https://ui.adsabs.harvard.edu/abs/2024ApJS..275...22S},
	doi = {10.3847/1538-4365/ad76ae},
	urldate = {2025-05-20},
	journal = {The Astrophysical Journal Supplement Series},
	author = {Soumagnac, Maayane T. and Nugent, Peter and Knop, Robert A. and Ho, Anna Y. Q. and Hohensee, William and Awbrey, Autumn and Andersen, Alexis and Aldering, Greg and Ventura, Matan and Aguilar, Jessica N. and Ahlen, Steven and Benzvi, Segev Y. and Brooks, David and Brout, Dillon and Claybaugh, Todd and Davis, Tamara M. and Dawson, Kyle and de la Macorra, Axel and Dey, Arjun and Dey, Biprateep and Doel, Peter and Douglass, Kelly A. and Forero-Romero, Jaime E. and Gaztañaga, Enrique and Gontcho A Gontcho, Satya and Graur, Or and Guy, Julien and Hahn, ChangHoon and Honscheid, Klaus and Howlett, Cullan and Kim, Alex G. and Kisner, Theodore and Kremin, Anthony and Lambert, Andrew and Landriau, Martin and Lang, Dustin and Le Guillou, Laurent and Manera, Marc and Meisner, Aaron and Miquel, Ramon and Moustakas, John and Myers, Adam D. and Nie, Jundan and Palmese, Antonella and Parkinson, David and Poppett, Claire and Prada, Francisco and Qin, Fei and Rezaie, Mehdi and Rossi, Graziano and Sanchez, Eusebio and Schlegel, David J. and Schubnell, Michael and Silber, Joseph H. and Tarlé, Gregory and Weaver, Benjamin A. and Zhou, Zhimin},
	month = dec,
	year = {2024},
	note = {Publisher: IOP
ADS Bibcode: 2024ApJS..275...22S},
	pages = {22},
}

@misc{frohmaier_tides_2025,
	title = {{TiDES}: {The} {4MOST} {Time} {Domain} {Extragalactic} {Survey}},
	shorttitle = {{TiDES}},
	url = {https://ui.adsabs.harvard.edu/abs/2025arXiv250116311F},
	doi = {10.48550/arXiv.2501.16311},
	urldate = {2025-05-19},
	publisher = {arXiv},
	author = {Frohmaier, C. and Vincenzi, M. and Sullivan, M. and Hönig, S. F. and Smith, M. and Addison, H. and Collett, T. and Dimitriadis, G. and Ellis, R. S. and Gandhi, P. and Graur, O. and Hook, I. and Kelsey, L. and Kim, Y. L. and Lidman, C. and Maguire, K. and Makrygianni, L. and Martin, B. and Möller, A. and Nichol, R. C. and Nicholl, M. and Schady, P. and Simmons, B. D. and Smartt, S. J. and Tempel, E. and Wiseman, P. and {the LSST Dark Energy Science Collaboration}},
	month = jan,
	year = {2025},
	note = {ADS Bibcode: 2025arXiv250116311F},
}

@article{villar_superraenn_2020,
	title = {{SuperRAENN}: {A} {Semisupervised} {Supernova} {Photometric} {Classification} {Pipeline} {Trained} on {Pan}-{STARRS1} {Medium}-{Deep} {Survey} {Supernovae}},
	volume = {905},
	issn = {0004-637X},
	shorttitle = {{SuperRAENN}},
	url = {https://ui.adsabs.harvard.edu/abs/2020ApJ...905...94V},
	doi = {10.3847/1538-4357/abc6fd},
	urldate = {2025-05-19},
	journal = {The Astrophysical Journal},
	author = {Villar, V. Ashley and Hosseinzadeh, Griffin and Berger, Edo and Ntampaka, Michelle and Jones, David O. and Challis, Peter and Chornock, Ryan and Drout, Maria R. and Foley, Ryan J. and Kirshner, Robert P. and Lunnan, Ragnhild and Margutti, Raffaella and Milisavljevic, Dan and Sanders, Nathan and Pan, Yen-Chen and Rest, Armin and Scolnic, Daniel M. and Magnier, Eugene and Metcalfe, Nigel and Wainscoat, Richard and Waters, Christopher},
	month = dec,
	year = {2020},
	note = {Publisher: IOP
ADS Bibcode: 2020ApJ...905...94V},
	pages = {94},
}

@misc{hinkle_double_2024,
	title = {On the {Double}: {Two} {Luminous} {Flares} from the {Nearby} {Tidal} {Disruption} {Event} {ASASSN}-22ci ({AT2022dbl}) and {Connections} to {Repeating} {TDE} {Candidates}},
	shorttitle = {On the {Double}},
	url = {https://ui.adsabs.harvard.edu/abs/2024arXiv241215326H},
	doi = {10.48550/arXiv.2412.15326},
	urldate = {2025-05-13},
	publisher = {arXiv},
	author = {Hinkle, Jason T. and Auchettl, Katie and Hoogendam, Willem B. and Payne, Anna V. and Holoien, Thomas W. -S. and Shappee, Benjamin J. and Tucker, Michael A. and Kochanek, Christopher S. and Stanek, K. Z. and Vallely, Patrick J. and Angus, Charlotte R. and Ashall, Chris and de Jaeger, Thomas and Desai, Dhvanil D. and Do, Aaron and Fausnaugh, Michael M. and Huber, Mark E. and Rickards Vaught, Ryan J. and Shi, Jennifer},
	month = dec,
	year = {2024},
	note = {ADS Bibcode: 2024arXiv241215326H},
}

@article{geha_least-luminous_2009,
	title = {The {Least}-{Luminous} {Galaxy}: {Spectroscopy} of the {Milky} {Way} {Satellite} {Segue} 1},
	volume = {692},
	issn = {0004-637X},
	shorttitle = {The {Least}-{Luminous} {Galaxy}},
	url = {https://ui.adsabs.harvard.edu/abs/2009ApJ...692.1464G},
	doi = {10.1088/0004-637X/692/2/1464},
	urldate = {2025-04-28},
	journal = {The Astrophysical Journal},
	author = {Geha, Marla and Willman, Beth and Simon, Joshua D. and Strigari, Louis E. and Kirby, Evan N. and Law, David R. and Strader, Jay},
	month = feb,
	year = {2009},
	note = {Publisher: IOP
ADS Bibcode: 2009ApJ...692.1464G},
	pages = {1464--1475},
}

@article{aihara_eighth_2011,
	title = {The {Eighth} {Data} {Release} of the {Sloan} {Digital} {Sky} {Survey}: {First} {Data} from {SDSS}-{III}},
	volume = {193},
	issn = {0067-0049},
	shorttitle = {The {Eighth} {Data} {Release} of the {Sloan} {Digital} {Sky} {Survey}},
	url = {https://ui.adsabs.harvard.edu/abs/2011ApJS..193...29A},
	doi = {10.1088/0067-0049/193/2/29},
	urldate = {2025-04-28},
	journal = {The Astrophysical Journal Supplement Series},
	author = {Aihara, Hiroaki and Allende Prieto, Carlos and An, Deokkeun and Anderson, Scott F. and Aubourg, Éric and Balbinot, Eduardo and Beers, Timothy C. and Berlind, Andreas A. and Bickerton, Steven J. and Bizyaev, Dmitry and Blanton, Michael R. and Bochanski, John J. and Bolton, Adam S. and Bovy, Jo and Brandt, W. N. and Brinkmann, J. and Brown, Peter J. and Brownstein, Joel R. and Busca, Nicolas G. and Campbell, Heather and Carr, Michael A. and Chen, Yanmei and Chiappini, Cristina and Comparat, Johan and Connolly, Natalia and Cortes, Marina and Croft, Rupert A. C. and Cuesta, Antonio J. and da Costa, Luiz N. and Davenport, James R. A. and Dawson, Kyle and Dhital, Saurav and Ealet, Anne and Ebelke, Garrett L. and Edmondson, Edward M. and Eisenstein, Daniel J. and Escoffier, Stephanie and Esposito, Massimiliano and Evans, Michael L. and Fan, Xiaohui and Femenía Castellá, Bruno and Font-Ribera, Andreu and Frinchaboy, Peter M. and Ge, Jian and Gillespie, Bruce A. and Gilmore, G. and González Hernández, Jonay I. and Gott, J. Richard and Gould, Andrew and Grebel, Eva K. and Gunn, James E. and Hamilton, Jean-Christophe and Harding, Paul and Harris, David W. and Hawley, Suzanne L. and Hearty, Frederick R. and Ho, Shirley and Hogg, David W. and Holtzman, Jon A. and Honscheid, Klaus and Inada, Naohisa and Ivans, Inese I. and Jiang, Linhua and Johnson, Jennifer A. and Jordan, Cathy and Jordan, Wendell P. and Kazin, Eyal A. and Kirkby, David and Klaene, Mark A. and Knapp, G. R. and Kneib, Jean-Paul and Kochanek, C. S. and Koesterke, Lars and Kollmeier, Juna A. and Kron, Richard G. and Lampeitl, Hubert and Lang, Dustin and Le Goff, Jean-Marc and Lee, Young Sun and Lin, Yen-Ting and Long, Daniel C. and Loomis, Craig P. and Lucatello, Sara and Lundgren, Britt and Lupton, Robert H. and Ma, Zhibo and MacDonald, Nicholas and Mahadevan, Suvrath and Maia, Marcio A. G. and Makler, Martin and Malanushenko, Elena and Malanushenko, Viktor and Mandelbaum, Rachel and Maraston, Claudia and Margala, Daniel and Masters, Karen L. and McBride, Cameron K. and McGehee, Peregrine M. and McGreer, Ian D. and Ménard, Brice and Miralda-Escudé, Jordi and Morrison, Heather L. and Mullally, F. and Muna, Demitri and Munn, Jeffrey A. and Murayama, Hitoshi and Myers, Adam D. and Naugle, Tracy and Neto, Angelo Fausti and Nguyen, Duy Cuong and Nichol, Robert C. and O'Connell, Robert W. and Ogando, Ricardo L. C. and Olmstead, Matthew D. and Oravetz, Daniel J. and Padmanabhan, Nikhil and Palanque-Delabrouille, Nathalie and Pan, Kaike and Pandey, Parul and Pâris, Isabelle and Percival, Will J. and Petitjean, Patrick and Pfaffenberger, Robert and Pforr, Janine and Phleps, Stefanie and Pichon, Christophe and Pieri, Matthew M. and Prada, Francisco and Price-Whelan, Adrian M. and Raddick, M. Jordan and Ramos, Beatriz H. F. and Reylé, Céline and Rich, James and Richards, Gordon T. and Rix, Hans-Walter and Robin, Annie C. and Rocha-Pinto, Helio J. and Rockosi, Constance M. and Roe, Natalie A. and Rollinde, Emmanuel and Ross, Ashley J. and Ross, Nicholas P. and Rossetto, Bruno M. and Sánchez, Ariel G. and Sayres, Conor and Schlegel, David J. and Schlesinger, Katharine J. and Schmidt, Sarah J. and Schneider, Donald P. and Sheldon, Erin and Shu, Yiping and Simmerer, Jennifer and Simmons, Audrey E. and Sivarani, Thirupathi and Snedden, Stephanie A. and Sobeck, Jennifer S. and Steinmetz, Matthias and Strauss, Michael A. and Szalay, Alexander S. and Tanaka, Masayuki and Thakar, Aniruddha R. and Thomas, Daniel and Tinker, Jeremy L. and Tofflemire, Benjamin M. and Tojeiro, Rita and Tremonti, Christy A. and Vandenberg, Jan and Vargas Magaña, M. and Verde, Licia and Vogt, Nicole P. and Wake, David A. and Wang, Ji and Weaver, Benjamin A. and Weinberg, David H. and White, Martin and White, Simon D. M. and Yanny, Brian and Yasuda, Naoki and Yeche, Christophe and Zehavi, Idit},
	month = apr,
	year = {2011},
	note = {Publisher: IOP
ADS Bibcode: 2011ApJS..193...29A},
	pages = {29},
}

@article{gonneau_x-shooter_2020,
	title = {The {X}-shooter {Spectral} {Library} ({XSL}): {Data} release 2},
	volume = {634},
	issn = {0004-6361},
	shorttitle = {The {X}-shooter {Spectral} {Library} ({XSL})},
	url = {https://ui.adsabs.harvard.edu/abs/2020A&A...634A.133G},
	doi = {10.1051/0004-6361/201936825},
	urldate = {2025-04-28},
	journal = {Astronomy and Astrophysics},
	author = {Gonneau, A. and Lyubenova, M. and Lançon, A. and Trager, S. C. and Peletier, R. F. and Arentsen, A. and Chen, Y. -P. and Coelho, P. R. T. and Dries, M. and Falcón-Barroso, J. and Prugniel, P. and Sánchez-Blázquez, P. and Vazdekis, A. and Verro, K.},
	month = feb,
	year = {2020},
	note = {Publisher: EDP
ADS Bibcode: 2020A\&A...634A.133G},
	pages = {A133},
}

@article{cappellari_full_2023,
	title = {Full spectrum fitting with photometry in {PPXF}: stellar population versus dynamical masses, non-parametric star formation history and metallicity for 3200 {LEGA}-{C} galaxies at redshift z ≈ 0.8},
	volume = {526},
	issn = {0035-8711},
	shorttitle = {Full spectrum fitting with photometry in {PPXF}},
	url = {https://ui.adsabs.harvard.edu/abs/2023MNRAS.526.3273C},
	doi = {10.1093/mnras/stad2597},
	urldate = {2025-04-28},
	journal = {Monthly Notices of the Royal Astronomical Society},
	author = {Cappellari, Michele},
	month = dec,
	year = {2023},
	note = {Publisher: OUP
ADS Bibcode: 2023MNRAS.526.3273C},
	pages = {3273--3300},
}

@article{kormendy_coevolution_2013,
	title = {Coevolution ({Or} {Not}) of {Supermassive} {Black} {Holes} and {Host} {Galaxies}},
	volume = {51},
	issn = {0066-4146},
	url = {https://ui.adsabs.harvard.edu/abs/2013ARA&A..51..511K},
	doi = {10.1146/annurev-astro-082708-101811},
	urldate = {2025-04-28},
	journal = {Annual Review of Astronomy and Astrophysics},
	author = {Kormendy, John and Ho, Luis C.},
	month = aug,
	year = {2013},
	note = {ADS Bibcode: 2013ARA\&A..51..511K},
	pages = {511--653},
}

@article{zhong_revisit_2022,
	title = {Revisit the {Rate} of {Tidal} {Disruption} {Events}: {The} {Role} of the {Partial} {Tidal} {Disruption} {Event}},
	volume = {933},
	issn = {0004-637X},
	shorttitle = {Revisit the {Rate} of {Tidal} {Disruption} {Events}},
	url = {https://ui.adsabs.harvard.edu/abs/2022ApJ...933...96Z},
	doi = {10.3847/1538-4357/ac71ad},
	urldate = {2025-04-16},
	journal = {The Astrophysical Journal},
	author = {Zhong, Shiyan and Li, Shuo and Berczik, Peter and Spurzem, Rainer},
	month = jul,
	year = {2022},
	note = {Publisher: IOP
ADS Bibcode: 2022ApJ...933...96Z},
	pages = {96},
}

@article{bortolas_partial_2023,
	title = {Partial stellar tidal disruption events and their rates},
	volume = {524},
	issn = {0035-8711},
	url = {https://ui.adsabs.harvard.edu/abs/2023MNRAS.524.3026B},
	doi = {10.1093/mnras/stad2024},
	urldate = {2025-04-16},
	journal = {Monthly Notices of the Royal Astronomical Society},
	author = {Bortolas, Elisa and Ryu, Taeho and Broggi, Luca and Sesana, Alberto},
	month = sep,
	year = {2023},
	note = {Publisher: OUP
ADS Bibcode: 2023MNRAS.524.3026B},
	pages = {3026--3038},
}

@article{metzger_bright_2016,
	title = {A bright year for tidal disruptions},
	volume = {461},
	issn = {0035-8711},
	url = {https://ui.adsabs.harvard.edu/abs/2016MNRAS.461..948M},
	doi = {10.1093/mnras/stw1394},
	urldate = {2025-04-16},
	journal = {Monthly Notices of the Royal Astronomical Society},
	author = {Metzger, Brian D. and Stone, Nicholas C.},
	month = sep,
	year = {2016},
	note = {Publisher: OUP
ADS Bibcode: 2016MNRAS.461..948M},
	pages = {948--966},
}

@article{lodato_stellar_2009,
	title = {Stellar disruption by a supermassive black hole: is the light curve really proportional to t-5/3?},
	volume = {392},
	issn = {0035-8711},
	shorttitle = {Stellar disruption by a supermassive black hole},
	url = {https://ui.adsabs.harvard.edu/abs/2009MNRAS.392..332L},
	doi = {10.1111/j.1365-2966.2008.14049.x},
	urldate = {2025-04-16},
	journal = {Monthly Notices of the Royal Astronomical Society},
	author = {Lodato, G. and King, A. R. and Pringle, J. E.},
	month = jan,
	year = {2009},
	note = {Publisher: OUP
ADS Bibcode: 2009MNRAS.392..332L},
	pages = {332--340},
}

@article{chen_light_2021,
	title = {Light {Curves} of {Partial} {Tidal} {Disruption} {Events}},
	volume = {914},
	issn = {0004-637X},
	url = {https://ui.adsabs.harvard.edu/abs/2021ApJ...914...69C},
	doi = {10.3847/1538-4357/abf9a7},
	urldate = {2025-04-16},
	journal = {The Astrophysical Journal},
	author = {Chen, Jin-Hong and Shen, Rong-Feng},
	month = jun,
	year = {2021},
	note = {Publisher: IOP
ADS Bibcode: 2021ApJ...914...69C},
	pages = {69},
}

@article{mainetti_fine_2017,
	title = {The fine line between total and partial tidal disruption events},
	volume = {600},
	issn = {0004-6361},
	url = {https://ui.adsabs.harvard.edu/abs/2017A&A...600A.124M},
	doi = {10.1051/0004-6361/201630092},
	urldate = {2025-04-16},
	journal = {Astronomy and Astrophysics},
	author = {Mainetti, Deborah and Lupi, Alessandro and Campana, Sergio and Colpi, Monica and Coughlin, Eric R. and Guillochon, James and Ramirez-Ruiz, Enrico},
	month = apr,
	year = {2017},
	note = {ADS Bibcode: 2017A\&A...600A.124M},
	pages = {A124},
}

@article{sharma_partial_2024,
	title = {Partial tidal disruption events: the elixir of life},
	volume = {532},
	issn = {0035-8711},
	shorttitle = {Partial tidal disruption events},
	url = {https://ui.adsabs.harvard.edu/abs/2024MNRAS.532...89S},
	doi = {10.1093/mnras/stae1455},
	urldate = {2025-04-16},
	journal = {Monthly Notices of the Royal Astronomical Society},
	author = {Sharma, Megha and Price, Daniel J. and Heger, Alexander},
	month = jul,
	year = {2024},
	note = {Publisher: OUP
ADS Bibcode: 2024MNRAS.532...89S},
	pages = {89--111},
}

@article{magnier_pan-starrs_2020,
	title = {The {Pan}-{STARRS} {Data}-processing {System}},
	volume = {251},
	issn = {0067-0049},
	url = {https://ui.adsabs.harvard.edu/abs/2020ApJS..251....3M},
	doi = {10.3847/1538-4365/abb829},
	urldate = {2025-04-13},
	journal = {The Astrophysical Journal Supplement Series},
	author = {Magnier, Eugene A. and Chambers, K. C. and Flewelling, H. A. and Hoblitt, J. C. and Huber, M. E. and Price, P. A. and Sweeney, W. E. and Waters, C. Z. and Denneau, L. and Draper, P. W. and Hodapp, K. W. and Jedicke, R. and Kaiser, N. and Kudritzki, R. -P. and Metcalfe, N. and Stubbs, C. W. and Wainscoat, R. J.},
	month = nov,
	year = {2020},
	note = {Publisher: IOP
ADS Bibcode: 2020ApJS..251....3M},
	pages = {3},
}

@article{tonry_atlas_2018,
	title = {{ATLAS}: {A} {High}-cadence {All}-sky {Survey} {System}},
	volume = {130},
	issn = {0004-6280},
	shorttitle = {{ATLAS}},
	url = {https://ui.adsabs.harvard.edu/abs/2018PASP..130f4505T},
	doi = {10.1088/1538-3873/aabadf},
	urldate = {2025-04-13},
	journal = {Publications of the Astronomical Society of the Pacific},
	author = {Tonry, J. L. and Denneau, L. and Heinze, A. N. and Stalder, B. and Smith, K. W. and Smartt, S. J. and Stubbs, C. W. and Weiland, H. J. and Rest, A.},
	month = jun,
	year = {2018},
	note = {Publisher: IOP
ADS Bibcode: 2018PASP..130f4505T},
	pages = {064505},
}

@article{albareti_13th_2017,
	title = {The 13th {Data} {Release} of the {Sloan} {Digital} {Sky} {Survey}: {First} {Spectroscopic} {Data} from the {SDSS}-{IV} {Survey} {Mapping} {Nearby} {Galaxies} at {Apache} {Point} {Observatory}},
	volume = {233},
	issn = {0067-0049},
	shorttitle = {The 13th {Data} {Release} of the {Sloan} {Digital} {Sky} {Survey}},
	url = {https://ui.adsabs.harvard.edu/abs/2017ApJS..233...25A},
	doi = {10.3847/1538-4365/aa8992},
	urldate = {2025-04-13},
	journal = {The Astrophysical Journal Supplement Series},
	author = {Albareti, Franco D. and Allende Prieto, Carlos and Almeida, Andres and Anders, Friedrich and Anderson, Scott and Andrews, Brett H. and Aragón-Salamanca, Alfonso and Argudo-Fernández, Maria and Armengaud, Eric and Aubourg, Eric and Avila-Reese, Vladimir and Badenes, Carles and Bailey, Stephen and Barbuy, Beatriz and Barger, Kat and Barrera-Ballesteros, Jorge and Bartosz, Curtis and Basu, Sarbani and Bates, Dominic and Battaglia, Giuseppina and Baumgarten, Falk and Baur, Julien and Bautista, Julian and Beers, Timothy C. and Belfiore, Francesco and Bershady, Matthew and Bertran de Lis, Sara and Bird, Jonathan C. and Bizyaev, Dmitry and Blanc, Guillermo A. and Blanton, Michael and Blomqvist, Michael and Bolton, Adam S. and Borissova, J. and Bovy, Jo and Brandt, William Nielsen and Brinkmann, Jonathan and Brownstein, Joel R. and Bundy, Kevin and Burtin, Etienne and Busca, Nicolás G. and Camacho Chavez, Hugo Orlando and Cano Díaz, M. and Cappellari, Michele and Carrera, Ricardo and Chen, Yanping and Cherinka, Brian and Cheung, Edmond and Chiappini, Cristina and Chojnowski, Drew and Chuang, Chia-Hsun and Chung, Haeun and Cirolini, Rafael Fernando and Clerc, Nicolas and Cohen, Roger E. and Comerford, Julia M. and Comparat, Johan and Correa do Nascimento, Janaina and Cousinou, Marie-Claude and Covey, Kevin and Crane, Jeffrey D. and Croft, Rupert and Cunha, Katia and Darling, Jeremy and Davidson, Jr., James W. and Dawson, Kyle and Da Costa, Luiz and Da Silva Ilha, Gabriele and Deconto Machado, Alice and Delubac, Timothée and De Lee, Nathan and De la Macorra, Axel and De la Torre, Sylvain and Diamond-Stanic, Aleksandar M. and Donor, John and Downes, Juan Jose and Drory, Niv and Du, Cheng and Du Mas des Bourboux, Hélion and Dwelly, Tom and Ebelke, Garrett and Eigenbrot, Arthur and Eisenstein, Daniel J. and Elsworth, Yvonne P. and Emsellem, Eric and Eracleous, Michael and Escoffier, Stephanie and Evans, Michael L. and Falcón-Barroso, Jesús and Fan, Xiaohui and Favole, Ginevra and Fernandez-Alvar, Emma and Fernandez-Trincado, J. G. and Feuillet, Diane and Fleming, Scott W. and Font-Ribera, Andreu and Freischlad, Gordon and Frinchaboy, Peter and Fu, Hai and Gao, Yang and Garcia, Rafael A. and Garcia-Dias, R. and Garcia-Hernández, D. A. and Garcia Pérez, Ana E. and Gaulme, Patrick and Ge, Junqiang and Geisler, Douglas and Gillespie, Bruce and Gil Marin, Hector and Girardi, Léo and Goddard, Daniel and Gomez Maqueo Chew, Yilen and Gonzalez-Perez, Violeta and Grabowski, Kathleen and Green, Paul and Grier, Catherine J. and Grier, Thomas and Guo, Hong and Guy, Julien and Hagen, Alex and Hall, Matt and Harding, Paul and Harley, R. E. and Hasselquist, Sten and Hawley, Suzanne and Hayes, Christian R. and Hearty, Fred and Hekker, Saskia and Hernandez Toledo, Hector and Ho, Shirley and Hogg, David W. and Holley-Bockelmann, Kelly and Holtzman, Jon A. and Holzer, Parker H. and Hu, Jian and Huber, Daniel and Hutchinson, Timothy Alan and Hwang, Ho Seong and Ibarra-Medel, Héctor J. and Ivans, Inese I. and Ivory, KeShawn and Jaehnig, Kurt and Jensen, Trey W. and Johnson, Jennifer A. and Jones, Amy and Jullo, Eric and Kallinger, T. and Kinemuchi, Karen and Kirkby, David and Klaene, Mark and Kneib, Jean-Paul and Kollmeier, Juna A. and Lacerna, Ivan and Lane, Richard R. and Lang, Dustin and Laurent, Pierre and Law, David R. and Leauthaud, Alexie and Le Goff, Jean-Marc and Li, Chen and Li, Cheng and Li, Niu and Li, Ran and Liang, Fu-Heng and Liang, Yu and Lima, Marcos and Lin, Lihwai and Lin, Lin and Lin, Yen-Ting and Liu, Chao and Long, Dan and Lucatello, Sara and MacDonald, Nicholas and MacLeod, Chelsea L. and Mackereth, J. Ted and Mahadevan, Suvrath and Maia, Marcio Antonio Geimba and Maiolino, Roberto and Majewski, Steven R. and Malanushenko, Olena and Malanushenko, Viktor and Mallmann, Nícolas Dullius and Manchado, Arturo and Maraston, Claudia and Marques-Chaves, Rui and Martinez Valpuesta, Inma and Masters, Karen L. and Mathur, Savita and McGreer, Ian D. and Merloni, Andrea and Merrifield, Michael R. and Mészáros, Szabolcs and Meza, Andres and Miglio, Andrea and Minchev, Ivan and Molaverdikhani, Karan and Montero-Dorta, Antonio D. and Mosser, Benoit and Muna, Demitri and Myers, Adam},
	month = dec,
	year = {2017},
	note = {Publisher: IOP
ADS Bibcode: 2017ApJS..233...25A},
	pages = {25},
}

@article{nicholl_at_2023,
	title = {{AT} 2022aedm and a {New} {Class} of {Luminous}, {Fast}-cooling {Transients} in {Elliptical} {Galaxies}},
	volume = {954},
	issn = {0004-637X},
	url = {https://ui.adsabs.harvard.edu/abs/2023ApJ...954L..28N},
	doi = {10.3847/2041-8213/acf0ba},
	urldate = {2025-04-13},
	journal = {The Astrophysical Journal},
	author = {Nicholl, M. and Srivastav, S. and Fulton, M. D. and Gomez, S. and Huber, M. E. and Oates, S. R. and Ramsden, P. and Rhodes, L. and Smartt, S. J. and Smith, K. W. and Aamer, A. and Anderson, J. P. and Bauer, F. E. and Berger, E. and de Boer, T. and Chambers, K. C. and Charalampopoulos, P. and Chen, T. -W. and Fender, R. P. and Fraser, M. and Gao, H. and Green, D. A. and Galbany, L. and Gompertz, B. P. and Gromadzki, M. and Gutiérrez, C. P. and Howell, D. A. and Inserra, C. and Jonker, P. G. and Kopsacheili, M. and Lowe, T. B. and Magnier, E. A. and McCully, C. and McGee, S. L. and Moore, T. and Müller-Bravo, T. E. and Newsome, M. and Gonzalez, E. Padilla and Pellegrino, C. and Pessi, T. and Pursiainen, M. and Rest, A. and Ridley, E. J. and Shappee, B. J. and Sheng, X. and Smith, G. P. and Terreran, G. and Tucker, M. A. and Vinkó, J. and Wainscoat, R. J. and Wiseman, P. and Young, D. R.},
	month = sep,
	year = {2023},
	note = {Publisher: IOP
ADS Bibcode: 2023ApJ...954L..28N},
	pages = {L28},
}

@article{hung_revisiting_2017,
	title = {Revisiting {Optical} {Tidal} {Disruption} {Events} with {iPTF16axa}},
	volume = {842},
	issn = {0004-637X},
	url = {https://ui.adsabs.harvard.edu/abs/2017ApJ...842...29H},
	doi = {10.3847/1538-4357/aa7337},
	urldate = {2025-04-13},
	journal = {The Astrophysical Journal},
	author = {Hung, T. and Gezari, S. and Blagorodnova, N. and Roth, N. and Cenko, S. B. and Kulkarni, S. R. and Horesh, A. and Arcavi, I. and McCully, C. and Yan, Lin and Lunnan, R. and Fremling, C. and Cao, Y. and Nugent, P. E. and Wozniak, P.},
	month = jun,
	year = {2017},
	note = {Publisher: IOP
ADS Bibcode: 2017ApJ...842...29H},
	pages = {29},
}

@article{bonnerot_long-term_2017,
	title = {Long-term stream evolution in tidal disruption events},
	volume = {464},
	issn = {0035-8711},
	url = {https://ui.adsabs.harvard.edu/abs/2017MNRAS.464.2816B},
	doi = {10.1093/mnras/stw2547},
	urldate = {2025-04-13},
	journal = {Monthly Notices of the Royal Astronomical Society},
	author = {Bonnerot, Clément and Rossi, Elena M. and Lodato, Giuseppe},
	month = jan,
	year = {2017},
	note = {Publisher: OUP
ADS Bibcode: 2017MNRAS.464.2816B},
	pages = {2816--2830},
}

@inproceedings{phinney_manifestations_1989,
	title = {Manifestations of a {Massive} {Black} {Hole} in the {Galactic} {Center}},
	volume = {136},
	url = {https://ui.adsabs.harvard.edu/abs/1989IAUS..136..543P},
	urldate = {2025-04-13},
	author = {Phinney, E. S.},
	month = jan,
	year = {1989},
	note = {ADS Bibcode: 1989IAUS..136..543P},
	pages = {543},
}

@article{uno_application_2020,
	title = {Application of {The} {Wind}-driven {Model} to a {Sample} of {Tidal} {Disruption} {Events}},
	volume = {905},
	issn = {0004-637X},
	url = {https://ui.adsabs.harvard.edu/abs/2020ApJ...905L...5U},
	doi = {10.3847/2041-8213/abca32},
	urldate = {2025-04-13},
	journal = {The Astrophysical Journal},
	author = {Uno, Kohki and Maeda, Keiichi},
	month = dec,
	year = {2020},
	note = {Publisher: IOP
ADS Bibcode: 2020ApJ...905L...5U},
	pages = {L5},
}

@article{van_velzen_radio_2016,
	title = {A radio jet from the optical and x-ray bright stellar tidal disruption flare {ASASSN}-14li},
	volume = {351},
	issn = {0036-8075},
	url = {https://ui.adsabs.harvard.edu/abs/2016Sci...351...62V},
	doi = {10.1126/science.aad1182},
	urldate = {2025-04-11},
	journal = {Science},
	author = {van Velzen, S. and Anderson, G. E. and Stone, N. C. and Fraser, M. and Wevers, T. and Metzger, B. D. and Jonker, P. G. and van der Horst, A. J. and Staley, T. D. and Mendez, A. J. and Miller-Jones, J. C. A. and Hodgkin, S. T. and Campbell, H. C. and Fender, R. P.},
	month = jan,
	year = {2016},
	note = {ADS Bibcode: 2016Sci...351...62V},
	pages = {62--65},
}

@article{lu_self-intersection_2020,
	title = {Self-intersection of the fallback stream in tidal disruption events},
	volume = {492},
	issn = {0035-8711},
	url = {https://ui.adsabs.harvard.edu/abs/2020MNRAS.492..686L},
	doi = {10.1093/mnras/stz3405},
	urldate = {2025-04-11},
	journal = {Monthly Notices of the Royal Astronomical Society},
	author = {Lu, Wenbin and Bonnerot, Clément},
	month = feb,
	year = {2020},
	note = {Publisher: OUP
ADS Bibcode: 2020MNRAS.492..686L},
	pages = {686--707},
}

@article{alexander_discovery_2016,
	title = {Discovery of an {Outflow} from {Radio} {Observations} of the {Tidal} {Disruption} {Event} {ASASSN}-14li},
	volume = {819},
	issn = {0004-637X},
	url = {https://ui.adsabs.harvard.edu/abs/2016ApJ...819L..25A},
	doi = {10.3847/2041-8205/819/2/L25},
	urldate = {2025-04-11},
	journal = {The Astrophysical Journal},
	author = {Alexander, K. D. and Berger, E. and Guillochon, J. and Zauderer, B. A. and Williams, P. K. G.},
	month = mar,
	year = {2016},
	note = {Publisher: IOP
ADS Bibcode: 2016ApJ...819L..25A},
	pages = {L25},
}

@article{roth_x-ray_2016,
	title = {The {X}-{Ray} through {Optical} {Fluxes} and {Line} {Strengths} of {Tidal} {Disruption} {Events}},
	volume = {827},
	issn = {0004-637X},
	url = {https://ui.adsabs.harvard.edu/abs/2016ApJ...827....3R},
	doi = {10.3847/0004-637X/827/1/3},
	urldate = {2025-04-11},
	journal = {The Astrophysical Journal},
	author = {Roth, Nathaniel and Kasen, Daniel and Guillochon, James and Ramirez-Ruiz, Enrico},
	month = aug,
	year = {2016},
	note = {Publisher: IOP
ADS Bibcode: 2016ApJ...827....3R},
	pages = {3},
}

@article{wevers_live_2023,
	title = {Live to {Die} {Another} {Day}: {The} {Rebrightening} of {AT} 2018fyk as a {Repeating} {Partial} {Tidal} {Disruption} {Event}},
	volume = {942},
	issn = {0004-637X},
	shorttitle = {Live to {Die} {Another} {Day}},
	url = {https://ui.adsabs.harvard.edu/abs/2023ApJ...942L..33W},
	doi = {10.3847/2041-8213/ac9f36},
	urldate = {2025-04-11},
	journal = {The Astrophysical Journal},
	author = {Wevers, T. and Coughlin, E. R. and Pasham, D. R. and Guolo, M. and Sun, Y. and Wen, S. and Jonker, P. G. and Zabludoff, A. and Malyali, A. and Arcodia, R. and Liu, Z. and Merloni, A. and Rau, A. and Grotova, I. and Short, P. and Cao, Z.},
	month = jan,
	year = {2023},
	note = {Publisher: IOP
ADS Bibcode: 2023ApJ...942L..33W},
	pages = {L33},
}

@misc{somalwar_first_2023,
	title = {The first systematically identified repeating partial tidal disruption event},
	url = {https://ui.adsabs.harvard.edu/abs/2023arXiv231003782S},
	doi = {10.48550/arXiv.2310.03782},
	urldate = {2025-04-11},
	publisher = {arXiv},
	author = {Somalwar, Jean J. and Ravi, Vikram and Yao, Yuhan and Guolo, Muryel and Graham, Matthew and Hammerstein, Erica and Lu, Wenbin and Nicholl, Matt and Sharma, Yashvi and Stein, Robert and van Velzen, Sjoert and Bellm, Eric C. and Coughlin, Michael W. and Groom, Steven L. and Masci, Frank J. and Riddle, Reed},
	month = oct,
	year = {2023},
	note = {ADS Bibcode: 2023arXiv231003782S},
}

@article{lin_unluckiest_2024,
	title = {The {Unluckiest} {Star}: {A} {Spectroscopically} {Confirmed} {Repeated} {Partial} {Tidal} {Disruption} {Event} {AT} 2022dbl},
	volume = {971},
	issn = {0004-637X},
	shorttitle = {The {Unluckiest} {Star}},
	url = {https://ui.adsabs.harvard.edu/abs/2024ApJ...971L..26L},
	doi = {10.3847/2041-8213/ad638e},
	urldate = {2025-04-11},
	journal = {The Astrophysical Journal},
	author = {Lin, Zheyu and Jiang, Ning and Wang, Tinggui and Kong, Xu and Li, Dongyue and He, Han and Wang, Yibo and Zhu, Jiazheng and Li, Wentao and Jiang, Ji-an and Singh, Avinash and Teja, Rishabh Singh and Sahu, D. K. and Jin, Chichuan and Maeda, Keiichi and Huang, Shifeng},
	month = aug,
	year = {2024},
	note = {Publisher: IOP
ADS Bibcode: 2024ApJ...971L..26L},
	pages = {L26},
}




\appendix

\begin{table}
    \centering
    \begin{threeparttable}
    \begin{tabular}{c|c|c|c}
    \hline
         Date & Phase & Telescope/Inst. & R ($\lambda/\Delta\lambda$) \\
         \hline
         25-01-2023 & -1 d~\tnote{\emph{a}} & P60/SEDM & 100 \\
         09-02-2023 & +15 d~\tnote{\emph{a}} & P60/SEDM & 100 \\ 
         11-02-2024 & -21 d~\tnote{\emph{b}} & Keck/LRIS & 5000\\
         09-03-2024 & +2 d~\tnote{\emph{b}} & LDT/DeVeny  & 2400\\
         07-07-2024 & +108 d~\tnote{\emph{b}} & Keck/LRIS & 5000\\
         \hline
    \end{tabular}
    \begin{tablenotes}
      \footnotesize
      \item[\emph{a}]{Measured with respect to the first peak.}
      \item[\emph{b}]{Measured with respect to the second peak.}
\end{tablenotes}
\end{threeparttable}
    \caption{Breakdown of the spectroscopic observations obtained of TDE\,2023adr, and presented in Figure \ref{fig:23ar_class}.}
    \label{tab:spec_obs_23adr}
\end{table}

\begin{figure}
	\includegraphics[width=1.0\columnwidth]{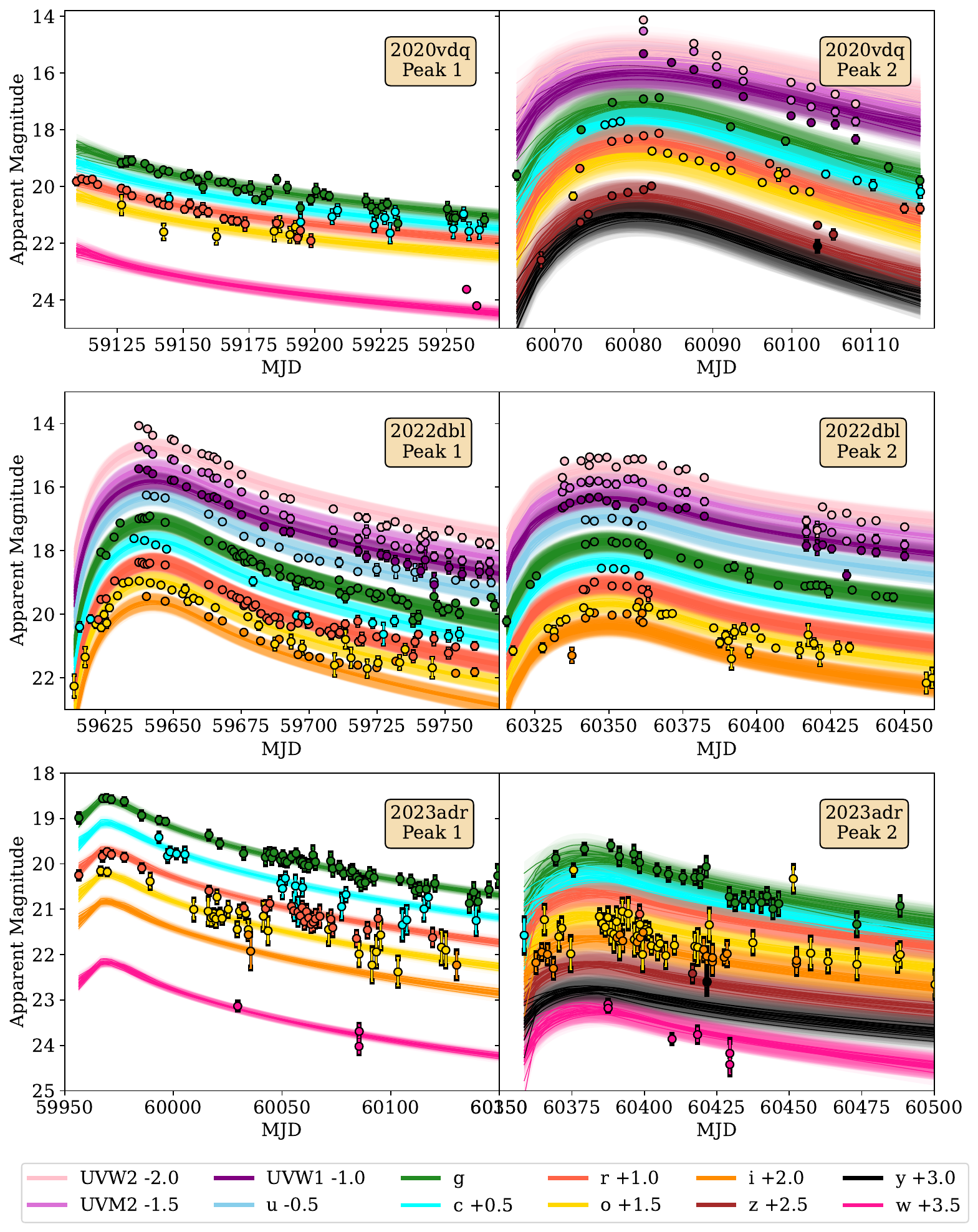}\\
    \caption{Light curve fits from {\tt{MOSFiT}} for the rpTDE flares. Bands are offset in magnitude space for clarity. }
    \label{fig:MOSFiT_LCs}
\end{figure}

\begin{figure}
	\includegraphics[width=1.0\columnwidth]{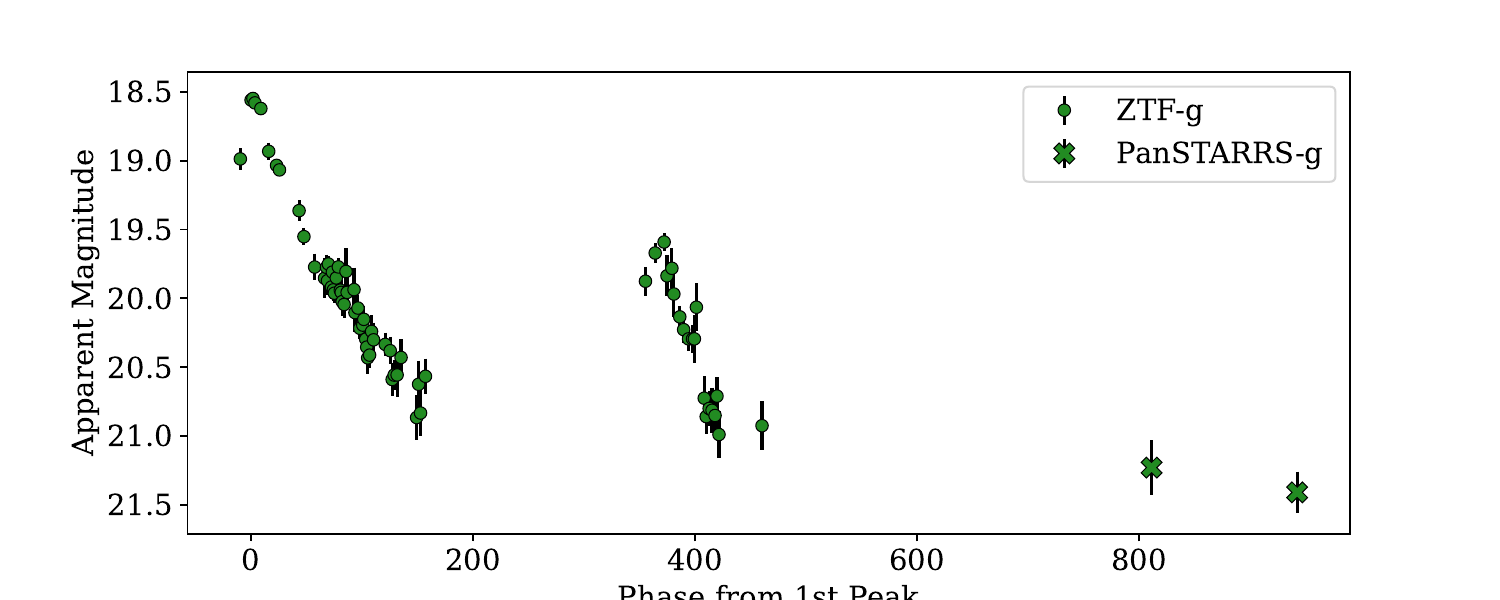}\\
    \caption{$g$-band light curve of TDE\,2023adr, plotted in the rest frame with respect to the first peak. The late-time $g$-band detections from PanSTARRS at +811 and +942 days from which we measure the plateau luminosity are marked with a cross. The observed luminosity appears to be consistent with the second flare having faded, although whether the TDE has truly reached its plateau luminosity is not clear.  }
    \label{fig:23adr_plateau}
\end{figure}

\bsp	
\label{lastpage}
\end{document}